\pdfoutput=1

\documentclass[%
 aip,
 pop,%
 amsmath,amssymb,
 reprint,%
]{revtex4-1}

\usepackage{graphicx}% Include figure files
\usepackage{dcolumn}% Align table columns on decimal point
\usepackage{bm}% bold math
\usepackage{caption}
\usepackage{subcaption}
\usepackage{amsfonts}

\usepackage{verbatim}

\begin{document}

\title[Collisionless plasma equilibria for twisted flux tubes]{Neutral and non-neutral collisionless plasma equilibria for twisted flux tubes:\\The Gold-Hoyle model in a background field}

\author{O. Allanson}
\email{oliver.allanson@st-andrews.ac.uk}
\author{F. Wilson}
%\email{fw237@st-andrews.ac.uk}
\author{T. Neukirch}
%\email{tn3@st-andrews.ac.uk}
\affiliation{School of Mathematics \& Statistics, University of St Andrews, United Kingdom, KY16 9SS}

\date{$17^{th}$ August 2016}% It is always \today, today,
             %  but any date may be explicitly specified

\title{Neutral and non-neutral collisionless plasma equilibria for twisted flux tubes:\\The Gold-Hoyle model in a background field} %Title of paper

\begin{abstract}
We calculate exact one-dimensional collisionless plasma equilibria for a continuum of flux tube models, for which the total magnetic field is made up of the `force-free' Gold-Hoyle magnetic flux tube embedded in a uniform and anti-parallel background magnetic field. For a sufficiently weak background magnetic field, the axial component of the total magnetic field reverses at some finite radius. The presence of the background magnetic field means that the total system is not exactly force-free, but by reducing its magnitude the departure from force-free can be made as small as desired. The distribution function for each species is a function of the three constants of motion; namely the Hamiltonian and the canonical momenta in the axial and azimuthal directions. Poisson's Equation and Amp\`{e}re's Law are solved exactly, and the solution allows either electrically neutral or non-neutral configurations, depending on the values of the bulk ion and electron flows. These equilibria have possible applications in various solar, space and astrophysical contexts, as well as in the laboratory.
\end{abstract}

\pacs{}% insert suggested PACS numbers in braces on next line

\maketitle %\maketitle must follow title, authors, abstract and \pacs

\section{Introduction}
There has been significant recent work on Vlasov-Maxwell (VM) equilibria that are consistent with nonlinear force-free \cite{Harrison-2009a, Harrison-2009b, Neukirch-2009, Wilson-2011, Abraham-Shrauner-2013, Kolotkov-2015, Allanson-2015PoP, Allanson-2016} and `nearly force-free' \cite{Artemyev-2011} magnetic fields in Cartesian geometry. Therein, force-free refers to a magnetic field for which the associated current density is exactly parallel, which is the definition we shall also use,
\begin{eqnarray}
\nabla\times\mathbf{B}&=&\mu_{0}\mathbf{j},\nonumber\\
\mathbf{j}\times\mathbf{B}&=&\mathbf{0}.\nonumber
\end{eqnarray}
These works consider one-dimensional (1D) collisionless current sheets, with Refs. \onlinecite{Harrison-2009a, Harrison-2009b, Neukirch-2009, Wilson-2011, Abraham-Shrauner-2013, Kolotkov-2015, Allanson-2015PoP, Allanson-2016} specifically calculating VM equilibrium distribution functions (DFs) that are self-consistent with a given specific magnetic field configuration. A natural question to consider is whether it is also possible to find self-consistent force-free (or nearly force-free) VM equilibria for other geometries, in particular cylindrical geometry. In this paper we shall present particular VM equilibria for 1D magnetic fields which are nearly force-free in cylindrical geometry, i.e. flux tubes/ropes 

Two of the archetypal field configurations in cylindrical geometry are the $z$-Pinch (with axial current and azimuthal magnetic field), a classical example of which is the Bennett Pinch\cite{Bennett-1934}; and the $\theta$-Pinch (azimuthal current and axial magnetic field). Consideration of `Vlasov-fluid' models of $z$-Pinch equilibria was given in Ref. \onlinecite{Channon-2001}, with Ref. \onlinecite{Mahajan-1989} calculating $z$-Pinch equilibria and an extension with azimuthal ion-currents. Others have also constructed kinetic models of the $\theta$-pinch, see Refs. \onlinecite{Nicholson-1963, Batchelor-1975} for examples. In the same year as Pfirsch \cite{Pfirsch-1962}, cylindrical kinetic equilibria with only azimuthal current were studied in Ref. \onlinecite{Komarov-1962} . For examples of treatments of the stability of fluid and kinetic linear pinches, see Refs \onlinecite{Newcomb-1960,Pfirsch-1962,Davidsonbook} respectively. 

Recently there have been studies on `tokamak-like' VM equilibria with flows \cite{Tasso-2007, Tasso-2014,Kuiroukidis-2015}, starting from the VM equation in cylindrical geometry and working towards Grad-Shafranov equations for the vector potential. We also note two Vlasov equilibrium DFs in the literature that are close in style to the one that we shall present. The first is described in a brief paper \cite{El-Nadi-1976}, with an equilibrium presented for a cylindrical pinch. However, their distribution describes a different magnetic field and the DF appears not to be positive over all phase space. The second DF is a very recent paper \cite{Vinogradov-2016} that actually describes a magnetic field much like the one that we discuss. Their DF is designed to model `ion-scale' flux tubes in the Earth's magnetosphere. Formally, their quasineutral model approaches a nonlinear force-free configuration in the limit of a vanishing electron to ion mass ratio. In their model, current is carried exclusively by electrons and the non-negativity of the DF depends on a suitable choice of microscopic parameters. Finally, we mention that in beam physics, much work on constructing cylindrical VM equilibria is done by looking for mono-energetic distributions with conserved angular momentum, see Refs. \onlinecite{Gratreau-1977, Uhm-1985, Hammer-1970, Morozov-1961} for some examples. 

Magnetic flux tubes and flux ropes are prevelant in the study of plasmas, with a wide variety of observed forms in nature and experiment, as well as uses and applications in numerical experiments and theory. Some examples of the environments and fields of study in which they feature include solar \cite{Priest-2002,Magara-2003}; solar wind \cite{Wangflux-1990, Borovsky-2008}; planetary magnetospheres \cite{Sato-1986, Pontius-1990} and magnetopauses \cite{Cowley-1989}; astrophysical plasmas \cite{Rogava-2000, Li-2006}; tokamak \cite{Bottino-2007, Ham-2016}, laboratory pinch experiments \cite{Rudakov-2000} and the basic study of energy release in magnetised plasmas \cite{Cowley-2015}, to give a small selection of references. 

One application of flux tubes is in the study of solar active regions \cite{Fan-2009} and the onset of solar flares and coronal mass ejections \cite{Torok-2003, Titov-2003,Hood-2016}. A classic magnetohydrodynamic (MHD) model for magnetic flux tubes was first presented by T. Gold and F. Hoyle (GH) \cite{Gold-1960}, initially intended for use in the study of solar flares. The GH model is an infinite, straight, 1D and nonlinear force-free magnetic flux tube with constant `twist'\cite{Birn-2007}. Mathematically, the GH magnetic field could be regarded as the cylindrical analogue\cite{Tassi-2008} of the Force-Free Harris sheet\cite{Harrison-2009b} (a planar current sheet model), as the Bennett Pinch  \cite{Bennett-1934} might be to the `original' Harris Sheet \cite{Harris-1962}.  

%Recent numerical modelling of `magnetohydrodynamic (MHD) avalanches' in the low-beta solar corona has used multiple flux ropes embedded in a uniform background magnetic field\cite{Hood-2016}. The magnetic field model used\cite{Hood-2009} is force-free and one-dimensional. 

It is typical to consider solar, space and astrophysical flux tubes within the framework of MHD, e.g. see Ref. \onlinecite{Priest-2014}. However, many of these plasmas can be weakly collisional or collisionless, with values of the collisional free path large against any fluid scale \cite{Marsh-2006}, making a description using collisionless kinetic theory necessary. It is the intention of this paper to study the GH flux tube model beyond the MHD description, since - apart from the very recent work in Ref. \onlinecite{Vinogradov-2016} -  we see no attempt in the literature of a microscopic description of the GH field. Other than any interesting theoretical advances, a possible application of the results of this study could be to implement the obtained model in kinetic (particle) numerical simulations. 

%In this paper, we work with collisionless kinetic theory, calculating Vlasov-Maxwell (VM) equilibria for a specific flux tube model, namely the GH field with a constant magnitude anti-parallel background field. A VM equilibrium is a fully self-consistent solution of the steady-state Vlasov equation and Maxwell's equations. There has been significant recent progress in the theory of collisionless VM equilibria for nonlinear force-free fields (see Refs. \onlinecite{Harrison-2009a, Harrison-2009b, Neukirch-2009, Wilson-2011, Abraham-Shrauner-2013, Kolotkov-2015, Allanson-2015PoP, Allanson-2016}), and `nearly force-free' fields (see Ref. \onlinecite{Artemyev-2011}) in Cartesian geometry. 

In Cartesian geometry, the work in Refs. \onlinecite{Harrison-2009a, Harrison-2009b, Neukirch-2009, Wilson-2011, Abraham-Shrauner-2013, Kolotkov-2015, Allanson-2015PoP, Allanson-2016} used the method proposed by Channell\cite{Channell-1976} to tackle the VM inverse problem, i.e. to determine self-consistent equilibrium DFs for a given magnetic field configuration. Channell described the extension of his work to cylindrical geometry as `not possible in a straightforward manner.' As explained in Ref. \onlinecite{Tasso-2014} (in which cylindrical coordinates are used to model a torus), this is due in part to the `toroidicity' of the problem, i.e. the $1/r$ factor in the equations. As we shall see in this paper, another potential complication is the need to allow -- at least in principle -- a non-zero charge density. The work in this paper does not present a generalised method for the VM inverse problem in cylindrical geometry, but instead some particular solutions for a specific given magnetic field.

%The references that derive force-free equilibria all used methods similar in spirit to those put forward in Ref. \onlinecite{Channell-1976}. Therein, Channell specifically tackled the `inverse problem in collisionless equilibria', namely that of finding an equilibrium distribution function (DF) for a given macroscopic configuration. He described the extension of this work to cylindrical geometry as `not possible in a straightforward manner.' As explained in Ref. \onlinecite{Tasso-2014}, this is due in part to the `toroidicity' of the problem, i.e. the $1/r$ factor in the equations. We see no attempt in the literature of a microscopic description of the GH field itself, nor of a formalised, general approach to the `inverse problem' of deriving an equilibrium distribution for a given magnetic field in cylindrical geometry. The work in this paper was motivated by the `challenge' set by Channell.  

The paper is structured as follows. In Section 2 we first review the theory of the equation of motion consistent with a collisionless DF in cylindrical geometry, and discuss the question of the possibility of 1D force-free equilibria. Then we introduce the magnetic field to be used. We note that whilst the work in this paper is applied to a particular magnetic field from Subsection 2.1 onwards, the steps taken to calculate the equilibrium DF seem as though they could be adaptable to other cases. In Section 3 we present the form of the DF that gives the required macroscopic equilibrium, and proceed to `fix' the parameters of the DF by explicitly solving Amp\`{e}re's Law and Poisson's Equation. Note that whilst we choose to consider a two-species plasma of ions and electrons, we see no obvious reason preventing the work in this paper being used to describe plasmas with a different composition. In Section 4. we present a preliminary analysis of the physical properties of the equilibrium. Particularly technical calculations are in the Appendices. Appendix A contains the zeroth and first order moment calculations, used to find the number densities and bulk flows directly, and in turn the charge and current densities. Appendix B contains the mathematical details of the existence and location of multiple maxima of the DF in velocity-space.

\section{General theory}
\subsection{The Vlasov equation and the equation of motion}
A collisionless equilibrium is characterised by the 1-particle distribution function, $f_s$, a solution of the steady-state Vlasov equation (e.g. see Ref. \onlinecite{Schindlerbook}). The Vlasov equation in cylindrical coordinates is
\begin{eqnarray}
&&\frac{\partial f_s}{\partial t}+v_{i}\frac{\partial f_s}{\partial x_i}+\frac{q_s}{m_s}\left(E^i+\varepsilon^{ijk}v_{j}B_k\right)\frac{\partial f_s}{\partial v^{i}}\nonumber\\
&&+\left[\frac{v_{\theta}^2}{r}\frac{\partial f_s}{\partial v_{r}}-\frac{v_{r}v_{\theta}}{r}\frac{\partial f_s}{\partial v_{\theta}}     \right]=0,\label{eq:vlasov}
\end{eqnarray}
see for example Refs. \onlinecite{Komarov-1962, Santini-1970, Tasso-2007}. Here $i,j$ and $k$ are used as `spatial' indices running over $\{ 1,2,3\}$, and $s$ is used as the particle species index. Individual particle positions and velocities are given by $(x^1,x^2,x^3)=(r,\theta,z)$ and $(v_{1},v_{2},v_{3})=(v_{r},v_{\theta},v_{z})$ respectively, for $r$ the horizontal distance from the $z$ axis, and $\theta$ the azimuthal angle. The totally antisymmetric unit tensor of rank 3 (the Levi-Civita tensor) is $\varepsilon^{ijk}$, and the Einstein summation convention is applied (such that repeated indices are summed over, with subscript and superscript indices used to describe co- and contravariant components respectively. \begin{comment}, with the cylindrical metric $g_{ij}$, such that $d\mathbf{x}^2=g_{ij}dx^idx^j$, used to raise and lower indices:
\begin{eqnarray}
\underline{\underline{g}}=\left(\begin{matrix}
  1 & 0 & 0 \\
  0 & r^2 & 0 \\
  0 & 0 & 1
 \end{matrix}\right).\nonumber
\end{eqnarray}
\end{comment}
The mass and charge of particle species $s$ are $m_s$ and $q_s$ respectively. The electric and magnetic fields are defined as $\mathbf{E}=-\nabla\phi$ and $\mathbf{B}=\nabla\times\mathbf{A}$, for $\phi$ the scalar potential and vector potential $\mathbf {A}$. 

The `fluid' equation of motion of a particular species $s$ is found by taking first-order velocity moments of the Vlasov equation.
After a routine but laborious moment-taking calculation, we see that - in equilibrium ($\partial /\partial t=0$), assuming a one-dimensional configuration with only radial dependence ($\partial /\partial \theta=\partial /\partial z=0)$, and letting $f_s$ be an even function of the radial velocity $v_r$  -  force balance for species $s$ is maintained according to
\begin{equation}
(\nabla\cdot\mathbf{P}_s)_r=(\mathbf{j}_s\times\mathbf{B})_r+\sigma_s \mathbf{E}+\frac{\rho_s}{r}u_{\theta s}^2.\label{eq:species}
\end{equation}
The pressure tensor for species $s$ is a rank-2 tensor and is defined by
\begin{equation*}
P_{ij,s}=\int w_{is}\,w_{js}\,f_s\,d^3v,
\end{equation*} 
where $v_{i}=u_{is}+w_{is}$, for $u_{is}$ the bulk velocity of species $s$ and $v_{i}$ the individual particle velocity. Note that the assumption of $f_{s}$ to be an even function of $v_{r}$ automatically implies that $u_{rs}=P_{r\theta}=P_{zr}=0$. Equation (\ref{eq:species}) can be summed over species to give
\begin{equation}
(\nabla\cdot\mathbf{P})_r=(\mathbf{j}\times\mathbf{B})_r+\sigma \mathbf{E}+\frac{1}{r}\mathcal{F}_{\text{c}}, \label{eq:CMHDE_alt}
\end{equation}
where
\begin{equation*}
\mathcal{F}_{\text{c}}=\rho_iu_{\theta i}^2+\rho_eu_{\theta e}^2
\end{equation*}
is the force density associated with the rotating bulk flows of the ions and electrons. Equation (\ref{eq:CMHDE_alt}) is a cylindrical analogue of the force balance equation in Cartesian geometry (e.g. see \onlinecite{Mynick-1979a}). There are `extra inertial terms' as compared to the case of Cartesian geometry. From the point of view of a particular magnetic field $\mathbf{B}$ (which is the point we take by specifying a particular macroscopic equilibrium), we see that equilibrium is maintained by a combination of density/pressure variations as in the case of Cartesian geometry, but with additional contributions from centrifugal forces and as an inevitable result of the resultant charge separation, an electric field. This clearly demonstrates that `sourcing' an exactly force-free macroscopic equilibrium with an equilibrium DF in a 1D cylindrical geometry is inherently a more difficult task than in the Cartesian case. The presence of `extra' positive definite inertial forces and, almost inevitably, forces associated with charge separation raises the question of whether exactly force-free equilibria are possible at all in this paradigm.

Before proceeding, we comment that given certain macroscopic constraints on the electromagnetic fields or fluid quantities - such as the force-free condition, or a specific given magnetic field (for example) - it is not \emph{a priori} known how to calculate a self-consistent Vlasov equilibrium, or if one even exists within the framework of the assumptions made. Hence one has to proceed more or less on a case by case basis, with the intention of achieving consistency with the required macroscopic conditions, upon taking moments of the DF.

\subsection{Methods for calculating an equilibrium DF}
In Refs. \onlinecite{Channell-1976, Harrison-2009b} for example, a method used to calculate a DF, given a prescribed 1D magnetic field was Inverse Fourier Transforms (IFT). A distribution function of the form 
\begin{equation}
f_s\propto e^{-\beta_sH_s}g_s(p_{xs},p_{ys}),
\end{equation}
was used, with $H_{s}$, $p_{xs}$ and $p_{ys}$ the conserved particle Hamiltonian and canonical momenta in the $x$ and $y$ directions, and $g_s$ an unknown function, to be determined. Since our problem is one of a 1D equilibrium with variation in the radial direction, the three constants of motion are the Hamiltonian, and the canonical momenta in the $\theta$ and $z$ directions:
\begin{eqnarray}
&&\mathcal{H}_s=\frac{m_s}{2}\left(v_{r}^2+v_{\theta}^2+v_{z}^2\right)+q_s\phi,\nonumber\\
&&p_{\theta s}=r\left(m_sv_{\theta}+q_sA_{\theta}\right),\hspace{3mm} p_{zs}=m_sv_{z}+q_sA_z.
\end{eqnarray}
A function of a subset of the constants of motion is automatically a solution of the VM equation (e.g. see Ref. \onlinecite{Schindlerbook}). One can try to calculate an equilibrium distribution for the Gold-Hoyle force-free flux tube without a background field by a similar method, assuming a DF of the form
\begin{equation}
f_s\propto e^{-\beta_sH_s}g_s(p_{\theta s},p_{zs}). \label{eq:fansatz}
\end{equation}
By exploiting the convolution in the definition of the current density, 
\begin{eqnarray}
\mathbf{j}(\mathbf{A},r)&=&\sum_sq_s\int \,\mathbf{v}\,f_s(H_s,p_{\theta s},p_{zs})\,d^3v,\nonumber\\
&=&r\sum_s \frac{q_s}{m_s^4}\int \,(\boldsymbol{\mathfrak{p}}_s-q_s\mathbf{A})\, f_s(H_s,r\mathfrak{p}_{\theta s},\mathfrak{p}_{z s})\,d^3\mathfrak{p}_s,\nonumber
\end{eqnarray}
Amp\`{e}re's law can be solved by IFT, with the quantity $\boldsymbol{\mathfrak{p}}_s$ defined by
\begin{equation}
\mathfrak{p}_{rs}=p_{rs},\hspace{3mm}\mathfrak{p}_{\theta s}=\frac{p_{\theta s}}{r},\hspace{3mm}\mathfrak{p}_{zs}=p_{zs}.\nonumber
\end{equation}
Notice how when written in this integral form, $\mathbf{j}$ is not only a function of $\mathbf{A}$, but - in contrast with the Cartesian case - also of the relevant spatial co-ordinate, $r$. In the case of zero scalar potential, the result of the calculation is to give a distribution function that is not a solution of the Vlasov equation as it is not a function of the constants of motion only. In essence, an additional $\exp (-r^2)$ factor is required in the DF to counter $\exp (r^2)$ terms that manifest by completing the square in the integration. The physical cause here would appear to be the inertial forces associated with the rotational bulk flow. 
 
If one assumes a non-zero scalar potential, then it seems impossible to satisfy Amp\`{e}re's Law. The physical cause seems to be that, in the case of force-free fields, one would require a `different' electrostatic potential to balance the inertial forces for the ions and electrons, which is of course nonsensical. Thus, our investigation seems to suggest that it is not possible to calculate a DF of the form of equation (\ref{eq:fansatz}) for the exact GH field. 

\subsection{The magnetic field: \\ A Gold-Hoyle flux tube plus a background field}
To make progress, we introduce a background field in the negative $z$ direction. The mathematical motivation for this change is to balance the `$\exp (r^2)$ problem'. Physically, it seems that the background field introduces an extra term (whose sign depends on species) into the force-balance, to allow for both the ion and electrons to be in force balance simultaneously, given one unique expression for the scalar potential.

The vector potential, magnetic field and current density used in this paper are as follows (GH+B):
\begin{eqnarray}
\mathbf{A}(\tilde{r})&=&\frac{B_0}{2\tau}\left(0,\frac{1}{\tilde{r}}\ln\left(1+\tilde{r}^2\right)-2k\tilde{r},-\ln\left(1+\tilde{r}^2\right)\right),\nonumber\\
&=&\mathbf{A}_{GH}-\left( 0,B_0k\tau^{-1}\tilde{r},0\right).\label{eq:vecfield}\\
\mathbf{B}(\tilde{r})&=&B_0\left(0,\frac{\tilde{r}}{1+\tilde{r}^2},\frac{1}{1+\tilde{r}^2}-2k\right),\nonumber\\
&=&\mathbf{B}_{GH}-(0,0,2kB_0)\label{eq:magfield}.\\
\mathbf{j}(\tilde{r})&=&2\frac{\tau B_0}{\mu_0}\left(0,\frac{\tilde{r}}{(1+\tilde{r}^2)^2},\frac{1}{(1+\tilde{r}^2)^2}\right),\nonumber\\
&=&\mathbf{j}_{GH}.\label{eq:current}
\end{eqnarray}
The magnetic permeability \emph{in vacuo} is given by $\mu_0$ and the characteristic magnetic field strength by $B_0$. The constant $\tau$ has units of inverse length, and we use $1/\tau$ to represent the characteristic length scale of the system ($\tilde{r}=\tau r$) (see Table 1 for a concise list of the dimensionless quantities used in this paper, all denoted with a tilde, $^{\tilde{}}$). The dimensionless constant $k>0$ controls the strength of the background field in the $z$ direction, and as a result there are now two different interpretations to be made. We could either consider the system as a GH flux tube of uniform twist embedded in an untwisted uniform background field, or consider the whole GH+B magnetic field as a non-uniformly twisted flux tube. We note that flux tubes embedded in an axially directed background field have recently been observed during reconnection events in the Earth's magnetotail, by the Cluster spacecraft \cite{Borg-2012}.

In the first interpretation, $\tau$ is a direct measure of the `twist' of the embedded flux tube (see Ref. \onlinecite{Birn-2007}), with the number of turns per unit length (in $z$) along a field line given by $\tau/(2\pi)$ \cite{Gold-1960}. In the second interpretation, we see that the system is not uniformly twisted, with the $z$ distance traversed when following a field line (e.g. \onlinecite{Marshbook}) given by 
\begin{equation*}
\int \frac{rB_z}{B_{\theta}}d\theta=\frac{1}{\tau}\left(1-2k(1+\tilde{r}^2)\right)\int d\theta.
\end{equation*}
The fact that this depends on $r$ demonstrates that the system as a whole has non-uniform twist. The number of turns per unit length in $z$ of the GH+B field: the `twist' is given by
\begin{equation*}
\left( \int_{\theta=0}^{\theta=2\pi} \frac{rB_z}{B_{\theta}}d\theta   \right)^{-1}=\frac{\tau}{2\pi}\left(\left(1-2k(1+\tilde{r}^2)\right)\right)^{-1},
\end{equation*}
and is plotted in Figure \ref{fig:twist} for three values of $k$. Since $k<1/2$ corresponds to the field-reversal regime, we see a mixture of positive and negative twists (Figure \ref{fig:twist1}). However, for $k\ge 1/2$ we see only negative values of the twist (Figures \ref{fig:twist2} and \ref{fig:twist3}), i.e. we travel in the negative $z$ direction as we wind round the GH+B flux tube in the anti-clockwise direction.

The magnetic field is plotted in Figures \ref{fig:0.3}-\ref{fig:0.5} for two values of $k$.  The $k=0.3$ case contains a reversal of the $\tilde{B}_z$ field direction and as such is akin to a Reversed Field Pinch (e.g. see Ref. \onlinecite{Escande-2015} for a laboratory interpretation): this configuration may be of use in the study of astrophysical jets, see Ref. \onlinecite{Li-2006} for example. The value $k=1/2$ gives zero $\tilde{B}_z$ at $\tilde{r}=0$, and as such is the value that distinguishes the two different classes of field configuration, namely unidirectional ($k\ge 1/2$) or including field reversal ($k<1/2$). The value of $\tilde{r}$ for which the $\tilde{B}_z$ field reverses is plotted in Figure \ref{fig:rev}. The magnitude of the GH+B magnetic field is plotted in Figure \ref{fig:bmag} for three values of $k$. For all values of $k$, $|\tilde{\mathbf{B}}|\to 2k$ for large $\tilde{r}$, i.e. to a potential field. 

The primary task of this paper is to calculate self-consistent collisionless equilibrium distribution functions for the GH+B field. This problem essentially reduces to solving Amp\`{e}re's Law such that equation (\ref{eq:vlasov}) is satisfied. We assume nothing about the electric field however, and in fact use that degree of freedom to solve Amp\`{e}re's Law. The resultant form of the scalar potential is then substituted into Poisson's equation, to establish the final relationships between the microscopic and macroscopic parameters of the equilibrium.

\section{The equilibrium Distribution Function}
Although the IFT method did not yield a self-consistent equilibrium DF for the GH field without a background field, the outcome of the calculation can still be used as an indication of possible forms for the DF for the GH+B field. Using trial and error we arrived at the distribution function
\begin{eqnarray}
&&f_s=\frac{n_{0s}}{(\sqrt{2\pi}v_{th,s})^3}\times\nonumber\\
&&\left[e^{-\left(\tilde{\mathcal{H}}_s-\tilde{\omega}_s\tilde{p}_{\theta s}-\tilde{{U}}_{zs}\tilde{p}_{zs}\right)}+{C}_se^{-\left(\tilde{\mathcal{H}}_s-\tilde{{V}}_{zs}\tilde{p}_{zs}\right)}\right],\label{eq:ansatz}
\end{eqnarray}
which is a superposition of two terms that are consistent macroscopically with a `Rigid-Rotor', see Ref. \cite{Davidsonbook} for example. A Rigid-Rotor is microscopically described by a DF of the form $F(\mathcal{H}-\omega p_{\theta}-Vp_z)$ (with $V=0$ in the second term of the DF in equation (\ref{eq:ansatz})). Each $F(H-\omega p_{\theta}-Vp_z)$ term corresponds to an average macroscopic motion of rigid rotation with angular frequency $\omega$, and rectilinear motion with velocity $V$.

The dimensionless constants $\tilde{\omega}_s$, $\tilde{{U}}_{zs}$, $\tilde{{V}}_{zs}$ and ${C}_s$ are yet to be determined, with ${C}_s>0$ for positivity of the distribution. Note that the thermal beta is $\beta_s=1/(k_BT_s)$ and $v_{th,s}$ is the thermal velocity of species $s$. The ratio of the thermal Larmor radius, $r_L=m_sv_{th,s}/(e|B|)$ (for $e=|q_s|$) to the macroscopic length scale of the system $L(=1/\tau)$, is given by
\begin{equation*}
\delta_s(r)=\frac{r_L}{L}=\frac{m_sv_{th,s}\tau}{eB(r)},
\end{equation*}    
typically known as the `magnetisation parameter' \cite{Fitzpatrickbook}. In our system, the magnitude of the magnetic field and hence $\delta_{s}$ itself is spatially variable. For the purposes of the calculations in this paper however, we set
\begin{equation*}
\frac{m_sv_{th,s}\tau}{eB_0}=\delta_s={\rm const.}
\end{equation*}
as a characteristic value (see Table \ref{tab:param} for a concise list of the micro and macroscopic parameters of the equilibrium).

\subsection{Maxwell's equations: Fixing the parameters of the DF}
By insisting on a specific magnetic field configuration (the GH+B field) we have made a statement on the macroscopic physics. In searching for the equilibrium DF, we are trying to understand the microscopic physics. In this sense we are tackling an `inverse problem'. Once an assumption on the form of the DF is made then -- should the assumed form be able to reproduce the correct moments -- this inverse problem reduces to establishing the relationships between the microscopic and macroscopic parameters of the equilibrium. In this section we `fix' the free parameters of the DF in equation (\ref{eq:ansatz}), such that Maxwell's equations are satisfied;
\begin{eqnarray}
\nabla\cdot\mathbf{E}&=&\frac{1}{\varepsilon_0}\sum_sq_s\int f_sd^3v,\\
\nabla\times\mathbf{B}&=&\mu_0\sum_sq_s\int \mathbf{v}f_sd^3v.
\end{eqnarray}
Note that the solenoidal constraint and Faraday's law are automatically satisfied for the GH+B field in equilibrium, since $\mathbf{B}=\nabla\times\mathbf{A}$ implies that $\nabla\cdot\mathbf{B}=0$ and $\mathbf{E}=-\nabla\phi$ implies that $\nabla\times\mathbf{E}=\mathbf{0}=-\frac{\partial\mathbf{B}}{\partial t}$.

\subsubsection{Amp\`{e}re's Law}
In Appendix \ref{app:moments} we have calculated the $j_{z}$ current density, found by summing first order moments in $v_z$  of the DF. We now substitute in the macroscopic expressions for $j_z(\tilde{r})$, $A_{\theta}(\tilde{r})$ and $A_z(\tilde{r})$ from (\ref{eq:current}) and (\ref{eq:vecfield}) into the expression for the $j_z$ current density of equation (\ref{eq:jzapp}). After this substitution, we can calculate a $\phi(r)$ that makes the system consistent. The substitution of the known expressions for $j_{z}$, $A_{z}$ and $A_{\theta}$ gives
\begin{widetext}
\begin{eqnarray}
&&j_{z}(\tilde{r})=\frac{2\tau B_0}{\mu_0}\frac{1}{(1+\tilde{r}^2)^2}=\sum_sn_{0s}q_sv_{th,s}e^{-q_s\beta_s\phi}\times\nonumber\\
&&\left({\tilde{{U}}_{zs}}e^{({\tilde{{U}}_{zs}}^2+\tilde{r}^2{\tilde{\omega}_{s}}^2)/2-\text{sgn}(q_s){\tilde{\omega}_{s}}\tilde{r}^2k/\delta_s}\left(1+\tilde{r}^2\right)^{\text{sgn}(q_s)({\tilde{\omega}_{s}}-{\tilde{{U}}_{zs}})/(2\delta_s)}+{\tilde{{V}}_{zs}}{C}_se^{{\tilde{{V}}_{zs}}^2/2}\left(1+\tilde{r}^2\right)^{-\text{sgn}(q_s){\tilde{{V}}_{zs}}/(2\delta_s)}\right)
\end{eqnarray}
In order to satisfy the above equality  we can construct a solution by introducing a `separation constant' $\gamma_1\ne 0, 1$. We multiply the above equation by $(1+\tilde{r}^2)^2$ which makes the left-hand side constant, whilst the right-hand side is a sum of two terms, one depending on ion parameters and the second depending on electron parameters. Then we can define $\gamma_1$ by
\begin{equation}
\frac{2\tau B_0}{\mu _0}=\frac{2\tau B_0}{\mu _0} (1-\gamma_1) + \frac{2\tau B_0}{\mu _0}\gamma_1 ,\label{eq:a1a2}
\end{equation}
associating the `ion term' with the first term on the right-hand side of (\ref{eq:a1a2}), and the `electron term' with the second term on the right-hand side of (\ref{eq:a1a2}). After some algebra we can rearrange these two associations to give two expressions for the scalar potential, one in terms of the ion parameters, and one in terms of the electron parameters:
\begin{eqnarray}
\phi(r)&=&\frac{1}{q_i\beta_i}\ln\left\{\frac{\mu_0n_{0i}q_iv_{th,i}}{2\tau B_0(1-\gamma_1)}\left[{\tilde{{U}}_{zi}}e^{({\tilde{{U}}_{zi}}^2+\tilde{r}^2{\tilde{\omega}_{i}}^2)/2-{\tilde{\omega}_{i}}\tilde{r}^2k/\delta_i}\left(1+\tilde{r}^2\right)^{2+({\tilde{\omega}_{i}}-{\tilde{{U}}_{zi}})/(2\delta_i)}+{\tilde{{V}}_{zi}}{C}_ie^{{\tilde{{V}}_{zi}}^2/2}\left(1+\tilde{r}^2\right)^{2-{\tilde{{V}}_{zi}}/(2\delta_i)}     \right]\right\}\nonumber\\
\phi(r)&=&\frac{1}{q_e\beta_e}\ln\left\{\frac{\mu_0n_{0e}q_ev_{th,e}}{2\tau B_0\gamma_1}\left[{\tilde{{U}}_{ze}}e^{({\tilde{{U}}_{ze}}^2+\tilde{r}^2{\tilde{\omega}_{e}}^2)/2+{\tilde{\omega}_{e}}\tilde{r}^2k/\delta_e}\left(1+\tilde{r}^2\right)^{2-({\tilde{\omega}_{e}}-{\tilde{{U}}_{ze}})/(2\delta_e)}+\tilde{{V}}_{ze}{C}_ee^{\tilde{{V}}_{ze}^2/2}\left(1+\tilde{r}^2\right)^{2+\tilde{{V}}_{ze}/(2\delta_e)}     \right]\right\}\nonumber
\end{eqnarray}
The two values of the scalar potential above must be made identical by a suitable choice of relationships between the ion and electron parameters. Given enough freedom in parameter space, we could say that the $z$ component of Amp\`{e}re's Law is \emph{implicitly} solved the above equations, in that one just needs to choose a consistent set of parameters. However, we seek a solution in an \emph{explicit} sense. 

In order to make progress we non-dimensionalise the above equations by multiplying both sides by $e\beta_r$ with 
\begin{equation*}
\beta_r=\frac{\beta_i\beta_e}{\beta_e+\beta_i}.
\end{equation*}
Once this is done we can write the scalar potential in the form
\begin{eqnarray}
e\beta_r\phi(r)&=&\ln\left\{\left[\text{ion terms}\right]^{\frac{e\beta_r}{q_i\beta_i}}\right\},\label{eq:phi3}\\
e\beta_r\phi(r)&=&\ln\left\{\left[\text{electron terms}\right]^{\frac{e\beta_r}{q_e\beta_e}}\right\}.\label{eq:phi4}
\end{eqnarray}
Specifically, equations (\ref{eq:phi3}) and (\ref{eq:phi4}) require the equality of the arguments of the logarithm to hold in order for a meaningful solution to be obtained for the scalar potential. A first step towards this is made by requiring consistent powers of the $1+\tilde{r}^2$ `profile' in the right-hand side of the above expression to allow factorisation. Hence
\begin{eqnarray}
({\tilde{\omega}_{i}}-{\tilde{{U}}_{zi}})/(2\delta_i)&=&-{\tilde{{V}}_{zi}}/(2\delta_i),\hspace{3mm} -({\tilde{\omega}_{e}}-{\tilde{{U}}_{ze}})/(2\delta_e)=\tilde{{V}}_{ze}/(2\delta_e),\nonumber\\
&\implies & {\tilde{\omega}_{i}}=\tilde{{U}}_{zi}-\tilde{{V}}_{zi},\hspace{3mm}\tilde{\omega}_e=\tilde{{U}}_{ze}-\tilde{{V}}_{ze},\label{eq:omegafix}
\end{eqnarray}
and hence the rigid-rotation, $\tilde{\omega}_s$, is fixed by the difference of the rectilinear motion, $\tilde{U}_{zs}-\tilde{V}_{zs}$. On top of this, we require that the power of the $1+\tilde{r}^2$ `profile' on the right-hand side is the same for both the ions and electrons, thus
\begin{equation}
 \frac{e\beta_r}{q_i\beta_i}\left(2-\tilde{{V}}_{zi}/(2\delta_i)\right)=\mathcal{E}=\frac{e\beta_r}{q_e\beta_e}\left(2+{\tilde{{V}}_{ze}}/(2\delta_e)\right).\label{eq:Gfix}
\end{equation}
This condition seems to be a statement on an average potential energy associated with the particles. Once more to allow factorisation of the $1+\tilde{r}^2$ `profile', we insist that net $\exp(r^2)$ terms cancel, i.e.
\begin{equation}
 \frac{{\tilde{\omega}_{i}}}{2}=\frac{k}{\delta_i}>0,\hspace{3mm}\frac{{\tilde{\omega}_{e}}}{2}=-\frac{k}{\delta_e} <0.\label{eq:kfix}
\end{equation}
The physical meaning of this condition seems to be that the frequencies of the rigid rotor for each species are matched according to the relevant magnetisation, and the background field magnitude. The remaining task is to ensure equality of the `coefficients'
\begin{eqnarray}
\left\{\frac{1}{4\delta_i(1-\gamma_1)} \frac{n_{0i}m_iv_{th,i}^2}{B_0^2/(2\mu_0)}\left[{\tilde{{U}}_{zi}}e^{{\tilde{{U}}_{zi}}^2/2}+\tilde{{V}}_{zi}{C}_ie^{\tilde{{V}}_{zi}^2/2}    \right]\right\}^{\frac{e\beta_r}{q_i\beta_i}}={\mathcal{D}}=\left\{- \frac{1}{4\delta_e\gamma_1 }\frac{n_{0e}m_ev_{th,e}^2}{B_0^2/(2\mu_0) }  \left[{\tilde{{U}}_{ze}}e^{{\tilde{{U}}_{ze}}^2/2}+\tilde{{V}}_{ze}{C}_ee^{\tilde{{V}}_{ze}^2/2} \right]\right\}^{\frac{e\beta_r}{q_e\beta_e}}\label{eq:D1fix}
\end{eqnarray}
These seem to be conditions on the ratios of the energy densities associated with the bulk rectilinear motion and the magnetic field respectively. Thus far we have 8 constraints and 12 unknowns (${\tilde{{U}}_{zs}}, {\tilde{{V}}_{zs}}, {\tilde{\omega}_{s}}, {C}_s, n_{0s}, \beta_{s}$) given fixed characteristic macroscopic parameters of the equilibrium $B_0$, $\tau$, and $k$. We can now write down an expression for $\phi$ that explicitly solves the $z$ component of Amp\`{e}re's law;
\begin{equation}
\phi(\tilde{r})=\frac{1}{e\beta_r}\mathcal{E}\ln\left(1+\tilde{r}^2\right)+\phi (0),\label{eq:scalarpot}
\end{equation}
with 
\[
\phi(0)=\frac{1}{e\beta_r}\ln {\mathcal{D}}.
\] 
Clearly, we require that $\mathcal{D}>0$ for the expression above to make sense. It is clear that the sign of $\gamma_1$ could, in principle, affect the sign of $\mathcal{D}$. It is seen from (\ref{eq:D1fix}) that positivity of $\mathcal{D}$ implies that
\begin{eqnarray}
\frac{1}{1-\gamma_1}\left[ \tilde{U}_{zi}e^{\tilde{U}_{zi}^2/2}+\tilde{V}_{zi}C_ie^{\tilde{V}_{zi}^2/2}    \right]&>&0,\\
\frac{1}{\gamma_1}\left[ \tilde{U}_{ze}e^{\tilde{U}_{ze}^2/2}+\tilde{V}_{ze}C_ee^{\tilde{V}_{ze}^2/2}    \right]&<&0.
\end{eqnarray}
By rearranging the above inequalities to make $C_s$ the subject, it can be seen after some algebra that positivity of $\mathcal{D}$ and $C_s$ is guaranteed when 
\[
\gamma_1>1, \hspace{3mm} \text{sgn}(\tilde{U}_{zs})=-\text{sgn}(\tilde{V}_{zs}).
\]
Note that these conditions are sufficient, but not necessary, i.e. it is possible to have $\mathcal{D}>0$ and $C_s>0$ for any value of $\gamma_1\ne 0,1$, and even for $\text{sgn}(\tilde{U}_{zs})=\text{sgn}(\tilde{V}_{zs})$ in the case of $\gamma_1<0$. 

Thus far we have only considered the $j_z$ component, and it is premature to consider all components of Amp\`{e}re's Law satisfied. Let us move on to consider the $\theta$ component. In a process similar to that above, we substitute in the macroscopic expressions for $j_\theta(\tilde{r})$, $A_{\theta}(\tilde{r})$ and $A_z(\tilde{r})$ for the GH+B field into the expression for the $j_\theta$ current density of equation (\ref{eq:jthetageneral}) in Appendix \ref{app:moments}. After this substitution, we can once more calculate the $\phi$ that makes the system consistent. The substitution gives
\begin{equation}
j_{\theta}=\frac{2\tau B_0}{\mu_0}=\sum_{s}n_{0s}q_sv_{th,s}\tilde{\omega}_se^{-q_s\beta_s\phi}e^{({\tilde{{U}}_{zs}}^2+\tilde{r}^2{\tilde{\omega}_{s}}^2)/2-\text{sgn}(q_{s)}{\tilde{\omega}_{s}}\tilde{r}^2k/\delta_s}\left(1+\tilde{r}^2\right)^{2+\text{sgn}(q_{s})({\tilde{\omega}_{s}}-{\tilde{{U}}_{zs}})/(2\delta_s)}
\end{equation}
Using the parameter relations as above, we determine that the scalar potential is again given in the form of (\ref{eq:scalarpot}), 
\begin{equation*}
\phi(\tilde{r})=\frac{1}{e\beta_r}\mathcal{E}\ln\left(1+\tilde{r}^2\right)+\phi(0).
\end{equation*} 
Hence, this form of the scalar potential is consistent provided
\begin{equation}
\left[\frac{1}{1-\gamma_2}\frac{1}{4\delta_i} \frac{n_{0i}m_iv_{th,i}\omega_i/\tau}{B_0^2/(2\mu_0)} e^{{\tilde{{U}}_{zi}}^2/2}   \right]^{\frac{e\beta_r}{q_i\beta_i}}={\mathcal{D}}=\left[ -\frac{1}{\gamma_2} \frac{1}{4\delta_e}\frac{n_{0e}m_ev_{th,e}\omega_e/\tau}{B_0^2/(2\mu_0)  } e^{{\tilde{{U}}_{ze}}^2/2}  \right]^{\frac{e\beta_r}{q_e\beta_e}}\label{eq:D2fix}
\end{equation}
for $\gamma_2\ne 1$ another separation constant. These seem to be conditions on the ratios of the energy densities associated with the bulk rotation and the magnetic field respectively. This has added two more constraints. 

Once again we must ensure that $\mathcal{D}>0$. Since $\omega_e<0$, the right-hand side of the above equation implies that $\gamma_2>0$ to ensure that $\mathcal{D}>0$. Whilst the left-hand side implies that $\gamma_2<1$ for positivity of $\mathcal{D}$ since $\omega_i>0$. Hence we can say that for positivity
\[
0<\gamma_2<1.
\]

We can now consider Amp\`{e}re's Law satsified, given a $\phi$ that solves Poisson's equation. As a result, the problem of consistency is now shifted to solving Poisson's Equation, where the remaining degrees of freedom lie. 

\subsubsection{Poisson's Equation}
The final step in `self-consistency' is to solve Poisson's Equation. Frequently in such equilibrium studies, this step is replaced by satisfying quasineutrality and in essence solving a first order approximation of Poisson's equation, see for example Refs. \onlinecite{Schindlerbook, Harrison-2009a,Tasso-2014}. Here we solve Poisson's equation exactly, i.e. to all orders. Poisson's equation in cylindrical coordinates with only radial dependence gives
\begin{equation}
\nabla\cdot\mathbf{E}=-\frac{1}{r}\frac{\partial}{\partial r}\left(r\frac{\partial \phi}{\partial r}\right)=\frac{\sigma}{\varepsilon_0}.\label{eq:Poiss}
\end{equation}
The electric field is calculated as $\mathbf{E}=-\nabla \phi$, giving
\begin{equation*}
E_r=-\partial_r\phi=-\frac{2\tau \mathcal{E}}{e\beta_r}\frac{\tilde{r}}{(1+\tilde{r}^2)}.
\end{equation*}
We can now take the divergence of the electric field $\nabla\cdot\mathbf{E}=\tau\tilde{r}^{-1}\partial_{\tilde{r}}(\tilde{r}E_{r})$ and so
\begin{equation}
\nabla\cdot\mathbf{E}=-\frac{4\tau^2\mathcal{E}}{e\beta_r}\frac{1}{(1+\tilde{r}^2)^2}\implies\sigma=-\frac{4\varepsilon_{0}\tau^2\mathcal{E}}{e\beta_r}\frac{1}{(1+\tilde{r}^2)^2}.\label{eq:Edivergence}
\end{equation}
This gives a non-zero net charge per unit length (in $z$) of    
\begin{equation}
\mathcal{Q}=\int_{\theta=0}^{\theta=2\pi}\int_{r=0}^{r=\infty} \,\sigma \,r\,dr\,d\theta= -\frac{4\pi\varepsilon_{0}\mathcal{E}}{e\beta_r}.\label{eq:coulomb}
\end{equation}
The charge density derived in equation (\ref{eq:Edivergence}) must equal the charge density calculated by taking the zeroth moment of the DF. The expression for the charge density calculated in (\ref{eq:sigmaapp}) gives
\begin{eqnarray}
\sigma=\sum_sq_sn_s&=&\sum_sn_{0s}q_se^{-q_s\beta_s\phi}\left(e^{({\tilde{{U}}_{zs}}^2+\tilde{r}^2{\tilde{\omega}_{s}}^2)/2}e^{{\tilde{{U}}_{zs}}\tilde{A}_{zs}}e^{{\tilde{\omega}_{s}}\tilde{r}\tilde{A}_{\theta s}}+{C}_se^{({\tilde{{U}}_{zs}}-{\tilde{\omega}_{s}})^2/2}e^{({\tilde{{U}}_{zs}}-{\tilde{\omega}_{s}})\tilde{A}_{zs}}\right) ,  \nonumber\\
&=&\sum_sn_{0s}q_se^{-q_s\beta_s\phi}\left(1+\tilde{r}^2\right)^{{\rm sgn}(q_s)({\tilde{\omega}_{s}}-{\tilde{{U}}_{zs}})/(2\delta_s)}\left(e^{{\tilde{{U}}_{zs}}^2/2}+{C}_se^{({\tilde{{U}}_{zs}}-{\tilde{\omega}_{s}})^2/2}\right),\nonumber\\
&=&\frac{1}{\left(1+\tilde{r}^2\right)^{2}}\sum_sn_{0s}q_s{\mathcal{D}}^{-\frac{q_s\beta_s}{e\beta_r}}\left(e^{{\tilde{{U}}_{zs}}^2/2}+{C}_se^{({\tilde{{U}}_{zs}}-{\tilde{\omega}_{s}})^2/2}\right).  \label{eq:chargedensity}
\end{eqnarray}
The second equality is found by substituting the form of the vector potential from equation (\ref{eq:vecfield}), and the final equality is reached by using the conditions derived in equations (\ref{eq:omegafix}) - (\ref{eq:scalarpot}).

We can now match equations (\ref{eq:Edivergence}) and (\ref{eq:chargedensity}) to get
\begin{equation}
\sigma=-\frac{4\varepsilon_0\tau^2\mathcal{E}}{e\beta_r}=\sum_{s}n_{0s}q_s  {\mathcal{D}}^{-\frac{q_s\beta_s}{e\beta_r}}\left(e^{{\tilde{{U}}_{zs}}^2/2}+{C}_se^{{\tilde{{V}}_{zs}}^2/2}\right).\label{eq:poisscon}
\end{equation}
We now have 12 physical parameters (${\tilde{{U}}_{zs}}, {\tilde{{V}}_{zs}}, {\tilde{\omega}_{s}}, {C}_s, n_{0s},\beta_{s}$) with 11 constraints (\ref{eq:omegafix}-\ref{eq:D1fix}), (\ref{eq:D2fix}) \& (\ref{eq:poisscon}). For example, if one picks $B_0$, $\tau$, $k$ and one microscopic parameter, say $\beta_i$, then the remaining parameters of the equilibrium, (${\tilde{{U}}_{zs}}, {\tilde{{V}}_{zs}}, {\tilde{\omega}_{s}}, {C}_s$, $n_{0s}$, $\beta_{e}$), are now determined. One could of course choose the values of a different set of parameters, and determine those that remain by using the constraints derived. Note that whilst the constants $\gamma_1\ne 0,1$ and $0< \gamma_2 <1 $ are system parameters, they are not physically meaningful as they only represent a change in the gauge of the scalar potential.

\end{widetext}

\section{Analysis of the equilibrium}
\subsection{Non-neutrality \& the electric field}
It is seen from equations (\ref{eq:Edivergence}) and (\ref{eq:coulomb}) that basic electrostatic properties of the equilibrium described by $f_s$ are encoded in $\mathcal{E}$. The equilibrium is electrically neutral only when $\mathcal{E}=0$, and non-neutral otherwise. Specifically, there is net negative charge when $\mathcal{E}>0$, and net positive charge when $\mathcal{E}<0$. This net charge is finite in the $(r,\theta)$ plane and given by $\mathcal{Q}$ in equation (\ref{eq:coulomb}).

Physically, the sign of $\mathcal{E}$ seems to be related to the respective magnitudes of the bulk rotation frequencies, $\tilde{\omega}_s$. From equations (\ref{eq:omegafix}) and (\ref{eq:Gfix}) we see that $\mathcal{E}>0$ implies that
\begin{eqnarray}
\tilde{\omega}_i>\omega_{i}^{\star}&=&\tilde{U}_{zi}-4\delta_i,\nonumber\\
|\tilde{\omega}_e|<\omega_{e}^{\star}&=&-\tilde{U}_{ze}-4\delta_e,\nonumber
\end{eqnarray}
and $\mathcal{E}<0$ implies that
\begin{eqnarray}
\tilde{\omega}_i<\omega_{i}^{\star}&=&\tilde{U}_{zi}-4\delta_i,\nonumber\\
|\tilde{\omega}_e|>\omega_{e}^{\star}&=&-\tilde{U}_{ze}-4\delta_e.\nonumber
\end{eqnarray}
Hence, $\mathcal{E}>0$ is seen to occur for `sufficiently large' bulk ion rotation frequencies, and `sufficiently small' (in magnitude) bulk electron rotation frequencies. A positive $\mathcal{E}$ corresponds to an electric field directed radially `inwards'. This seems to make sense physically, by the following argument. A `larger' ($\tilde{\omega}_i>\omega_{i}^{\star}$) bulk ion rotation freqency gives a `larger' centrifugal force, and a `smaller' ($ |\tilde{\omega}_e|<\omega_{e}^{\star} $) bulk electron rotation frequency gives a `smaller' centrifugal force. For a dynamic interpretation, at a fixed $r$, the ions are forced to a slightly larger radius than the electrons, i.e. a charge separation manifests on small scales. This charge separation results in an inward electric field, $E_r<0$. An equally valid interpretation is to say that for an equilibrium to exist, an electric field must exist to counteract the differences in the centrifugal forces associated with the bulk ion and electron rotational flows.

In a similar manner, $\mathcal{E}<0$ is seen to occur for `sufficiently small' ($\tilde{\omega}_i<\omega_{i}^{\star}$) bulk ion rotation frequencies, and `sufficiently large' ($ |\tilde{\omega}_e|>\omega_{e}^{\star} $) bulk electron rotation frequencies. A negative $\mathcal{E}$ corresponds to an electric field directed radially `outwards'. We can then interpret these result physically, in a manner like that above.

Finally, we can interpret the neutral case, $\mathcal{E}=0$, as the intermediary between the two circumstances considered above. That is to say that the equilibrium is neutral when the bulk rotation flows are just matched accordingly, such that there is no charge separation and hence no electric field.

\subsection{The equation of state and the plasma beta}
For certain considerations, e.g. the solar corona, it would be advantageous if the DF had the capacity to describe plasmas with sub-unity values of the plasma beta: the ratio of the thermal energy density to the magnetic energy density
\begin{equation}
\beta_{pl}(\tilde{r})=\frac{2\mu _0k_B}{B^2}\sum _sn_sT_s\label{eq:beta}.
\end{equation}
For our configuration, the number density is seen to be proportional to the $rr$ component of the pressure tensor, $P_{rr,s}=n_sk_BT_s$. This is demonstrated by the following calculation. In order to calculate $P_{rr}$, we must consider the integral
\begin{equation}
P_{rr}=\sum_s m_s\int_{-\infty}^\infty \, w_{rs}\,w_{rs}\,f_s\,d^3v.\label{eq:Prr}
\end{equation}
However, we do not have to consider a bulk velocity in the $r$ direction here $(u_{rs}=0)$, since $f_s$ is an even function of $v_r$. Using the fact that
\begin{eqnarray}
\int_{-\infty}^{\infty} v_r^2e^{-v_r^2/(2v_{th,s}^2)}dv_r&=&v_{th,s}^2\int_{-\infty}^{\infty} e^{-v_r^2/(2v_{th,s}^2)}dv_r,\nonumber
\end{eqnarray}
and by consideration of equations (\ref{eq:Prr}) and the number density, we see that
\begin{eqnarray}
P_{rr,s}=m_sv_{th,s}^2n_s,\label{eq:temp}
\end{eqnarray}
that is to say that $k_BT_s=m_sv_{th,s}^2$. Note that if $n_i=n_e:= n$ and hence $\mathcal{E}=0$ (neutrality), then we have an equation of state given by
\begin{equation*}
P_{rr}=\frac{\beta_e+\beta_i}{\beta_e\beta_i}n.
\end{equation*}
This resembles expressions found in the Cartesian case, in Refs. \onlinecite{Channell-1976, Neukirch-2009, Allanson-2015PoP} for example. Incidentally, we can use the connection between $n_s$ and $P_{rr}$ to give an expression for the $\beta_{pl}$ that is perhaps more typically seen,
\begin{equation*}
\beta_{pl}(\tilde{r})=\frac{2\mu _0}{B^2}\sum _sP_{rr,s}.
\end{equation*}
The square magnitude of the magnetic field (equation (\ref{eq:magfield})) is given by
\begin{equation*}
B^2=\frac{B_0^2}{(1+\tilde{r}^2)}\left(1-4k+4k^2(1+\tilde{r}^2)\right).
\end{equation*}
Using the number density from equation (\ref{eq:numdensity}) in the definition of the plasma beta from equation (\ref{eq:beta}), as well as the equilibrium conditions (\ref{eq:omegafix}) - (\ref{eq:scalarpot}) gives
\begin{eqnarray}
&&\beta_{pl}(\tilde{r})=\frac{2\mu _0}{B_0^2(1+\tilde{r}^2)\left(1-4k+4k^2(1+\tilde{r}^2)\right)}\times\nonumber\\
&&\sum _s\frac{n_{0s}}{\beta_s}{\mathcal{D}}^{-\frac{q_s\beta_s}{e\beta_r}}\left(e^{{\tilde{{U}}_{zs}}^2/2}+{C}_se^{{\tilde{{V}}_{zs}}^2/2}\right).
\end{eqnarray}
It is not immediately obvious from the above equation what values $\beta_{pl}$ can have. However it is readily seen that as $\tilde{r}\to\infty$ then $\beta_{pl}\to 0$, essentially since the number density is vanishing at large radii. On the central axis of the tube we see that 
\begin{eqnarray}
&&\beta_{pl}(0)=\frac{2\mu _0}{B_0^2\left(1-4k+4k^2\right)}\times\nonumber\\
&&\sum _s\frac{n_{0s}}{\beta_s}{\mathcal{D}}^{-\frac{q_s\beta_s}{e\beta_r}}\left(e^{{\tilde{{U}}_{zs}}^2/2}+{C}_se^{{\tilde{{V}}_{zs}}^2/2}\right),
\end{eqnarray}
suggesting that for a suitable choice of parameters, it should be possible to attain any value of $\beta_{pl}$ on the axis.

\subsection{Plots of the DF}
A characteristic that one immediately looks for in a new DF is the existence of multiple maxima in velocity space, which are a direct indication of non-thermalisation, relevant for the existence of micro-instabilities (e.g. see \cite{Gary-2005}). Using an analysis very similar to that in \cite{Neukirch-2009}, we can derive - for a given value of $\tilde{\omega}_{s}$ - conditions on $\tilde{r}$ and either $\tilde{v}_{z}$ or $\tilde{v}_{\theta}$, for the existence of multiple maxima in the $\tilde{v}_{\theta}$ or $\tilde{v}_{z}$ direction respectively. We present these calculations in Appendices \ref{app:vthetamax} and \ref{app:vzmax}. The most readily understood results are that multiple maxima in the $\tilde{v}_{\theta}$ direction can only occur for $\tilde{r}>2/|\tilde{\omega}_{s}|$, and in the $\tilde{v}_z$ direction for $|\tilde{\omega}_s|>2$. Given these necessary conditions, one can then calculate that multiple maxima of $f_s$ will occur in the $\tilde{v}_{\theta}$ direction for $\tilde{v}_z$ bounded above and below, and vice versa. 

In Figures (\ref{fig:4}-\ref{fig:7}) we present plots of the DFs over a range of parameter values. Figures (\ref{fig:4}) and (\ref{fig:5}) show the ion DFs for $k=0.1$ and $k=1$ respectively, for all combinations of $\tilde{\omega}_{i}=1,3$, $\tilde{r}=0.5,2$ and $C_s=0.1, 1$, and with the magnetisation parameter $\delta_i=1$. As a graphical confirmation of the above discussion, we can only see multiple maxima in the $\tilde{v}_{\theta}$ direction for $\tilde{r}>2/|\tilde{\omega}_{s}|$, and in the $\tilde{v}_z$ direction for $|\tilde{\omega}_s|>2$, with the appropriate bounds marked by the horizontal/vertical white lines.  

Aside from multiple maxima in the orthogonal directions, the DF can also be `two-peaked'. That is, the DF can have two isolated peaks in $(\tilde{v}_z,\tilde{v}_{\theta})$ space. This is seen to occur for figures (\ref{fig:5d}, \ref{fig:5g}, \ref{fig:5h}). Hence, $f_i$ is seen to be `two-peaked' when $k=1$ for both $\tilde{r}>2/\tilde{\omega}_i$ and $\tilde{r}<2/\tilde{\omega}_i$. However, we do not see a two-peaked DF for $k=0.1$. This seems to suggest that the stronger guide field ($k=1$) correlates with multiple peaks. Physically, this may correspond to the fact that a homogeneous guide field is consistent with a Maxwellian DF centred on the origin in $(\tilde{v}_z,\tilde{v}_{\theta})$ space, given that a Maxwellian contributes zero current. Hence, if the `main' part/peak of the DF is centred away from the origin, then the Maxwellian contribution from the guide field could contribute a secondary peak. These secondary peaks are seen to be more pronounced when $\tilde{C}_i$ is larger, i.e. the contribution from the second term from the DF is greater.  

Figures (\ref{fig:6}) and (\ref{fig:7}) show the electron DFs for $k=0.1$ and $k=1$ respectively, for all combinations of $\tilde{\omega}_{e}=1,3$; $\tilde{r}=0.5,2$, and $C_e=0.1,1$, and with the magnetistaion parameter $\delta_e=\delta_i\sqrt{m_e/m_i}\approx 1/\sqrt{1836}$. This choice of magnetisation corresponds to $T_i=T_e$. In general we see DFs with fewer multiple maxima in velocity space than the ion plots, which is physically consistent with the electrons being more magnetised, i.e. more `fluid-like'. In particular we see no multiple maxima in figure \ref{fig:7}, the case with the stronger background field. 

Note that when the electrons to have the same magnetisation as the ions, i.e. $\delta_e=\delta_i=1$, then these marked differences in the velocity-space plots disappear, and we observe a qualitative symmetry $f_i(\tilde{v}_\theta,\tilde{v}_z,r)\propto f_e(-\tilde{v}_\theta,-\tilde{v}_z,r)$.

\section{Summary}
In this paper we have calculated one-dimensional collisionless equilibria for a continuum of magnetic field models based on the Gold-Hoyle flux tube, with an additional constant background field in the axial direction. This study was motivated by a desire to extend the existing methods for solutions of the `inverse problem in Vlasov equilibria' in Cartesian geometry, to cylindrical geometry. Initial efforts focussed on solving for the exact force-free Gold-Hoyle field, but this seems impossible due to the positive definite centrifugal forces. The Gold-Hoyle field in particular was chosen as it represents the `natural' analogue of the Force-Free Harris Sheet in cylindrical geometry, a magnetic field whose VM equilibria have been the subject of recent study, \cite{Harrison-2009b, Neukirch-2009, Wilson-2011, Allanson-2015PoP, Kolotkov-2015, Allanson-2016}.

A background field was introduced, and an equilibrium distribution function was found that reproduces the required magnetic field, i.e. solves Amp\`{e}re's Law. It is the presence of the background field that allows us to solve Vlasov's equation and Amp\`{e}re's Law, and it appears physically necessary as it introduces an `asymmetry'; namely an extra term into the equation of motion whose sign depends explicitly on species. In contrast to the `demands' of insisting on a particular magnetic field, no condition was made on the electric field. The distribution function allows both electrically neutral and non-neutral configurations, and in the case of non-neutrality we find an exact and explicit solution to Poisson's equation for an electric field that decays like $1/r$ far from the axis. We note here that the type of solutions derived in this paper could - after a Galilean transformation - be interpreted as 1D BGK modes with finite magnetic field (see Refs. \onlinecite{Abraham-Shrauner-1968,Ng-2005,Grabbe-2005,Ng-2006} for example, to provide some context).

An analysis of the physical properties of the DF was given in Section IV, with some detailed calculations in Appendix B. The dependence of the sign of the charge density (and hence the electric field) on the bulk ion and electron rotational flows was analysed, with a physical interpretation given. Essentially the argument states that the electric field exists in order to balance the difference in the centrifugal forces between the two species. The DF was found to be able to give sub-unity values of the plasma beta, should this be required/desirable given the relevant physical system that it is intended to model. The final part of the analysis focussed on plotting the DF in velocity space, for certain parameter values, and at different radii. Mathematical conditions were found that determine whether or not the DF could have multiple maxima in the orthogonal directions in velocity space, and these are corroborated by the plots of the distribution functions. For certain parameter values, the DF was also seen to have two separate, isolated peaks. This non-thermalisation suggests the existence of microinstabilities, for a certain choice of parameters.

Further work could involve a deeper anlysis of the properties of the distribution functions and their stability. This work has also raised a fundamental question: `is it possible to describe a one-dimensional force-free collisionless equilibrium in cylindrical geometry?' Preliminary investigations seem to suggest that it is not possible.

% If in two-column mode, this environment will change to single-column format so that long equations can be displayed. 
% Use only when necessary.
%\begin{widetext}
%$$\mbox{put long equation here}$$
%\end{widetext}

% Figures should be put into the text as floats. 
% Use the graphics or graphicx packages (distributed with LaTeX2e).
% See the LaTeX Graphics Companion by Michel Goosens, Sebastian Rahtz, and Frank Mittelbach for examples. 
%
% Here is an example of the general form of a figure:
% Fill in the caption in the braces of the \caption{} command. 
% Put the label that you will use with \ref{} command in the braces of the \label{} command.
%
% \begin{figure}
% \includegraphics{}%
% \caption{\label{}}%
% \end{figure}

% Tables may be be put in the text as floats.
% Here is an example of the general form of a table:
% Fill in the caption in the braces of the \caption{} command. Put the label
% that you will use with \ref{} command in the braces of the \label{} command.
% Insert the column specifiers (l, r, c, d, etc.) in the empty braces of the
% \begin{tabular}{} command.
%
% \begin{table}
% \caption{\label{} }
% \begin{tabular}{}
% \end{tabular}
% \end{table}

\begin{acknowledgments}
O.A. would like to thank both Professor A.W. Hood of the University of St Andrews and Professor P.K. Browning of the University of Manchester for encouraging discussions. The authors gratefully acknowledge the support of the Science and Technology Facilities Council Consolidated Grants ST/K000950/1 and ST/N000609/1, as well as Doctoral Training Grant ST/K502327/1. We also gratefully acknowledge funding from Leverhulme Trust Research Project Grant F/00268/BB.  
\end{acknowledgments}

\appendix
\section{Moments of the DF}\label{app:moments}
In this appendix we calculate the zeroth and first order velocity space moments of the DF, necessary for the charge density and the current density respectively. See Table 1 for a clarification of all dimensionless quantities denoted by a tilde, $^{\tilde{}}$.
\begin{comment}
\subsection{Zeroth order moments}\label{app:number}
\end{comment}
The number density of species $s$ is given by the zeroth moment of the DF;
\begin{comment}
\begin{eqnarray}
&&n_s=\int f_sd^3v_s=\frac{n_{0s}}{(\sqrt{2\pi})^3}\times\nonumber\\
&&\int e^{-\tilde{\mathcal{H}}_s}\bigg(e^{{\tilde{{U}}_{zs}} \tilde{p}_{zs}}e^{{\tilde{\omega}_{s}}\tilde{p}_{\theta s}}+{C}_se^{\tilde{{V}}_{zs} \tilde{p}_{zs}}\bigg)d^3\tilde{v}_s\label{eq:nbasic}\\
&&=\frac{n_{0s}}{(\sqrt{2\pi})^2}e^{-\tilde{\phi}_s}\bigg[e^{\left(\tilde{{U}}_{zs}^2+\tilde{r}^2\tilde{\omega}_{s}^2\right)/2}e^{{\tilde{{U}}_{zs}}\tilde{A}_{zs}}e^{{\tilde{\omega}_{s}}\tilde{r}\tilde{A}_{\theta s}}\times\nonumber\\
&&\int_{-\infty}^{\infty} e^{-\left(\tilde{v}_{zs}-\tilde{{U}}_{zs}\right)^2/2}d\tilde{v}_{zs}\int_{-\infty}^{\infty} e^{-\left(\tilde{v}_{\theta s}-\tilde{\omega}_{s}\tilde{r}\right)^2/2}d\tilde{v}_{\theta s}\nonumber \\
&&+{C}_s\sqrt{2\pi}e^{\tilde{{V}}_{zs}^2/2}e^{\tilde{{V}}_{zs}\tilde{A}_{zs}}\int_{-\infty}^{\infty} e^{-\left(\tilde{v}_{zs}-\tilde{{V}}_{zs}\right)^2/2}d\tilde{v}_{zs}\bigg] \nonumber\\
&&=n_{0s}e^{-\tilde{\phi}_s}\times\nonumber\\
&&\left[e^{\left(\tilde{{U}}_{zs}^2+\tilde{r}^2\tilde{\omega}_{s}^2\right)/2}e^{{\tilde{{U}}_{zs}}\tilde{A}_{zs}}e^{{\tilde{\omega}_{s}}\tilde{r}\tilde{A}_{\theta s}}+{C}_se^{\tilde{{V}}_{zs}^2/2}e^{\tilde{{V}}_{zs}\tilde{A}_{zs}}\right] \label{eq:numdensity}
\end{eqnarray}
\end{comment}
\begin{eqnarray}
&&n_s=\int f_sd^3v_s=n_{0s}e^{-\tilde{\phi}_s}\times\nonumber\\
&&\left[e^{\left(\tilde{{U}}_{zs}^2+\tilde{r}^2\tilde{\omega}_{s}^2\right)/2}e^{{\tilde{{U}}_{zs}}\tilde{A}_{zs}}e^{{\tilde{\omega}_{s}}\tilde{r}\tilde{A}_{\theta s}}+{C}_se^{\tilde{{V}}_{zs}^2/2}e^{\tilde{{V}}_{zs}\tilde{A}_{zs}}\right] \label{eq:numdensity}
\end{eqnarray}
The following sum gives the charge density,
\begin{eqnarray}
&&\sigma=\sum_sq_sn_s=\sum_s n_{0s}q_se^{-\tilde{\phi}_s}\times\nonumber\\
&&\left[e^{\left(\tilde{{U}}_{zs}^2+\tilde{r}^2\tilde{\omega}_{s}^2\right)/2}e^{{\tilde{{U}}_{zs}}\tilde{A}_{zs}}e^{{\tilde{\omega}_{s}}\tilde{r}\tilde{A}_{\theta s}}+{C}_se^{\tilde{{V}}_{zs}^2/2}e^{\tilde{{V}}_{zs}\tilde{A}_{zs}}\right]\label{eq:sigmaapp}
\end{eqnarray}

\begin{comment}
\subsection{First order moments}\label{app:jz}
\end{comment}
We take the $v_{z}$ moment of the DF to calculate the $z-$ component of the bulk velocity,
\begin{comment}
\begin{eqnarray}
&&u_{zs}=\frac{v_{th,s}^4}{n_s}\int\tilde{v}_{z s} f_s d^3\tilde{v}_s,\nonumber\\
&&=\frac{v_{th,s}}{n_s}\frac{n_{0s}}{(\sqrt{2\pi})^2}e^{-\tilde{\phi}_s}\bigg[e^{\left(\tilde{{U}}_{zs}^2+\tilde{r}^2\tilde{\omega}_{s}^2\right)/2}e^{{\tilde{{U}}_{zs}}\tilde{A}_{zs}}e^{{\tilde{\omega}_{s}}\tilde{r}\tilde{A}_{\theta s}}\nonumber\\
&&\int_{-\infty}^{\infty} \tilde{v}_{zs}e^{-\left(\tilde{v}_{zs}-\tilde{{U}}_{zs}\right)^2/2}d\tilde{v}_{zs}\int_{-\infty}^{\infty} e^{-\left(\tilde{v}_{\theta s}-\tilde{\omega}_{s}\tilde{r}\right)^2/2}d\tilde{v}_{\theta s}\nonumber \\
&&+{C}_s\sqrt{2\pi}e^{\tilde{{V}}_{zs}^2/2}e^{\tilde{{V}}_{zs}\tilde{A}_{zs}}\int_{-\infty}^{\infty} \tilde{v}_{zs}e^{-\left(\tilde{v}_{zs}-\tilde{{V}}_{zs}\right)^2/2}d\tilde{v}_{zs}\bigg] \nonumber\\
&&=\frac{n_{0s}v_{th,s}}{n_s}e^{-\tilde{\phi}_s}\bigg[{\tilde{{U}}_{zs}}e^{{\tilde{{U}}_{zs}}\tilde{A}_{zs}}e^{\left(\tilde{{U}}_{zs}^2+\tilde{r}^2\tilde{\omega}_{s}^2\right)/2}e^{{\tilde{\omega}_{s}}\tilde{r}\tilde{A}_{\theta s}}\nonumber\\
&&+\tilde{{V}}_{zs}{C}_se^{\tilde{{V}}_{zs}^2/2}e^{\tilde{{V}}_{zs}\tilde{A}_{zs}}\bigg],
\end{eqnarray}
\end{comment}
\begin{eqnarray}
&&u_{zs}=\frac{v_{th,s}^4}{n_s}\int\tilde{v}_{z s} f_s d^3\tilde{v}_s,\nonumber\\
&&=\frac{n_{0s}v_{th,s}}{n_s}e^{-\tilde{\phi}_s}\bigg[{\tilde{{U}}_{zs}}e^{{\tilde{{U}}_{zs}}\tilde{A}_{zs}}e^{\left(\tilde{{U}}_{zs}^2+\tilde{r}^2\tilde{\omega}_{s}^2\right)/2}e^{{\tilde{\omega}_{s}}\tilde{r}\tilde{A}_{\theta s}}\nonumber\\
&&+\tilde{{V}}_{zs}{C}_se^{\tilde{{V}}_{zs}^2/2}e^{\tilde{{V}}_{zs}\tilde{A}_{zs}}\bigg],
\end{eqnarray}
for $n_s$ the number density. The following sum gives the $z-$ component of the current density,
\begin{eqnarray}
&&j_z=\sum_sq_sn_su_{zs}=\sum_s n_{0s}q_sv_{th,s}e^{-\tilde{\phi}_s}\times\nonumber\\
&&\bigg({\tilde{{U}}_{zs}}e^{{\tilde{{U}}_{zs}}\tilde{A}_{zs}}e^{\left(\tilde{{U}}_{zs}^2+\tilde{r}^2\tilde{\omega}_{s}^2\right)/2}e^{{\tilde{\omega}_{s}}\tilde{r}\tilde{A}_{\theta s}}\nonumber\\
&&+\tilde{{V}}_{zs}{C}_se^{\tilde{{V}}_{zs}^2/2}e^{\tilde{{V}}_{zs}\tilde{A}_{zs}}\bigg).\label{eq:jzapp}
\end{eqnarray}

By taking the $v_{\theta}$ moment of the DF we can calculate the $\theta-$ component of the bulk velocity,
\begin{comment}
\begin{eqnarray}
&&u_{\theta s}=\frac{v_{th,s}^4}{n_s}\int\tilde{v}_{\theta s} f_s d^3\tilde{v}_s,\nonumber\\
&&=\frac{v_{th,s}}{n_s}\frac{n_{0s}}{(\sqrt{2\pi})^2}e^{-\tilde{\phi}_s}\bigg[e^{\left(\tilde{{U}}_{zs}^2+\tilde{r}^2\tilde{\omega}_{s}^2\right)/2}e^{{\tilde{{U}}_{zs}}\tilde{A}_{zs}}e^{{\tilde{\omega}_{s}}\tilde{r}\tilde{A}_{\theta s}}\times\nonumber\\
&&\int_{-\infty}^{\infty} e^{-\left(\tilde{v}_{zs}-\tilde{{U}}_{zs}\right)^2/2}d\tilde{v}_{zs}\int_{-\infty}^{\infty}\tilde{v}_{\theta s} e^{-\left(\tilde{v}_{\theta s}-\tilde{\omega}_{s}\tilde{r}\right)^2/2}d\tilde{v}_{\theta s}\nonumber \\
&&=\frac{\tilde{r}\tilde{\omega}_sn_{0s}v_{th,s}e^{-\tilde{\phi}_s}}{n_s}e^{\left(\tilde{{U}}_{zs}^2+\tilde{r}^2\tilde{\omega}_{s}^2\right)/2}e^{{\tilde{{U}}_{zs}}\tilde{A}_{zs}}e^{{\tilde{\omega}_{s}}\tilde{r}\tilde{A}_{\theta s}} ,
\end{eqnarray}
\end{comment}
\begin{eqnarray}
&&u_{\theta s}=\frac{v_{th,s}^4}{n_s}\int\tilde{v}_{\theta s} f_s d^3\tilde{v}_s,\nonumber\\
&&=\frac{\tilde{r}\tilde{\omega}_sn_{0s}v_{th,s}e^{-\tilde{\phi}_s}}{n_s}e^{\left(\tilde{{U}}_{zs}^2+\tilde{r}^2\tilde{\omega}_{s}^2\right)/2}e^{{\tilde{{U}}_{zs}}\tilde{A}_{zs}}e^{{\tilde{\omega}_{s}}\tilde{r}\tilde{A}_{\theta s}} ,
\end{eqnarray}
for $n_s$ the number density. The following sum gives the $\theta-$ component of the current density,
\begin{eqnarray}
&&j_\theta=\sum_sq_sn_su_{\theta s}=\sum_s n_{0s}q_sv_{th,s}\tilde{r}\tilde{\omega}_se^{-\tilde{\phi}_s}\times\nonumber\\
&&e^{{\tilde{{U}}_{zs}}\tilde{A}_{zs}}e^{\left(\tilde{{U}}_{zs}^2+\tilde{r}^2\tilde{\omega}_{s}^2\right)/2}e^{{\tilde{\omega}_{s}}\tilde{r}\tilde{A}_{\theta s}}.\label{eq:jthetageneral}
\end{eqnarray}

\section{Looking for multiple maxima}
\subsection{Maxima of the DF in $v_{\theta}$ space}\label{app:vthetamax}
The $\tilde{p}_{rs}$ dependence of the DF is irrelevant to our discussion, and as such can be integrated out. We can also neglect the scalar potential $\phi$. The reduced DF, $\tilde{F}_s$, in dimensionless form is 
\begin{eqnarray}
\tilde{F}_s=((\sqrt{2\pi}v_{th,s})^2/n_{0s})\,e^{\tilde{\phi}_s}\,\int_{-\infty}^{\infty} \,f_s\, dv_{r},\nonumber
\end{eqnarray}
which then reads
\begin{eqnarray}
&&\tilde{F}_s=\exp\left\{-\frac{1}{2}\left[\left(\frac{\tilde{p}_{\theta s}}{\tilde{r}}-\tilde{A}_{\theta s}\right)^2+\left(\tilde{p}_{zs}-\tilde{A}_{zs}\right)^2\right]\right\}\times\nonumber\\
&&\left[\exp\left(\tilde{\omega}_s\tilde{p}_{\theta s}+\tilde{{U}}_{zs}\tilde{P}_{zs}  \right)+{C}_s\exp\left(\tilde{{V}}_{zs}\tilde{P}_{zs}   \right)\right].\label{eq:reduced}
\end{eqnarray}
We have written $\tilde{F}_s$ in terms of the canonical momenta, and so we search for stationary points given by $\partial \tilde{F}_s/\partial \tilde{p}_{\theta s}=0$, equivalent to $\partial \tilde{F}_s/\partial \tilde{v}_{\theta s}=0$.
Setting $\partial \tilde{F}_s/\partial \tilde{p}_{\theta s}=0$ gives
\begin{eqnarray}
\tilde{p}_{\theta s}-\tilde{r}\tilde{A}_{\theta s}&=&\frac{\tilde{\omega}_s\tilde{r}^2}{1+{C}_se^{-\tilde{\omega}_s\tilde{p}_{z s}}e^{-\tilde{\omega}_s\tilde{p}_{\theta s}}}\nonumber\\
&=&\frac{A}{1+Be^{-\tilde{\omega}_s\tilde{p}_{\theta s}}}:=R(\tilde{p}_{\theta s}).\label{eq:lineartheta}
\end{eqnarray}
To derive a necessary condition for multiple maxima, we analyse the RHS of equation (\ref{eq:lineartheta}), $R(\tilde{p}_{\theta s})$. This function is bounded between 0 and A, and is monotonically increasing. Hence, using techniques similar to those in \cite{Neukirch-2009}, a necessary condition for multiple maxima in the DF is that 
\begin{eqnarray}
\max_{\tilde{p}_{\theta s}}R^{\prime}(\tilde{p}_{\theta s})>1.
\end{eqnarray} 
This condition can be shown to be equivalent to $A\tilde{\omega}_s/4>1$ and so 
\begin{equation}
\tilde{\omega}_s^2>4\tilde{r}^{-2}\iff\tilde{r}>2/|\tilde{\omega}_s|%=\tilde{r}^\star .
\label{eq:rcond}
\end{equation}
This demonstrates that for sufficiently small $\tilde{r}$, there cannot exist multiple maxima. Equivalently, this condition will always be satisfied for some $\tilde{r}$, and as such is just a condition on the domain, in $\tilde{r}$, for which multiple maxima can occur. This condition is not sufficient however, as it could still be the case that there exists only one point of intersection (and hence one maximum), depending on the value of $B$. It is seen that $R$ has unit slope at 
\begin{eqnarray}
&&\tilde{p}_{\theta s}^{\pm}=\frac{1}{\tilde{\omega}_s}\times\nonumber\\
&&\left[\ln\left(2B\right)-\ln\left(A\tilde{\omega}_s-2\pm\sqrt{A\tilde{\omega}_s\left(A\tilde{\omega}_s-4\right)}\right)\right].
\label{eq:pthetastar}    
\end{eqnarray}
Clearly $R$ has unit slope for two values of $\tilde{p}_{\theta s}$. After some graphical consideration of the problem, it becomes apparent that $B$ should be bounded above and below for multiple maxima. After elementary consideration of the functional form of (\ref{eq:lineartheta}), for example with graph plotting software, we see that multiple maxima in the $\tilde{v}_{\theta}$ direction can only occur, for a given $\tilde{r}$, when $B$ (and hence $\tilde{v}_{z}$) satisfies these inequalities for ions
\begin{eqnarray}
&&\tilde{p}_{\theta i}^{ +}-R(\tilde{p}_{\theta i}^{ +})-\tilde{r}\tilde{A}_{\theta i}>0,\nonumber\\
&&\tilde{p}_{\theta i}^{ -}-R(\tilde{p}_{\theta i}^{ -})-\tilde{r}\tilde{A}_{\theta i}<0,
\end{eqnarray}
and these for electrons
\begin{eqnarray}
&&\tilde{p}_{\theta e}^{ +}-R(\tilde{p}_{\theta e}^{ +})-\tilde{r}\tilde{A}_{\theta e}<0,\nonumber\\
&&\tilde{p}_{\theta e}^{ -}-R(\tilde{p}_{\theta e}^{ -})-\tilde{r}\tilde{A}_{\theta e}>0.
\end{eqnarray}

\subsection{Maxima of the DF in $v_{z}$ space}\label{app:vzmax}
We shall once again use the reduced DF defined in equation (\ref{eq:reduced}) in our analysis. Thus, we shall consider $\partial \tilde{F}_s/\partial \tilde{p}_{z s}=0$, which is equivalent to $\partial \tilde{F}_s/\partial \tilde{v}_{z s}=0$. Setting $\partial \tilde{F}_s/\partial \tilde{p}_{z s}=0$ gives
\begin{eqnarray}
\tilde{p}_{z s}-\tilde{A}_{z s}&=&\frac{\tilde{{U}}_{zs}+{C}_s\tilde{{V}}_{zs}e^{-\tilde{\omega}_s(\tilde{p}_{zs}+\tilde{p}_{\theta s})}}{1+{C}_se^{-\tilde{\omega}_s(\tilde{p}_{zs}+\tilde{p}_{\theta s})}}\nonumber\\
&=&\frac{A_1}{1+B_1e^{-D_1\tilde{p}_{z s}}}+\frac{A_2}{1+B_2e^{-D_2\tilde{p}_{z s}}}\nonumber\\
&:=&R_1(\tilde{p}_{z s})+R_2(\tilde{p}_{z s})=R(\tilde{p}_{zs})\label{eq:linearz},\nonumber
\end{eqnarray}
such that
\begin{eqnarray}
A_1&=&\tilde{{U}}_{zs},\hspace{3mm}A_2=\tilde{{V}}_{zs},\nonumber\\
B_1&=&{C}_se^{-\tilde{\omega}_s\tilde{p}_{\theta s}}=B_2^{-1},\hspace{3mm}D_1=\tilde{\omega}_s=-D_2.\nonumber
\end{eqnarray}
 To derive a necessary condition for multiple maxima, we analyse the RHS of equation (\ref{eq:linearz}). Each $R$ function is bounded and  monotonic. Once again using techniques similar to those in \cite{Neukirch-2009}, a necessary condition for multiple maxima in the DF is that 
\begin{eqnarray}
\max_{\tilde{p}_{z s}}\left(R_1^{\prime}(\tilde{p}_{z s})+R_2^{\prime}(\tilde{p}_{z s})\right)>1.
\end{eqnarray} 
After some algebra this condition can be shown to be equivalent to $\tilde{\omega}_s^2/4>1$ and so 
\begin{eqnarray}
|\tilde{\omega}_s|>2.
\label{eq:omegacond}
\end{eqnarray}
This condition is not sufficient however, as it could still be the case that there exists only one point of intersection, depending on the value of $B_1(=1/B_2)$. The transition between 3 points of intersection and one occurs at the value of $B_1$ for which the straight line of slope unity through $\tilde{p}_{z s}=0$ just touches $R_1(\tilde{p}_{z s})+R_2(\tilde{p}_{z s})$ at the point where it also has unit slope. It is readily seen that $R_1+R_2$ has unit slope at 
\begin{eqnarray}
&&\tilde{p}_{z s}^{\pm}=\frac{1}{\tilde{\omega}_s}\times\nonumber\\
&&\left[\ln\left(  2B_{1} \right)      -   \ln\left(     \tilde{\omega}_s^2-2 \pm \sqrt{\tilde{\omega}_{s}^{2}(  \tilde{\omega}_{s}^{2}-4   )}    \right)     \right].
\end{eqnarray}
We see again that $R$ has unit slope for two values of $\tilde{p}_{z s}$. Once again, after some graphical consideration of the problem, it becomes apparent that $B_{1}$ should be bounded above and below for multiple maxima. After elementary consideration of the functional form of (\ref{eq:linearz}), for example with graph plotting software we see that multiple maxima in the $\tilde{v}_{z}$ direction can only occur, for a given $\tilde{r}$, when $B_1$ (and hence $\tilde{v}_{\theta}$) satisfies these inequalities for ions
\begin{eqnarray}
&&\tilde{p}_{z i}^{ +}-R(\tilde{p}_{z i}^{ +})-\tilde{A}_{z i}>0,\nonumber\\
&&\tilde{p}_{z i}^{ -}-R(\tilde{p}_{z i}^{ -})-\tilde{A}_{z i}<0,
\end{eqnarray}
and these for electrons
\begin{eqnarray}
&&\tilde{p}_{z e}^{ +}-R(\tilde{p}_{z e}^{ +})-\tilde{A}_{z e}<0,\nonumber\\
&&\tilde{p}_{z e}^{ -}-R(\tilde{p}_{z e}^{ -})-\tilde{A}_{z e}>0.
\end{eqnarray}

\section*{Tables \& Figures follow}
\clearpage
\begin{widetext}

\begin{table*}
 \centering
 \begin{minipage}{140mm}
  \caption{Dimensionless form of some important variables.\\ The $s$ subscript refers to particles of species $s$.}
  \begin{tabular}{cc}
  \hline
       Variable & Dimensionless form\\
       \hline
  Particle Hamiltonian      & $ \tilde{\mathcal{H}}_s=\beta_s\mathcal{H}_s       $  \\
  Particle angular momentum     &$      \tau p_{\theta s}=m_sv_{th,s}\tilde{p}_{\theta s}  $     \\
  Particle $z$-Momentum          &$          p_{zs}=m_sv_{th,s}\tilde{p}_{zs}   $ \\
 Vector potential  &        $ q_s\mathbf{A}=m_sv_{th,s}\tilde{\mathbf{A}}_s   $ \\
Scalar Potential  &         $ \tilde{\phi}_s=q_s\beta_s\phi  $\\
Bulk rectilinear flows  &    $      v_{th,s}\tilde{{U}}_{zs}={U}_{zs},\hspace{3mm}v_{th,s}\tilde{{V}}_{zs}={V}_{zs}$    \\
Bulk angular frequency  &     $      \tau v_{th,s}\tilde{\omega}_s=\omega_{s}     $ \\
Particle velocity  &       $\mathbf{v}=v_{th,s}\tilde{\mathbf{v}}_s       $ \\
\hline
\label{tab:norm}
\end{tabular}
\end{minipage}
\end{table*}

\begin{table*}
 \centering
 \begin{minipage}{140mm}
  \caption{The fundamental parameters of the equilibrium.\\ The $s$ subscript refers to particles of species $s$.}
  \begin{tabular}{cccc}
  \hline
Macroscopic&&Microscopic& \\
parameter&Meaning&parameter&Meaning\\
\hline
$B_0$&Characteristic magnetic field strength&$m_s$& Mass of particle\\
$\tau$&Measure of the twist of flux tube&$q_s$, $q$& Charge, magnitude of charge\\
$k$&Strength of the background field&$\beta_s=1/(k_BT_s)$&  Thermal beta\\
$\gamma_1\ne 0,1$, $0<\gamma_2 <1$&Gauge for scalar potential&$v_{th,s}$&  Thermal velocity\\
${U}_{zs}, {V}_{zs}$&Bulk rectilinear flows&$\delta_s(r), \delta_s$&   Magnetisation parameters\\
$ \omega_{s}$&Bulk angular frequency&$n_{0s}$&Normalistaion of particle number\\
 
\hline
\label{tab:param}
\end{tabular}
\end{minipage}
\end{table*}

\begin{figure}
    \centering
    \begin{subfigure}[b]{0.25\textwidth}
        \includegraphics[width=\textwidth]{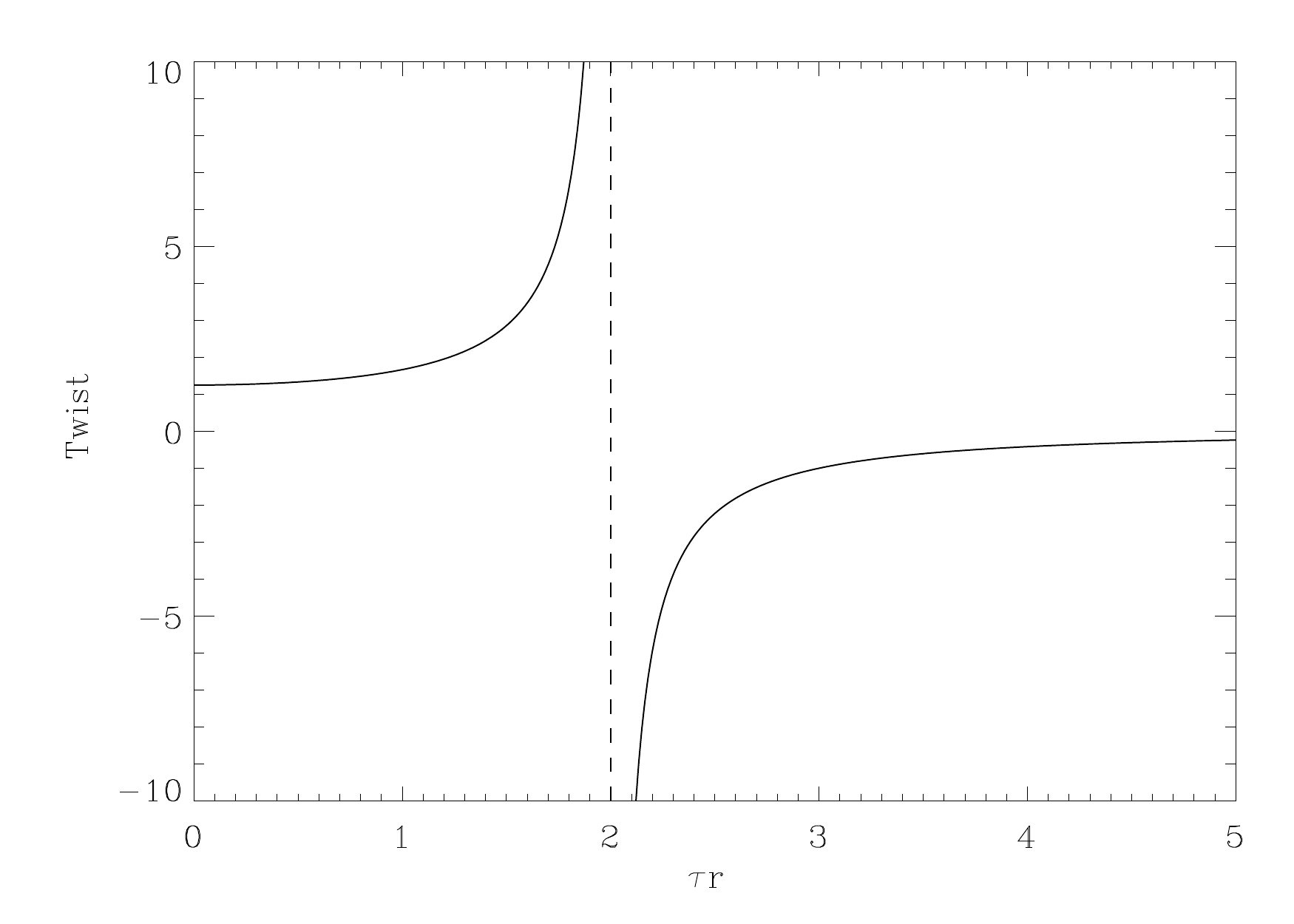}
        \caption{$k=0.1$}
        \label{fig:twist1}
    \end{subfigure}
       \begin{subfigure}[b]{0.25\textwidth}
        \includegraphics[width=\textwidth]{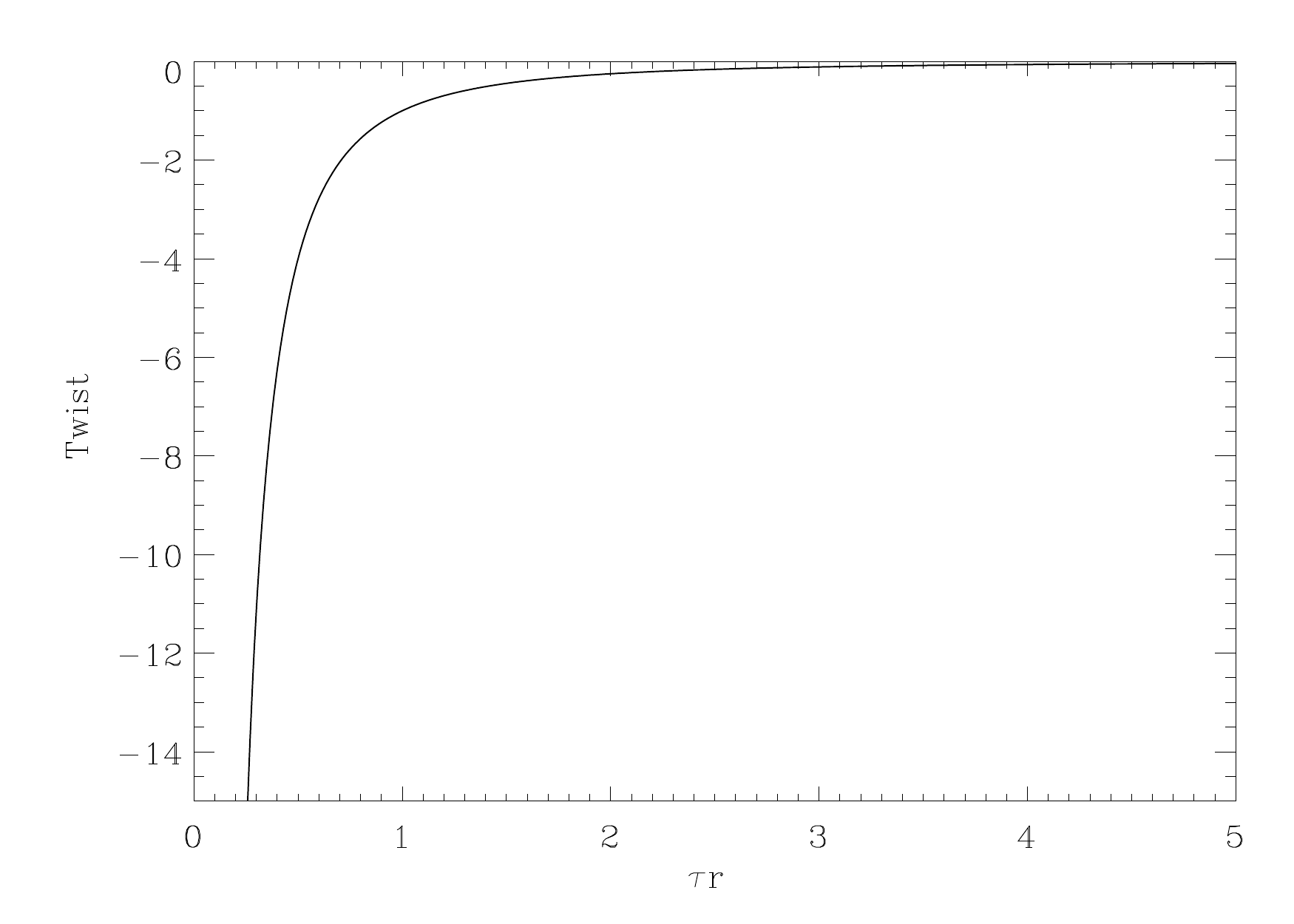}
        \caption{$k=0.5$}
        \label{fig:twist2}
    \end{subfigure}
        \begin{subfigure}[b]{0.25\textwidth}
        \includegraphics[width=\textwidth]{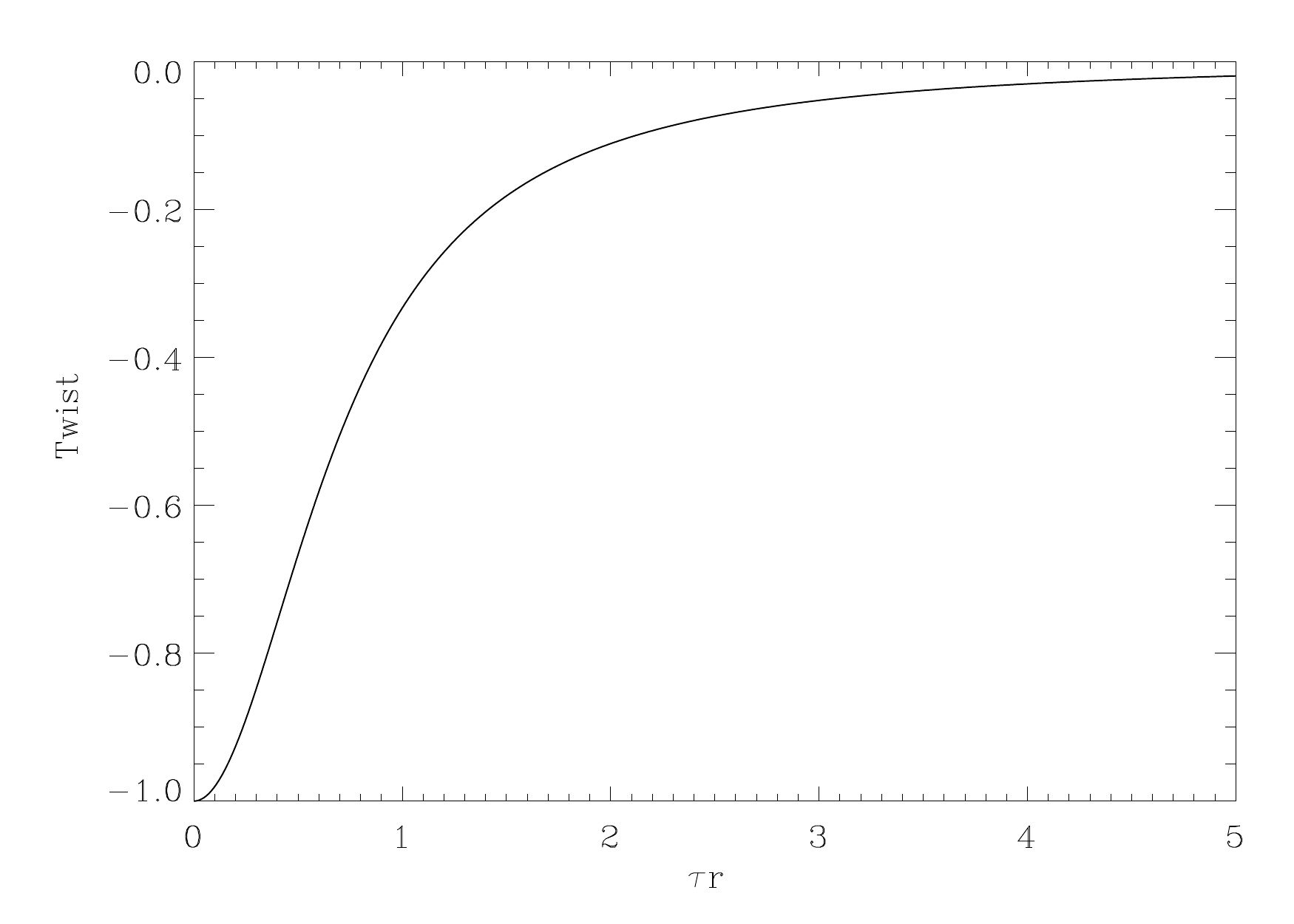}
        \caption{$k=1$}
        \label{fig:twist3}
    \end{subfigure}
    \caption{The twist (normalised by $\tau/(2\pi)$) of the GH+B field for three values of $k$. \ref{fig:twist1} shows the twist for $k<1/2$, and as such there are both negative and positive twists, due to the field reversal. \ref{fig:twist2} and \ref{fig:twist3} both show negative twist, since there is no magnetic field reversal. }\label{fig:twist}
\end{figure}

\begin{figure}
    \centering
    \begin{subfigure}[b]{0.17\textwidth}
        \includegraphics[width=\textwidth]{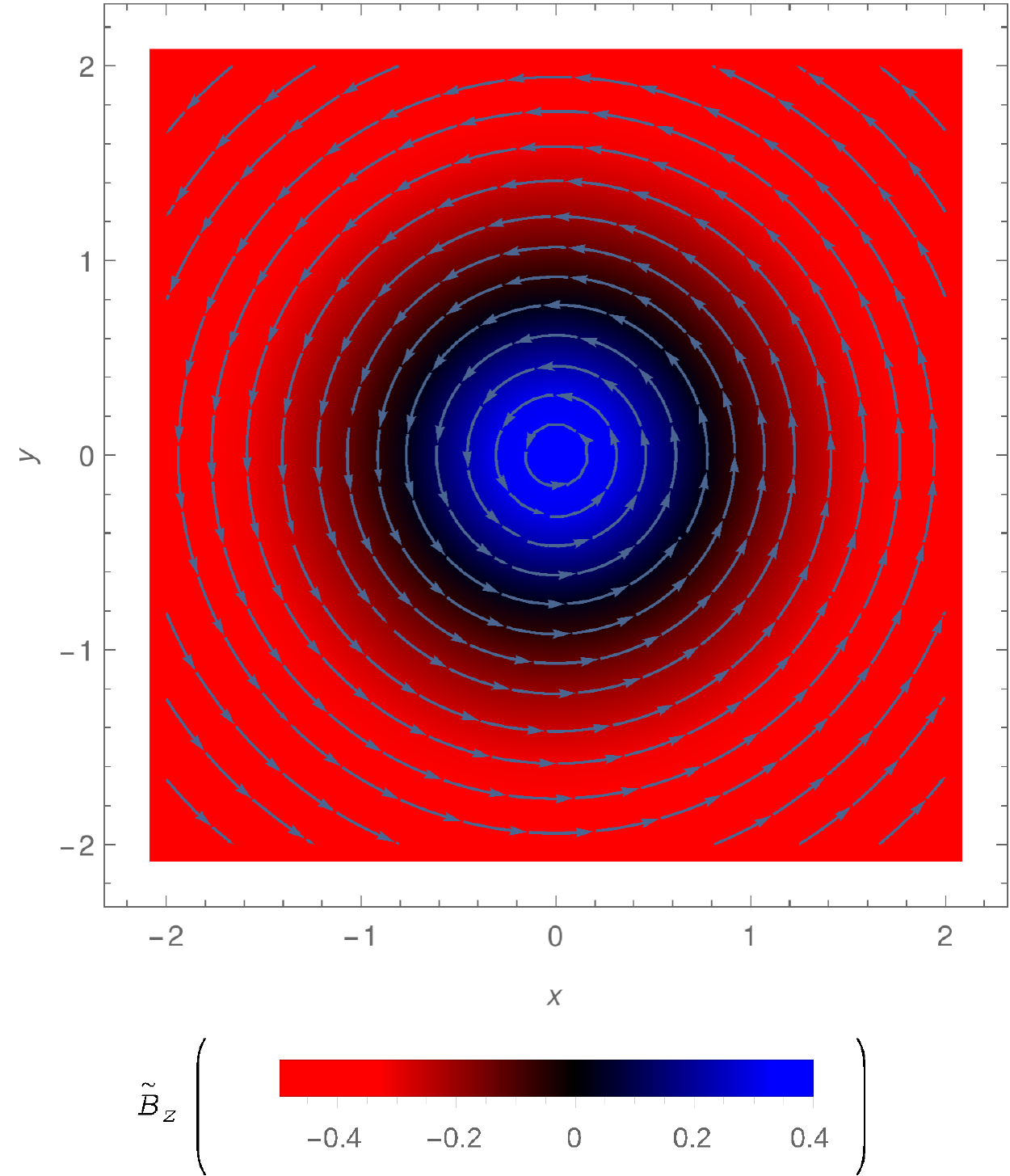}
        \caption{$\mathbf{B}$ for $k=0.3$}
        \label{fig:0.3}
    \end{subfigure}
       \begin{subfigure}[b]{0.17\textwidth}
        \includegraphics[width=\textwidth]{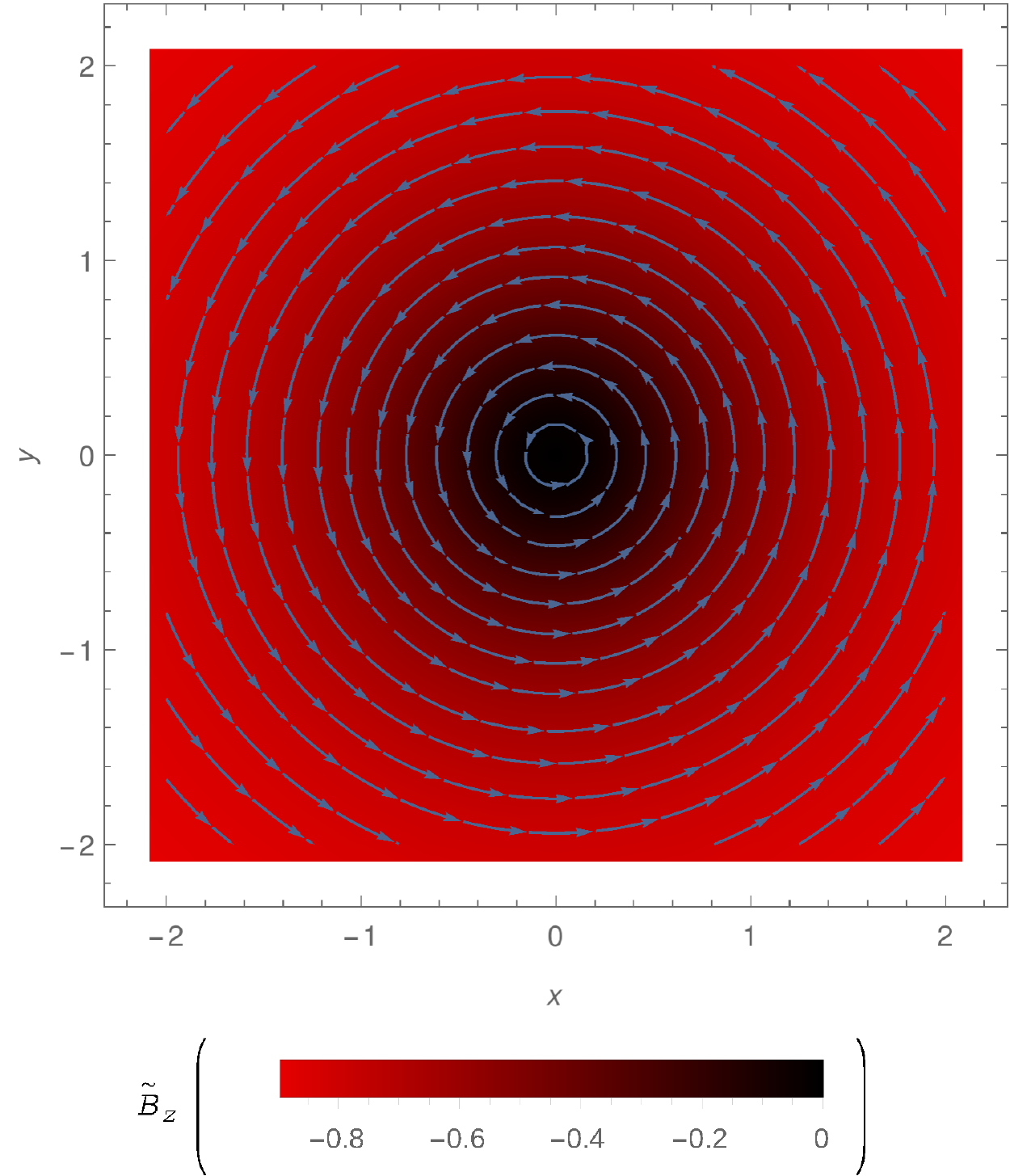}
        \caption{$\mathbf{B}$ for $k=0.5$}
        \label{fig:0.5}
    \end{subfigure}
        \begin{subfigure}[b]{0.25\textwidth}
        \includegraphics[width=\textwidth]{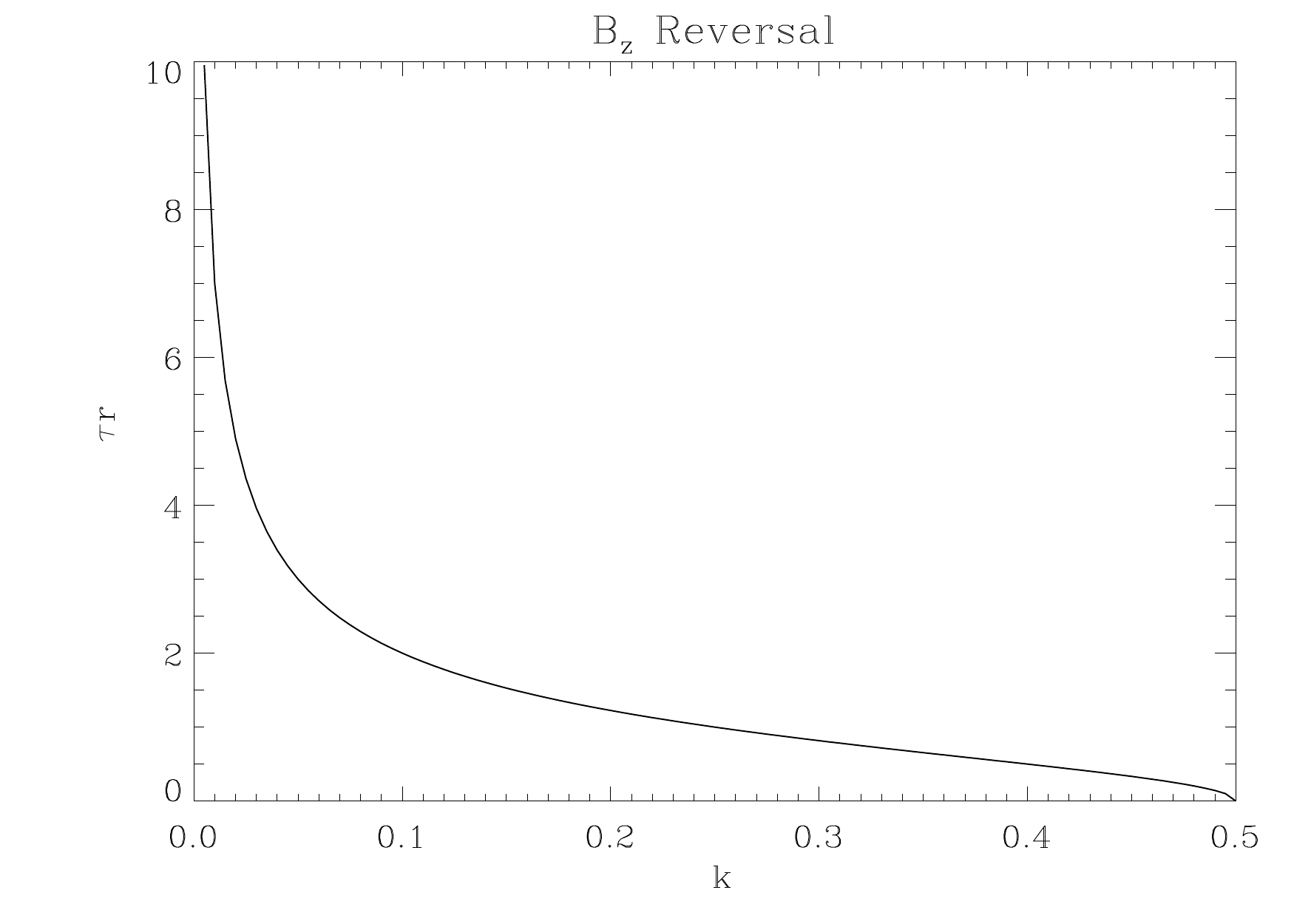}
        \caption{Radius of $B_z$ reversal, given $0<k<1/2$}
        \label{fig:rev}
    \end{subfigure}
    \caption{\ref{fig:0.3} and \ref{fig:0.5} show the GH+B magnetic field in the $xy$ plane, for two values of $k$. The curved arrows indicate the direction of the $\tilde{B}_\theta$ components, whilst the blue-black-red shading denotes the magnitude and direction of the $\tilde{B}_z$ component. The $k=0.3$ case contains a reversal of the $\tilde{B}_z$ field direction and as such is a Reversed Field Pinch whilst $k=0.5$ gives zero $\tilde{B}_z$ at $\tilde{r}=0$. \ref{fig:rev} shows the radius at which $\tilde{B}_z$ changes its direction, for $0<k<1/2$. $\tilde{B}_z$ does not reverse for $k\ge 1/2$. }\label{fig:magfield}
\end{figure}

\begin{figure}
    \centering
    \begin{subfigure}[b]{0.25\textwidth}
        \includegraphics[width=\textwidth]{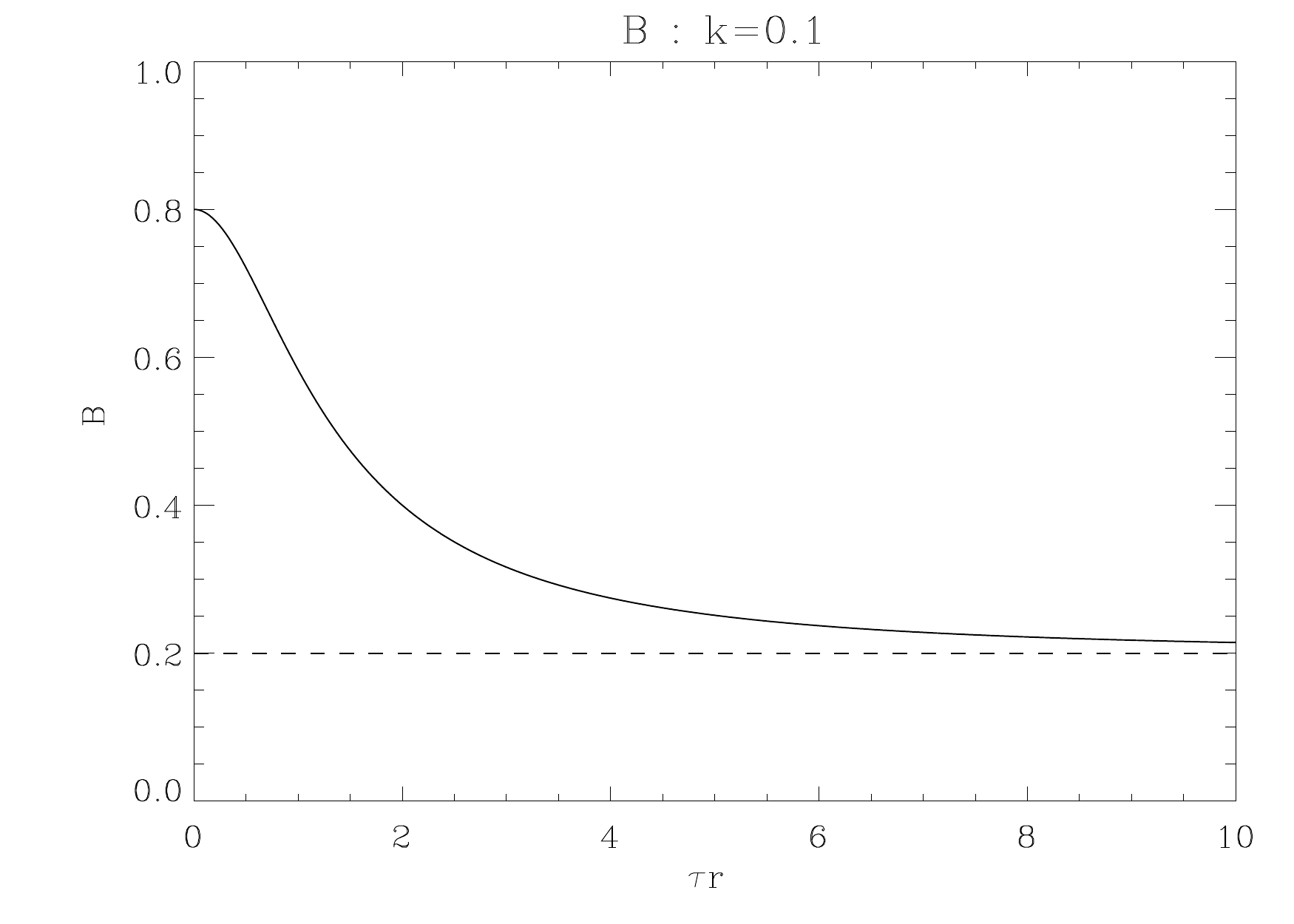}
        \caption{$|\mathbf{B}|$ for $k=0.1$}
        \label{fig:bmag1}
    \end{subfigure}
       \begin{subfigure}[b]{0.25\textwidth}
        \includegraphics[width=\textwidth]{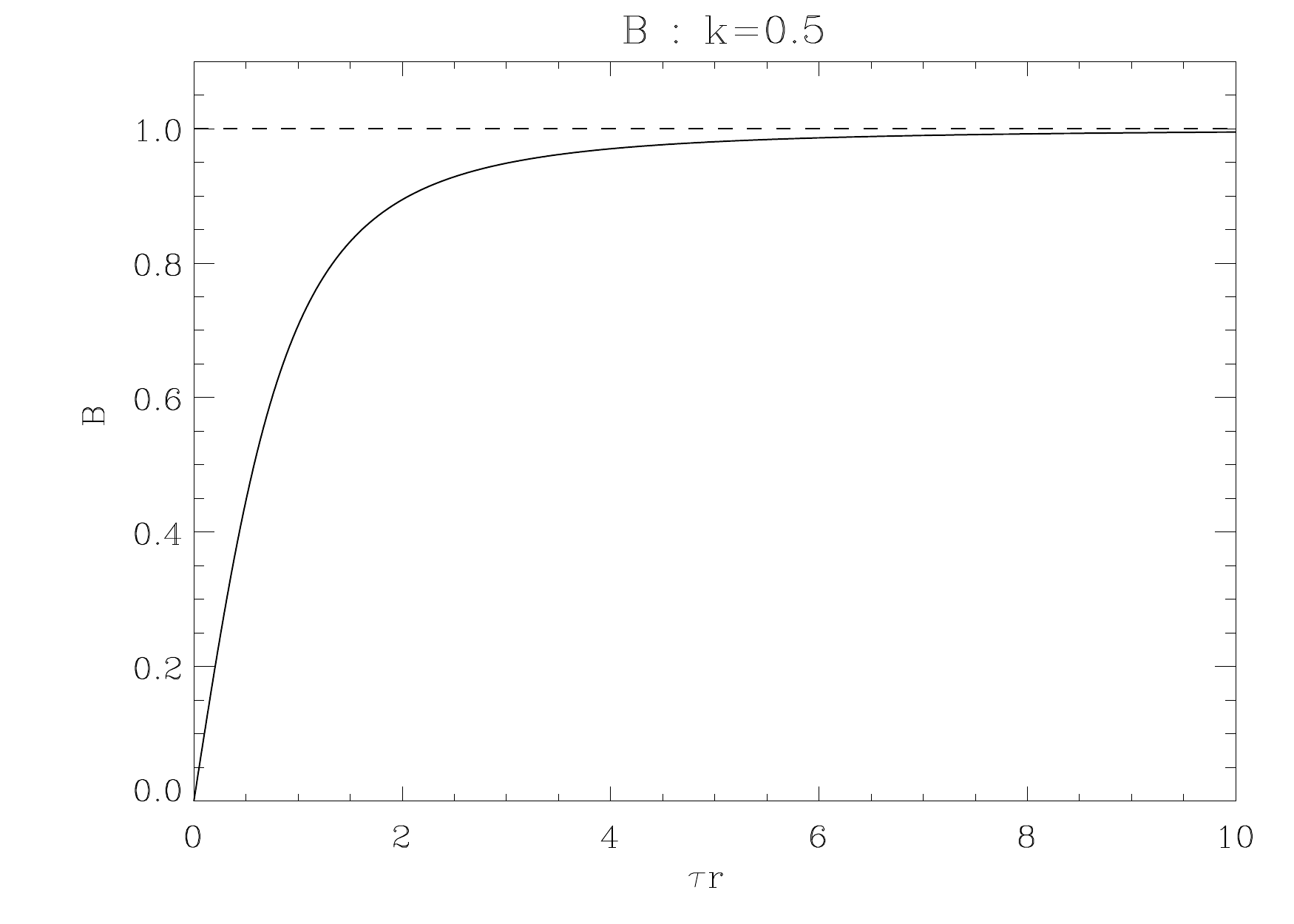}
        \caption{ $|\mathbf{B}|$ for $k=0.5$ }
        \label{fig:bmag2}
    \end{subfigure}
        \begin{subfigure}[b]{0.25\textwidth}
        \includegraphics[width=\textwidth]{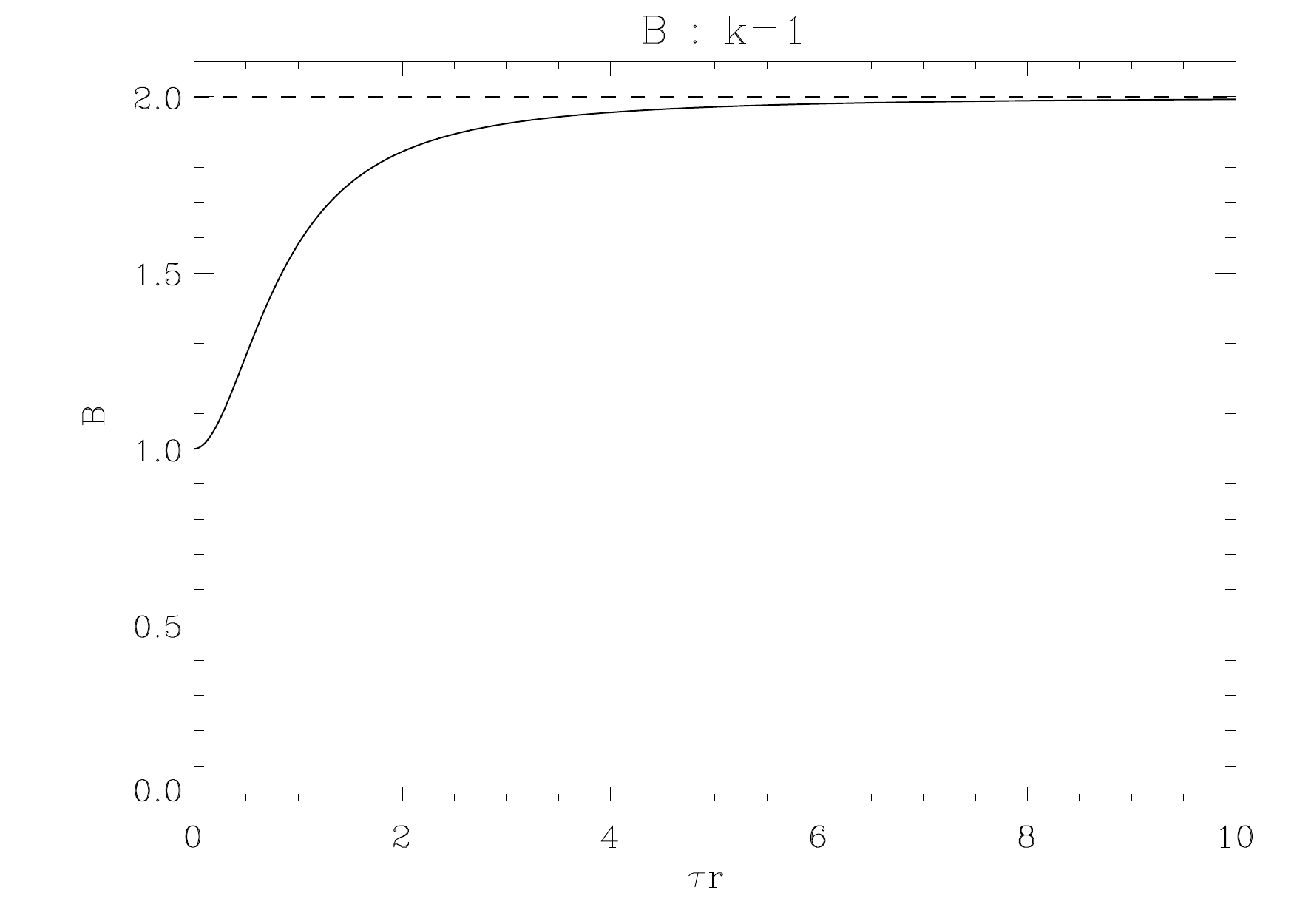}
        \caption{ $|\mathbf{B}|$ for $k=1$  }
        \label{fig:bmag3}
    \end{subfigure}
    \caption{\ref{fig:bmag1}-\ref{fig:bmag3} show the magnitude of the GH+B magnetic field for $k=0.1, 0.5$ and $k=1$ respectively, normalised by $B_0$. For $k<0.5$, $|\tilde{\mathbf{B}}|\to 2k$ from above, whereas for $k\ge 1/2$, $|\tilde{\mathbf{B}}|\to 2k$ from below.}\label{fig:bmag}
\end{figure}

\begin{figure}
    \centering
    \begin{subfigure}[b]{0.245\textwidth}
        \includegraphics[width=\textwidth]{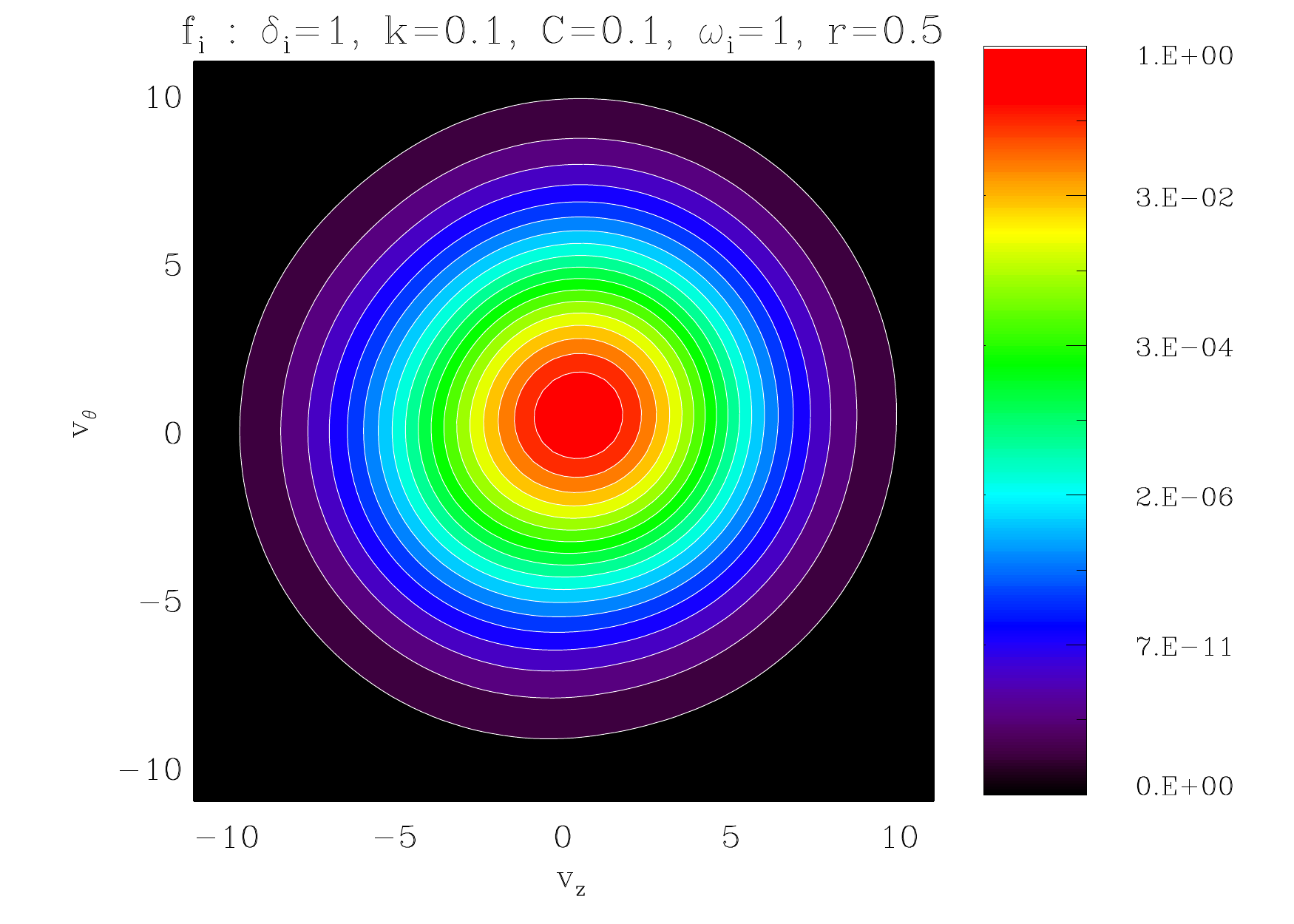}
        \caption{$(\tilde{\omega}_i,\tilde{r},C_i)=(1,0.5,0.1)$}
        \label{fig:4a}
    \end{subfigure}
       \begin{subfigure}[b]{0.245\textwidth}
        \includegraphics[width=\textwidth]{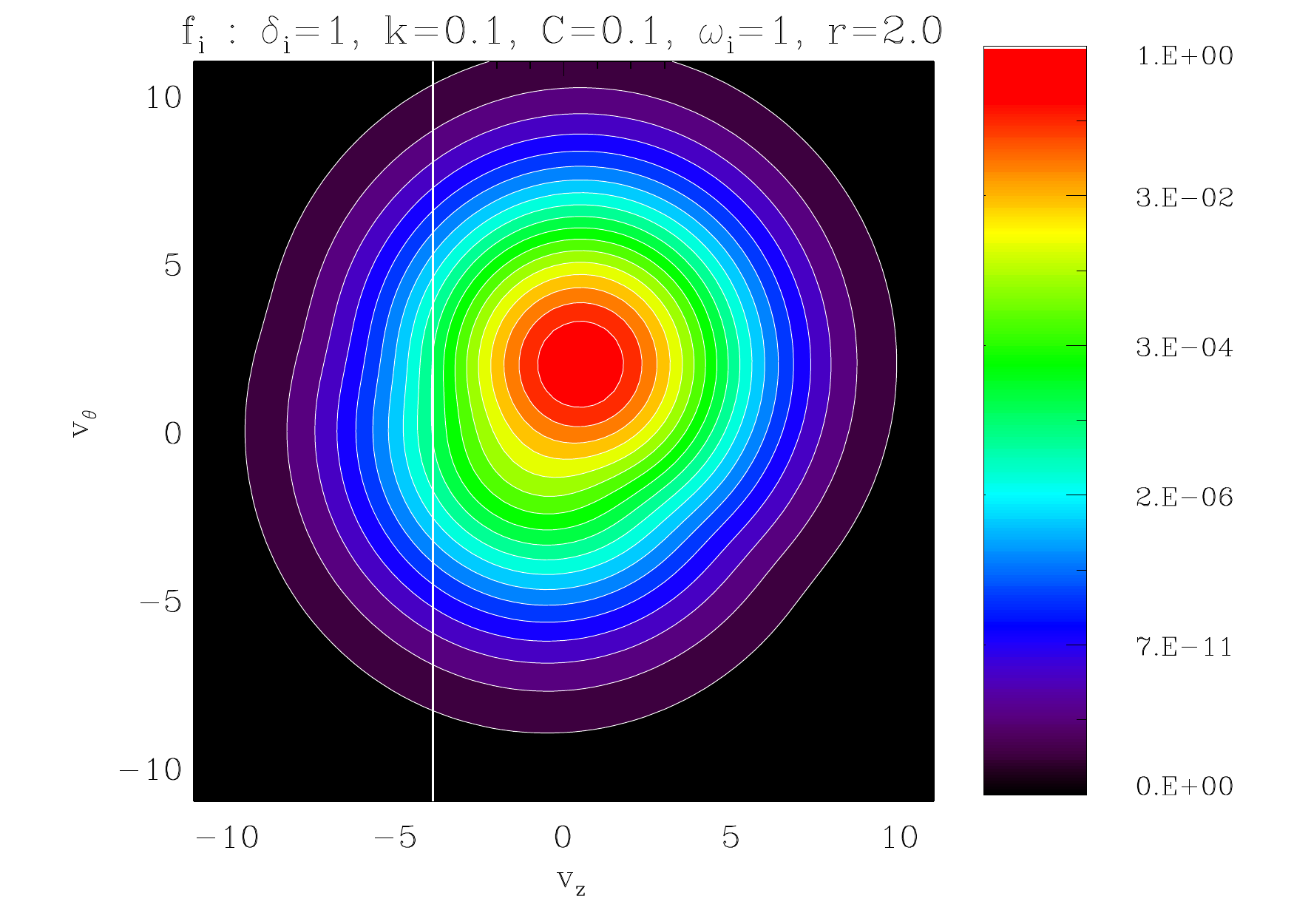}
        \caption{$(\tilde{\omega}_i,\tilde{r},C_i)=(1,2,0.1)$}
        \label{fig:4b}
    \end{subfigure}
        \begin{subfigure}[b]{0.245\textwidth}
        \includegraphics[width=\textwidth]{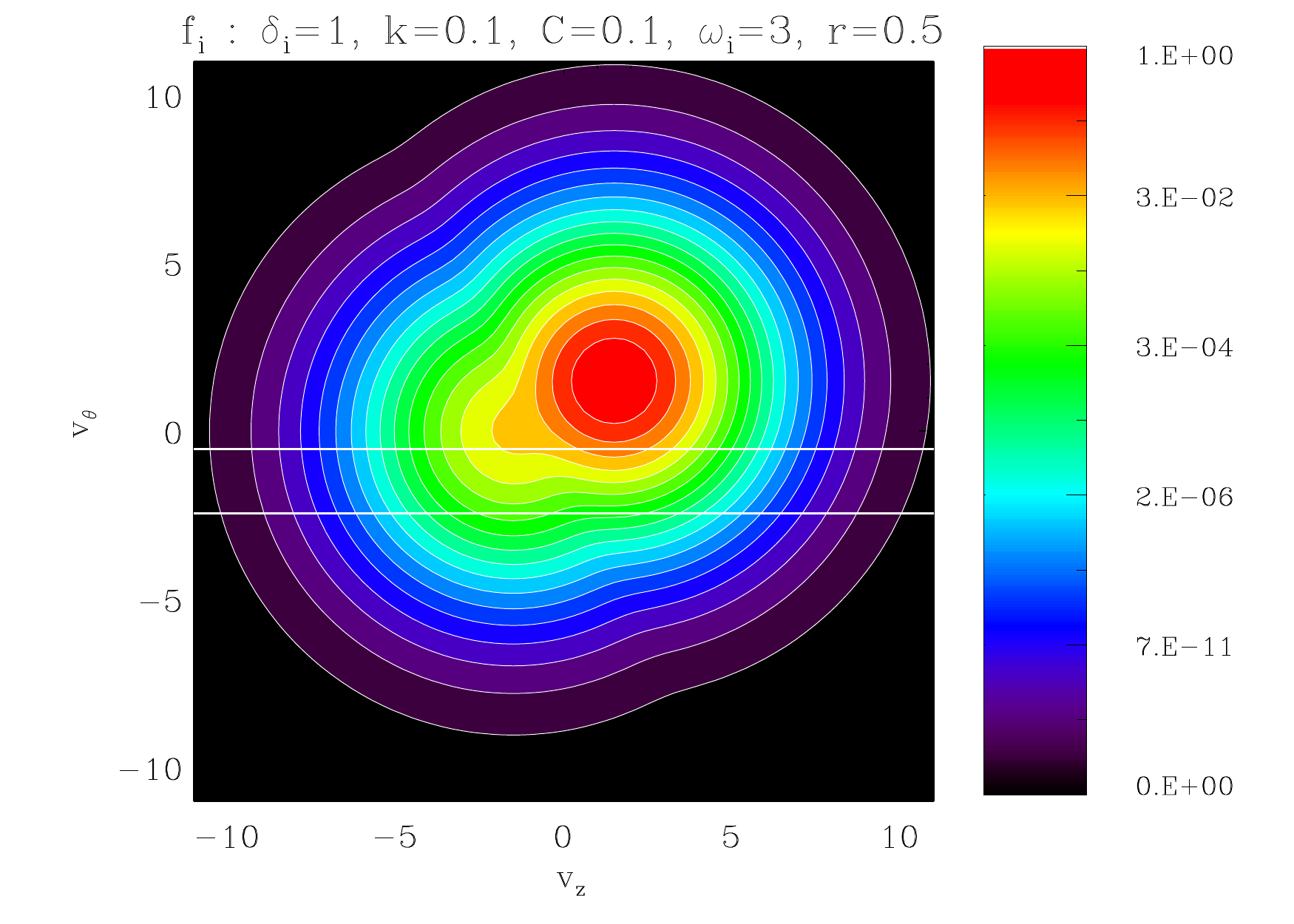}
        \caption{$(\tilde{\omega}_i,\tilde{r},C_i)=(3,0.5,0.1)$}
        \label{fig:4c}
    \end{subfigure}
    \begin{subfigure}[b]{0.245\textwidth}
        \includegraphics[width=\textwidth]{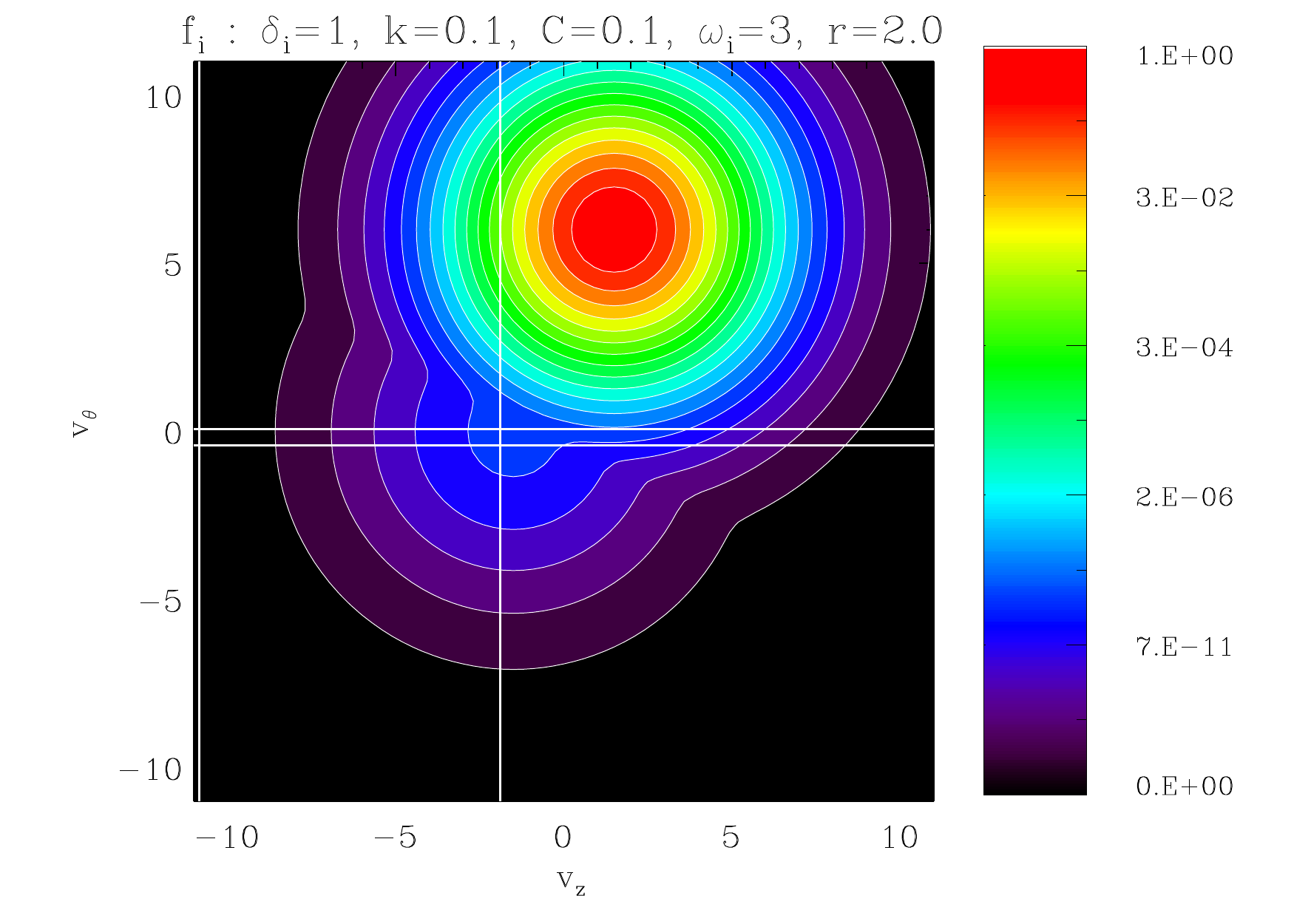}
        \caption{$(\tilde{\omega}_i,\tilde{r},C_i)=(3,2,0.1)$}
        \label{fig:4d}
    \end{subfigure}
        \begin{subfigure}[b]{0.245\textwidth}
        \includegraphics[width=\textwidth]{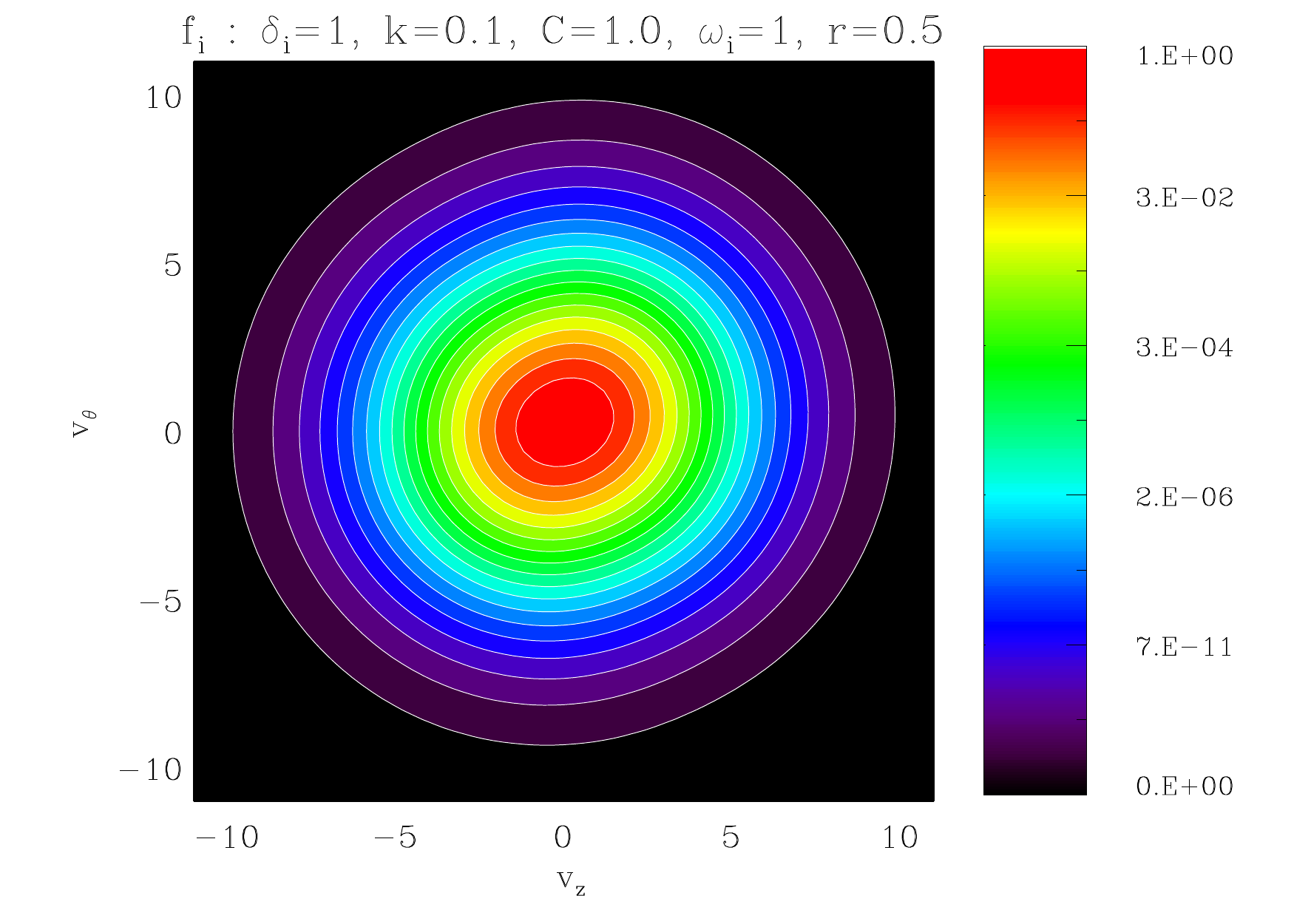}
        \caption{$(\tilde{\omega}_i,\tilde{r},C_i)=(1,0.5,1)$}
        \label{fig:4e}
    \end{subfigure}
    \begin{subfigure}[b]{0.245\textwidth}
        \includegraphics[width=\textwidth]{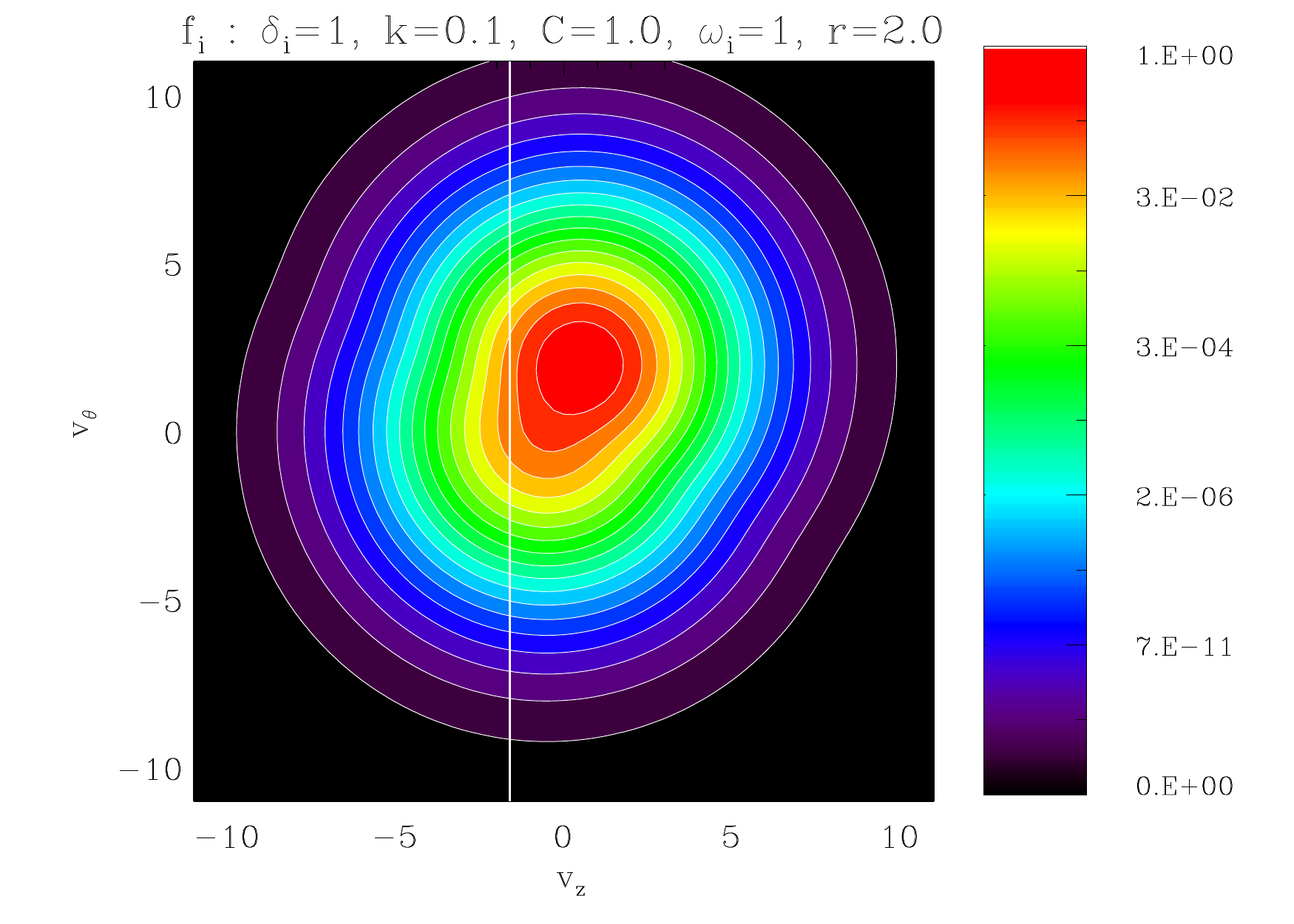}
        \caption{$(\tilde{\omega}_i,\tilde{r},C_i)=(1,2,1)$}
        \label{fig:4f}
    \end{subfigure}
        \begin{subfigure}[b]{0.245\textwidth}
        \includegraphics[width=\textwidth]{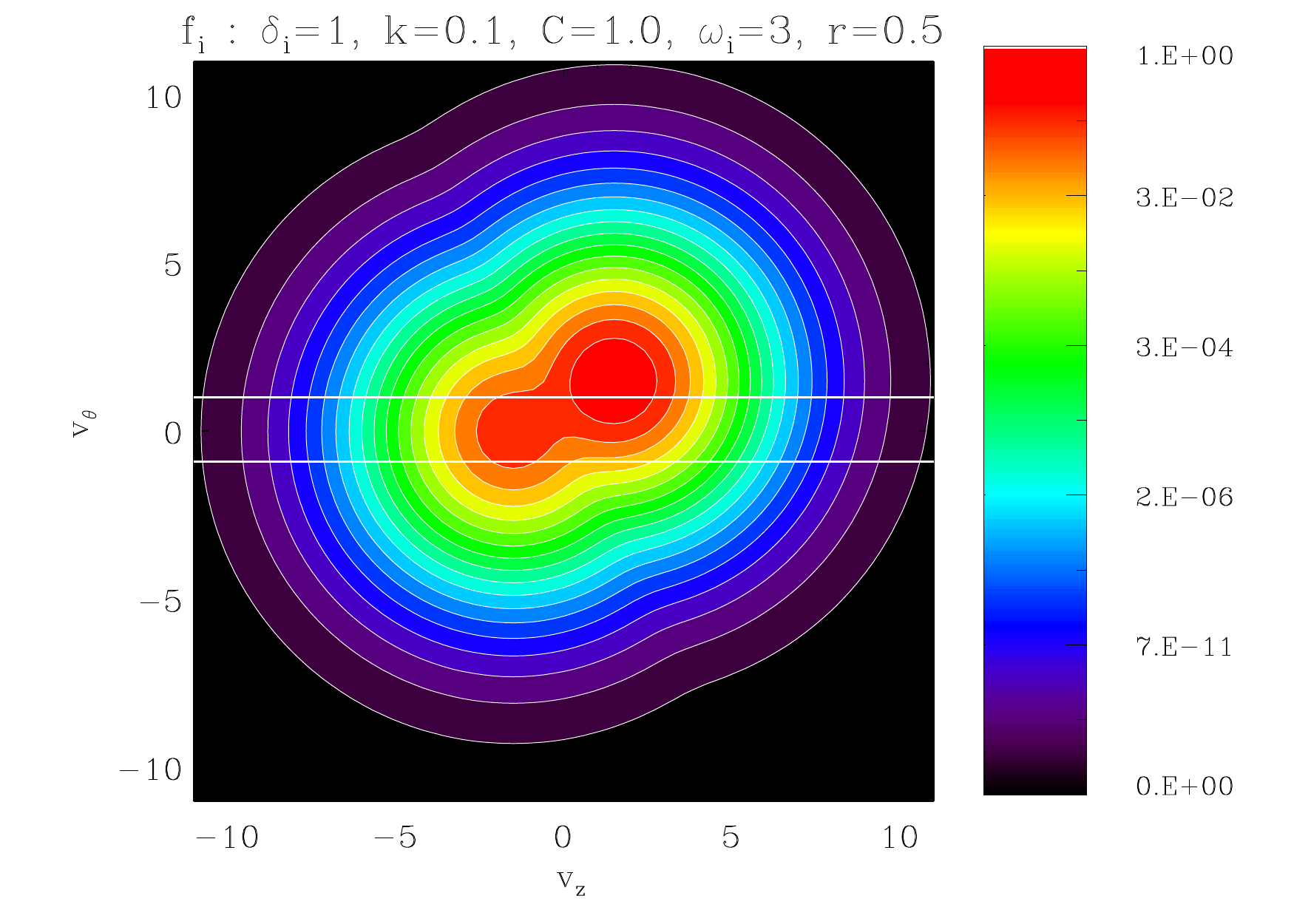}
        \caption{$(\tilde{\omega}_i,\tilde{r},C_i)=(3,0.5,1)$}
        \label{fig:4g}
    \end{subfigure}
     \begin{subfigure}[b]{0.245\textwidth}
        \includegraphics[width=\textwidth]{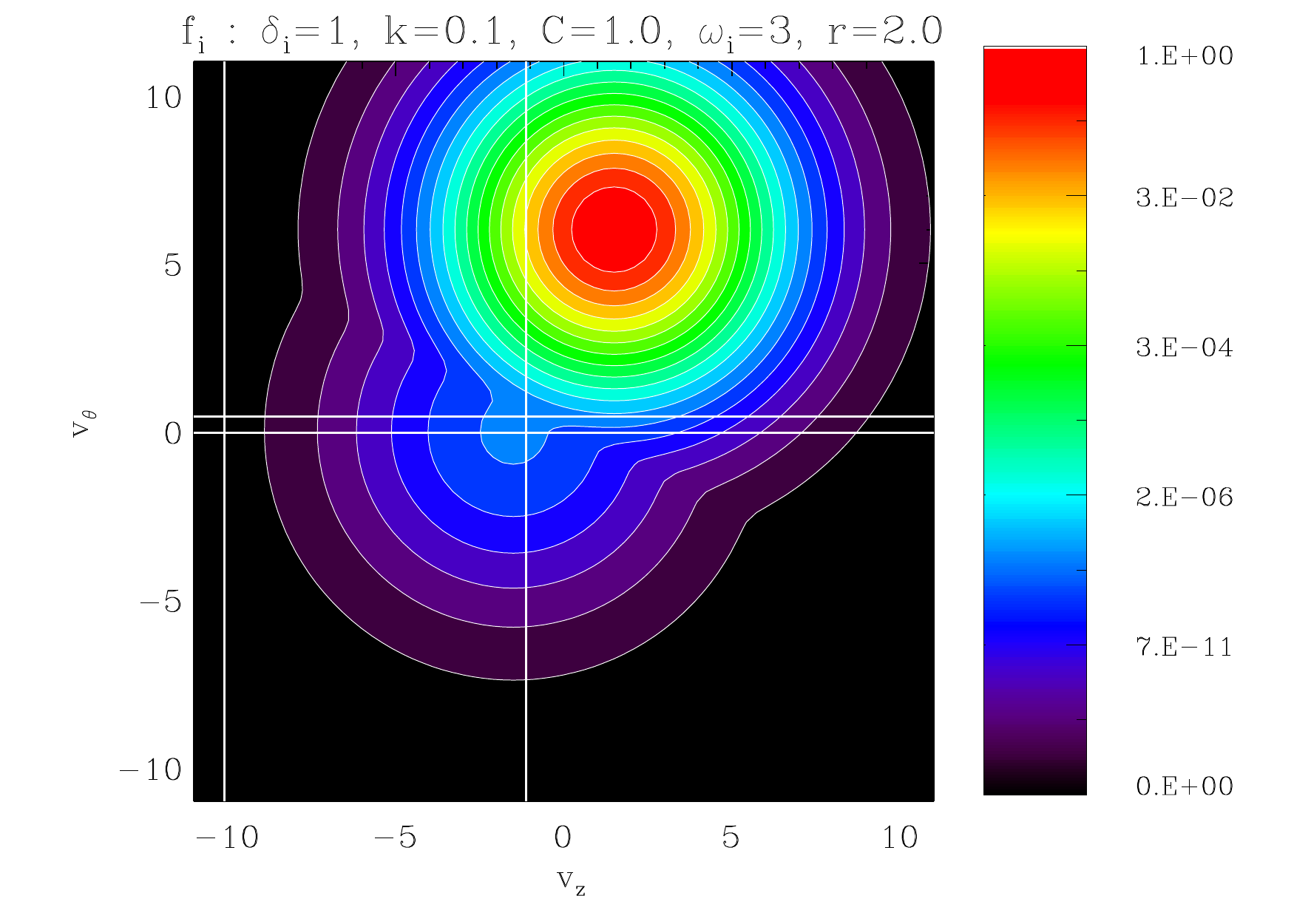}
        \caption{$(\tilde{\omega}_i,\tilde{r},C_i)=(3,2,1)$}
        \label{fig:4h}
    \end{subfigure}
    \caption{Contour plots of the $f_i$ in $(\tilde{v}_z,\tilde{v}_{\theta})$ space for an equilibrium with field reversal ($k=0.1<0.5$), for a variety of parameters ($\tilde{\omega}_i,\tilde{r}, C_i$) and $\delta_i=1$. The white horizontal/vertical lines indicate the regions in which multiple maxima in either the $\tilde{v}_z$ or $\tilde{v}_{z}$ directions can occur, if at all. A single line indicates that the `region' is a line.   }\label{fig:4}
\end{figure}

\begin{figure}
    \centering
    \begin{subfigure}[b]{0.245\textwidth}
        \includegraphics[width=\textwidth]{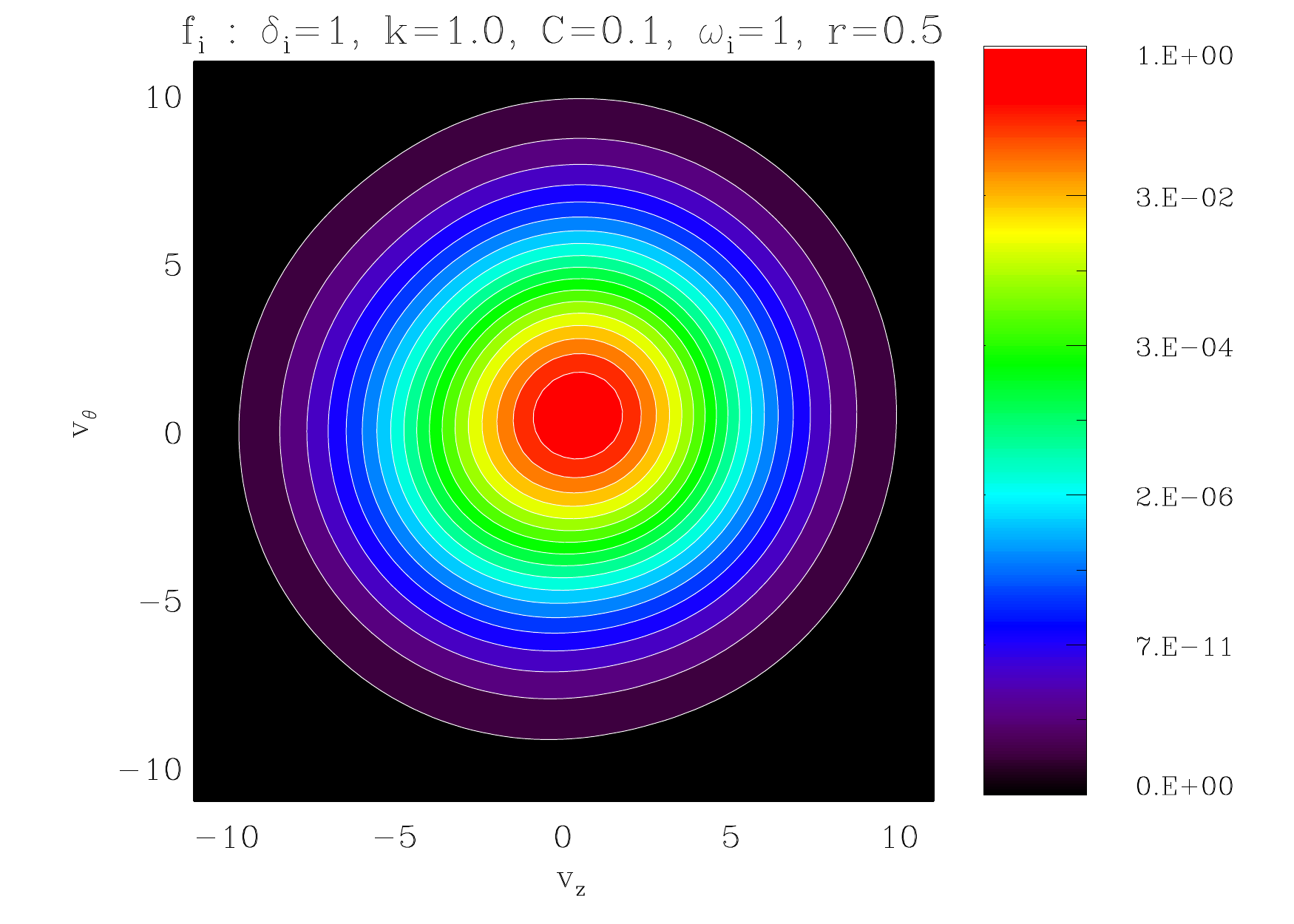}
        \caption{$(\tilde{\omega}_i,\tilde{r},C_i)=(1,0.5,0.1)$}
        \label{fig:5a}
    \end{subfigure}
       \begin{subfigure}[b]{0.245\textwidth}
        \includegraphics[width=\textwidth]{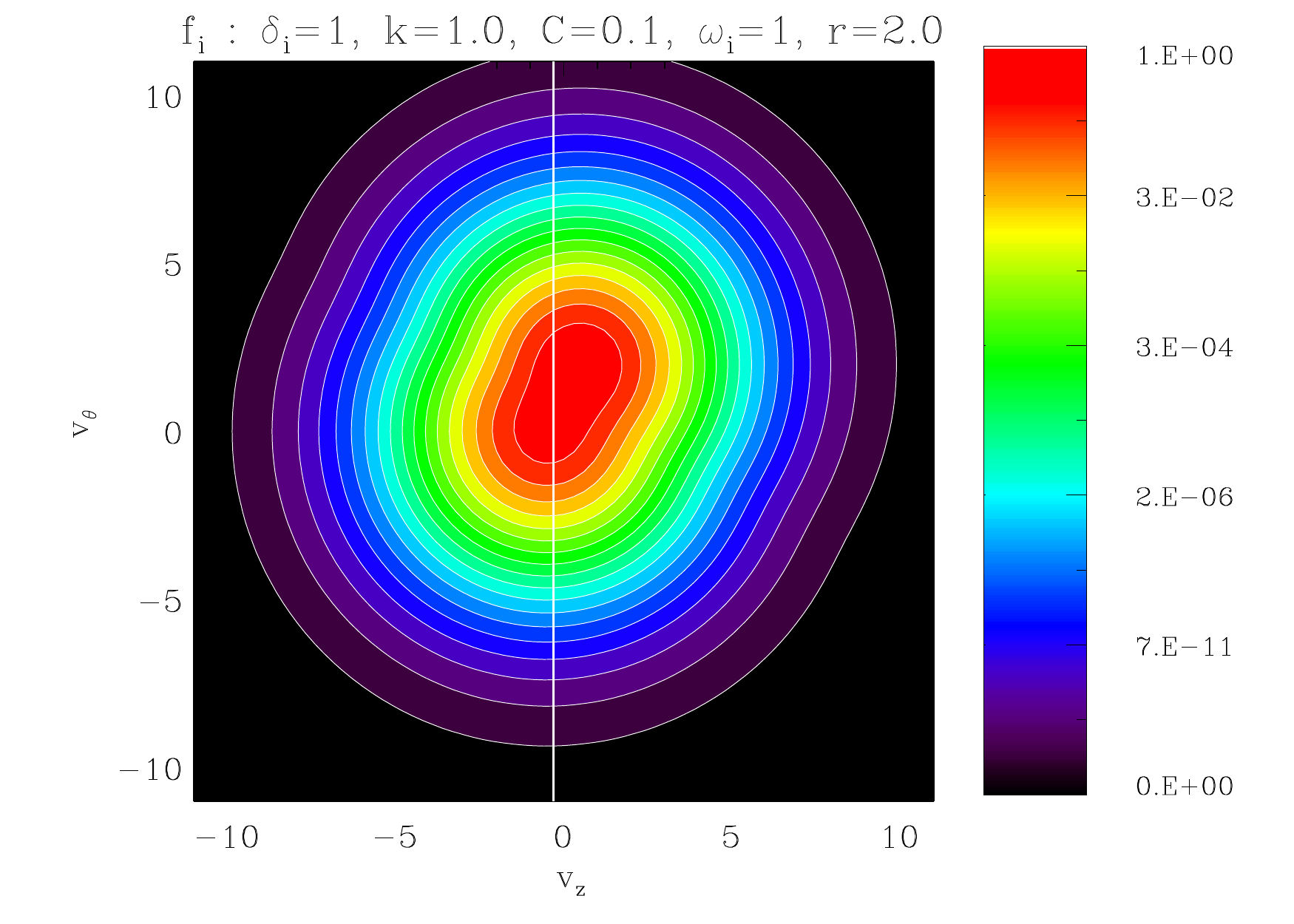}
        \caption{$(\tilde{\omega}_i,\tilde{r},C_i)=(1,2,0.1)$}
        \label{fig:5b}
    \end{subfigure}
        \begin{subfigure}[b]{0.245\textwidth}
        \includegraphics[width=\textwidth]{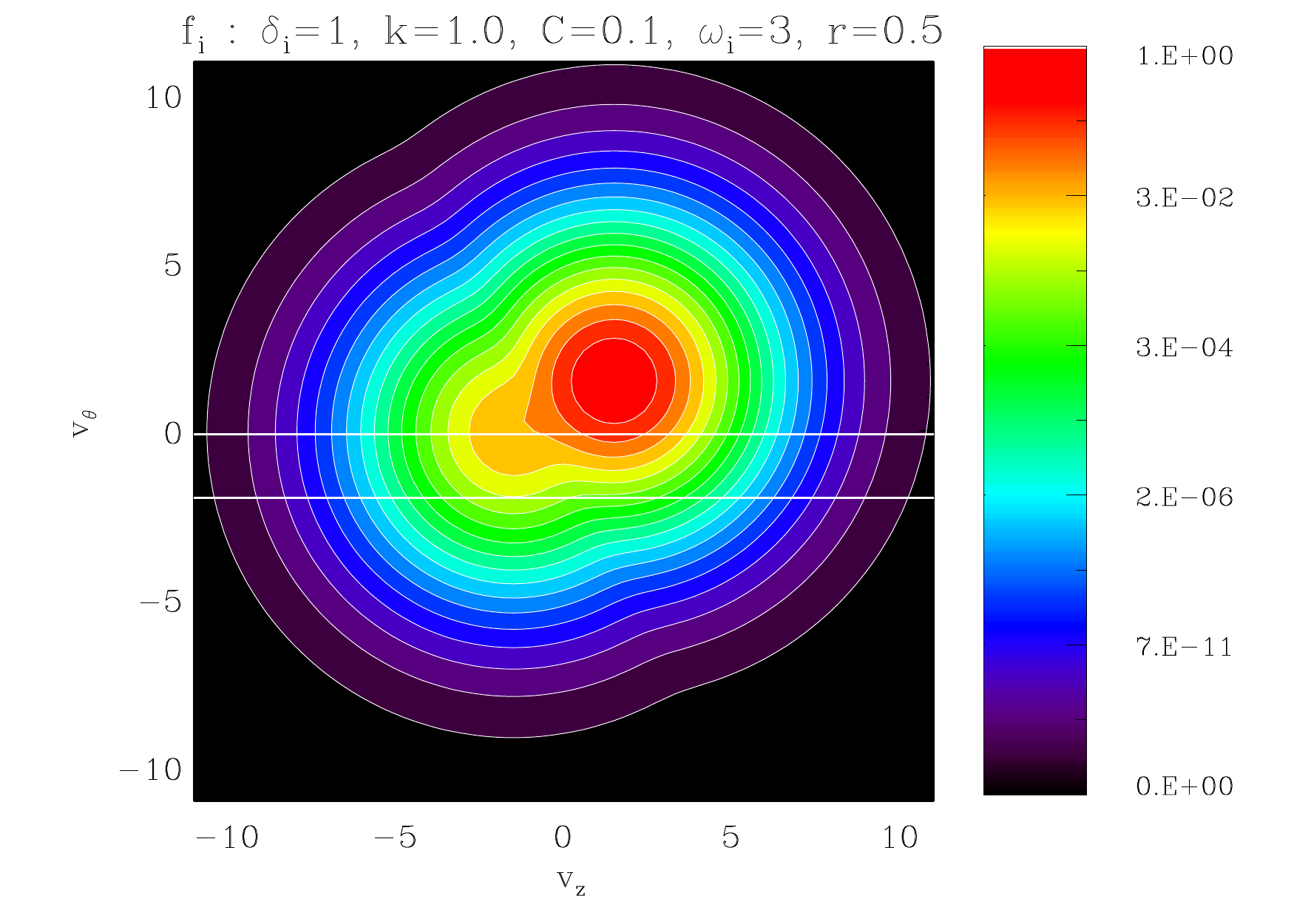}
        \caption{$(\tilde{\omega}_i,\tilde{r},C_i)=(3,0.5,0.1)$}
        \label{fig:5c}
    \end{subfigure}
    \begin{subfigure}[b]{0.245\textwidth}
        \includegraphics[width=\textwidth]{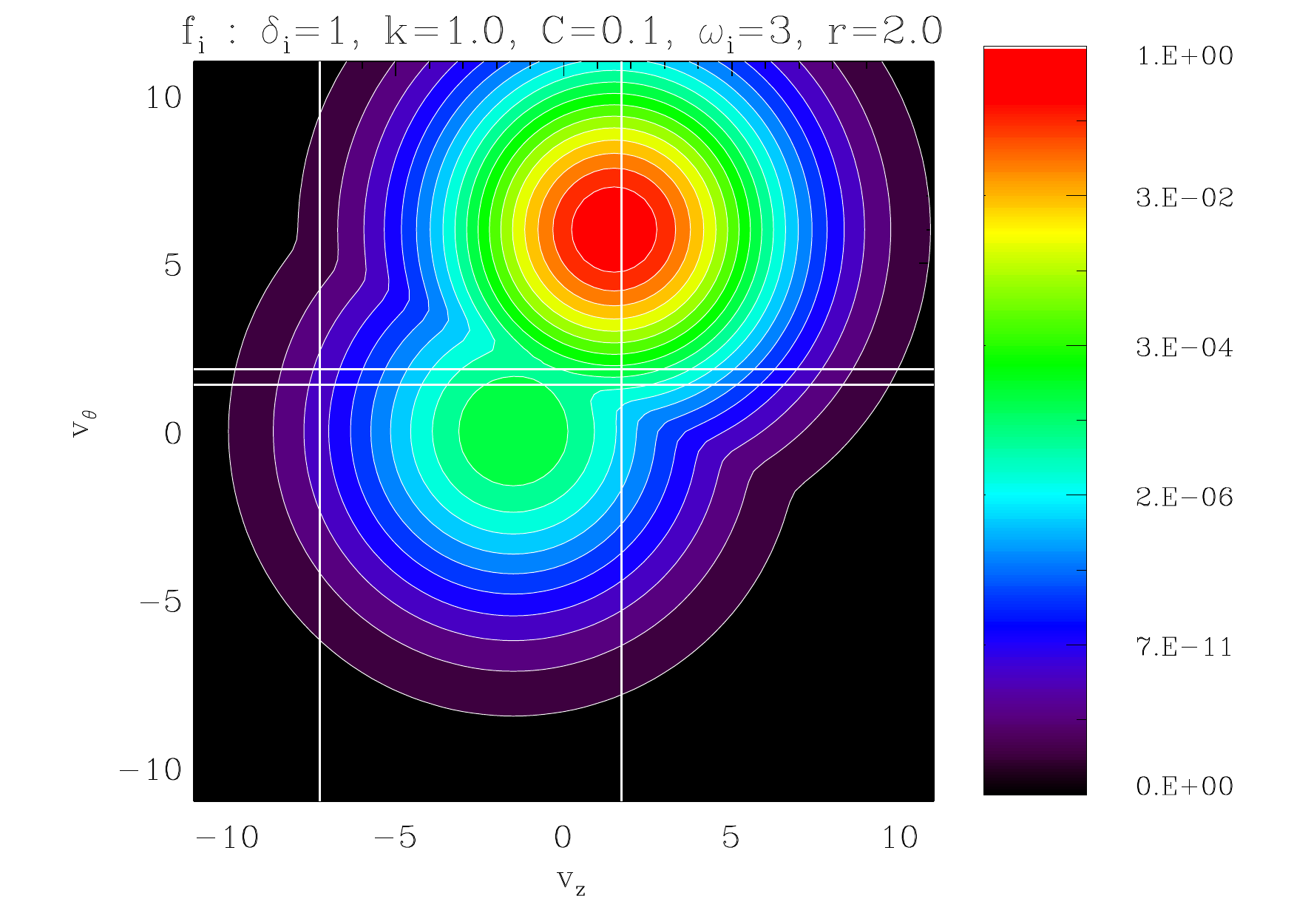}
        \caption{$(\tilde{\omega}_i,\tilde{r},C_i)=(3,2,0.1)$}
        \label{fig:5d}
    \end{subfigure}
        \begin{subfigure}[b]{0.245\textwidth}
        \includegraphics[width=\textwidth]{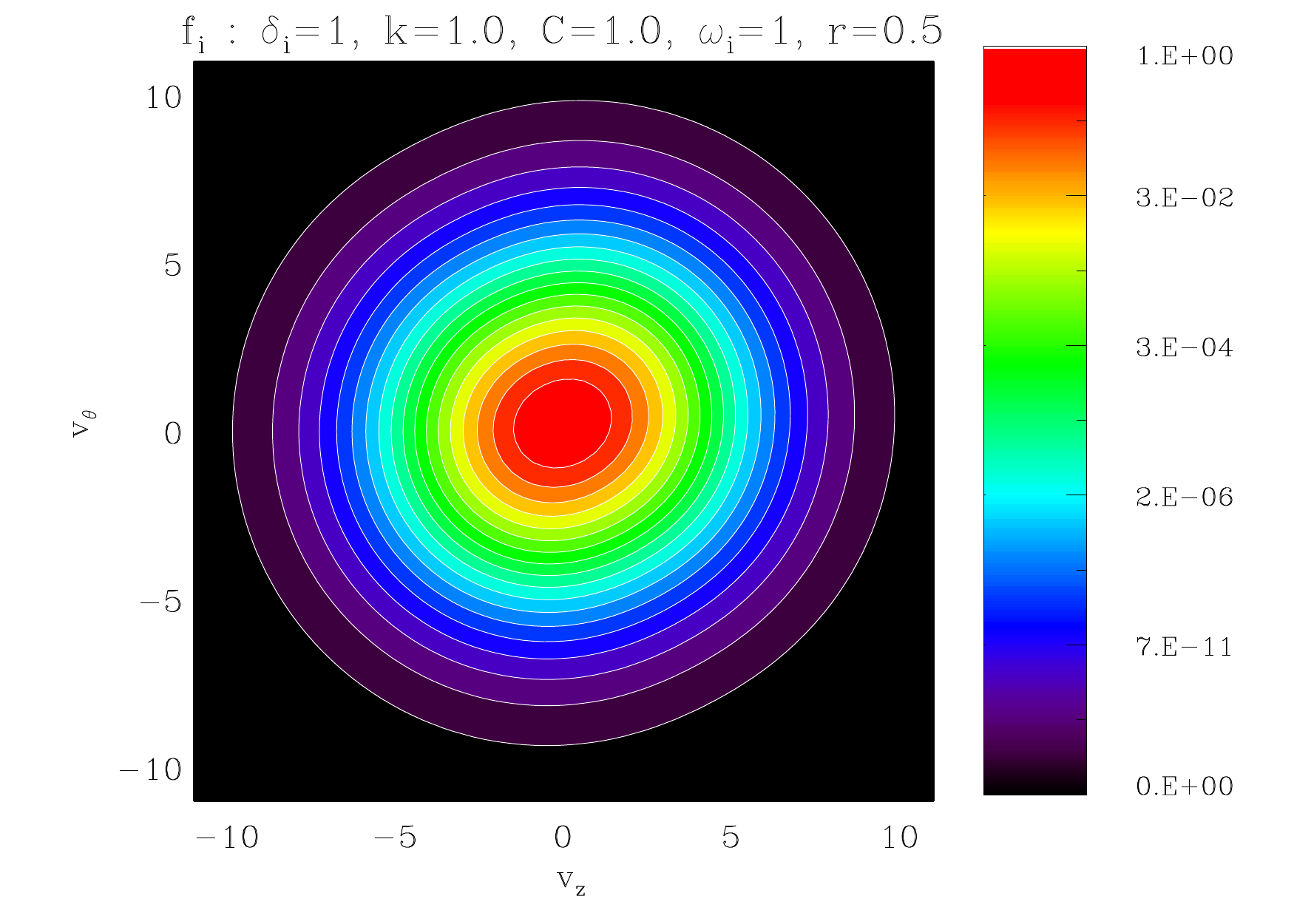}
        \caption{$(\tilde{\omega}_i,\tilde{r},C_i)=(1,0.5,1)$}
        \label{fig:5e}
    \end{subfigure}
    \begin{subfigure}[b]{0.245\textwidth}
        \includegraphics[width=\textwidth]{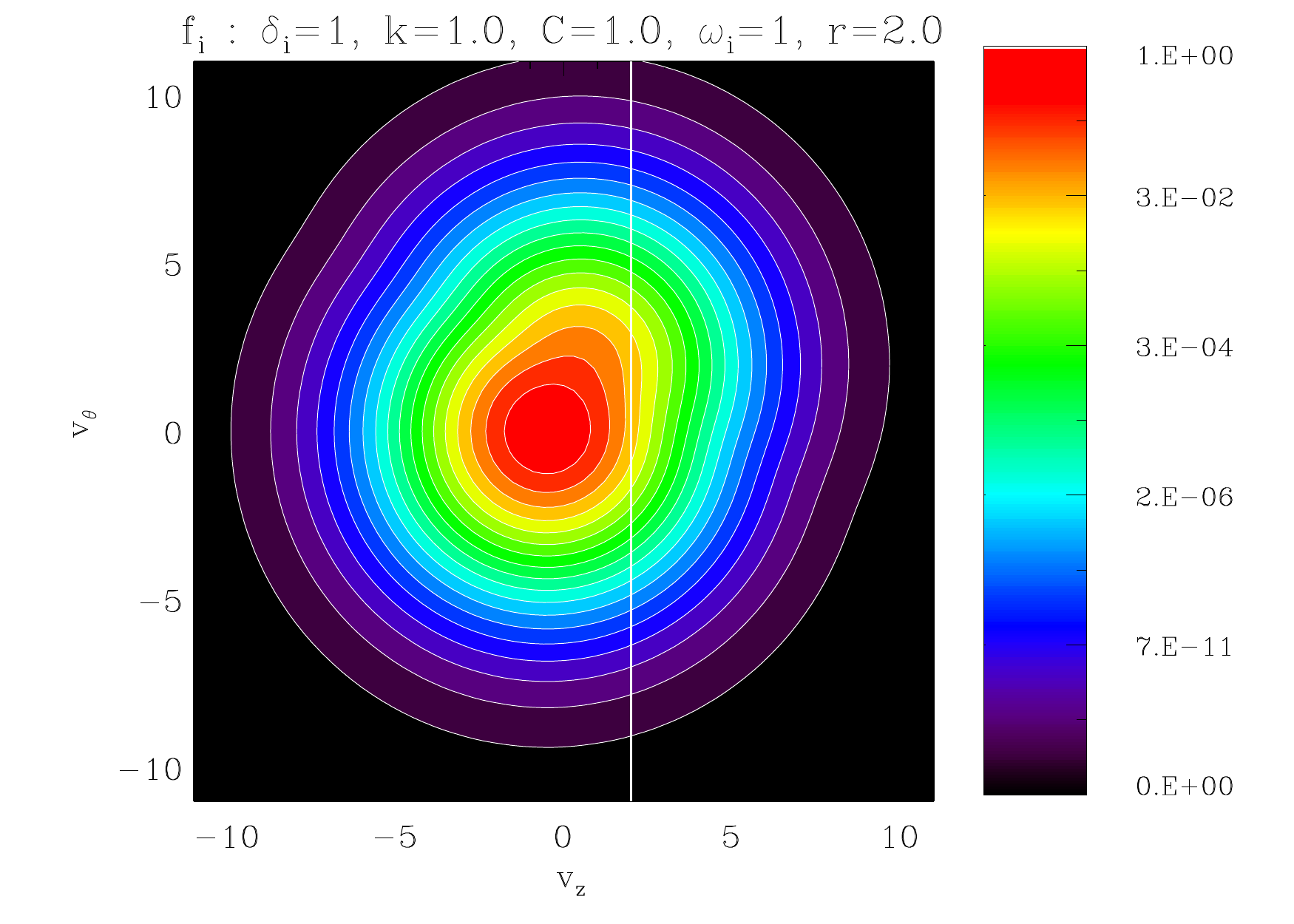}
        \caption{$(\tilde{\omega}_i,\tilde{r},C_i)=(1,2,1)$}
        \label{fig:5f}
    \end{subfigure}
        \begin{subfigure}[b]{0.245\textwidth}
        \includegraphics[width=\textwidth]{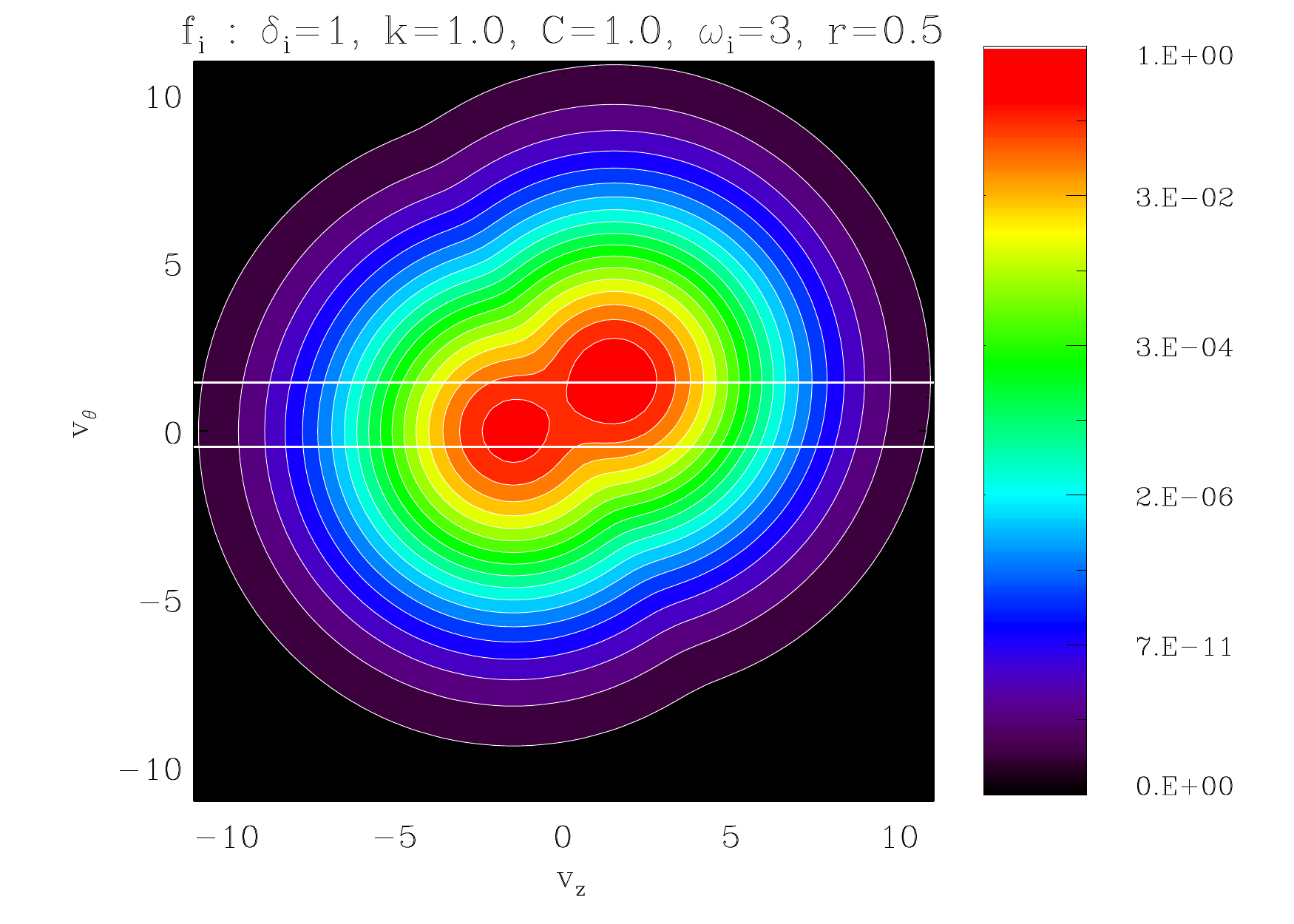}
        \caption{$(\tilde{\omega}_i,\tilde{r},C_i)=(3,0.5,1)$}
        \label{fig:5g}
    \end{subfigure}
     \begin{subfigure}[b]{0.245\textwidth}
        \includegraphics[width=\textwidth]{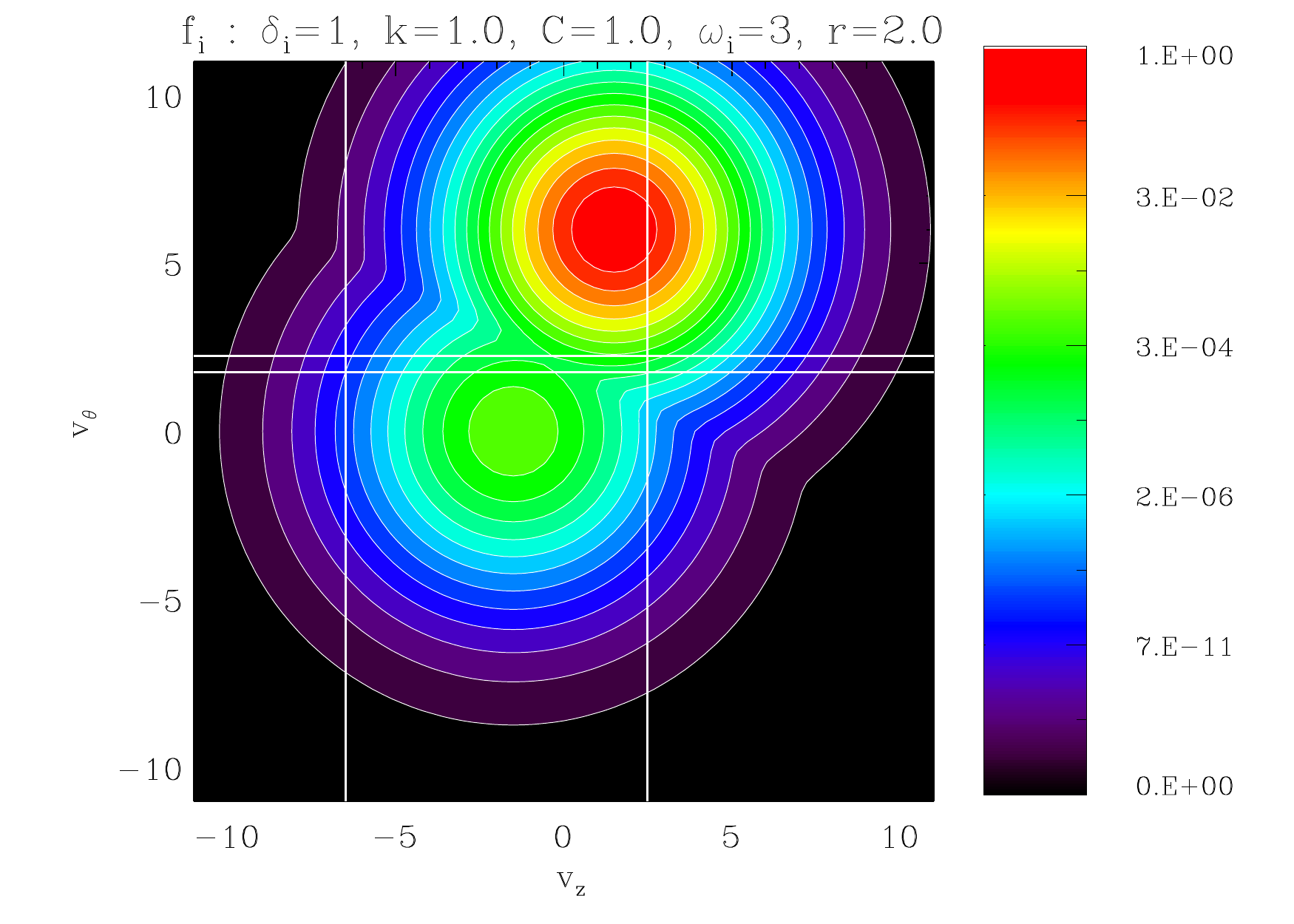}
        \caption{$(\tilde{\omega}_i,\tilde{r},C_i)=(3,2,1)$}
        \label{fig:5h}
    \end{subfigure}
    \caption{Contour plots of $f_i$ in $(\tilde{v}_z,\tilde{v}_{\theta})$ space for an equilibrium without field reversal ($k=1>0.5$), for a variety of parameters ($\tilde{\omega}_i,\tilde{r},C_i$) and $\delta_i=1$. The white horizontal/vertical lines indicate the regions in which multiple maxima in either the $\tilde{v}_z$ or $\tilde{v}_{z}$ directions can occur, if at all. A single line indicates that the `region' is a line.  }\label{fig:5}
\end{figure}

\begin{figure}
    \centering
    \begin{subfigure}[b]{0.245\textwidth}
        \includegraphics[width=\textwidth]{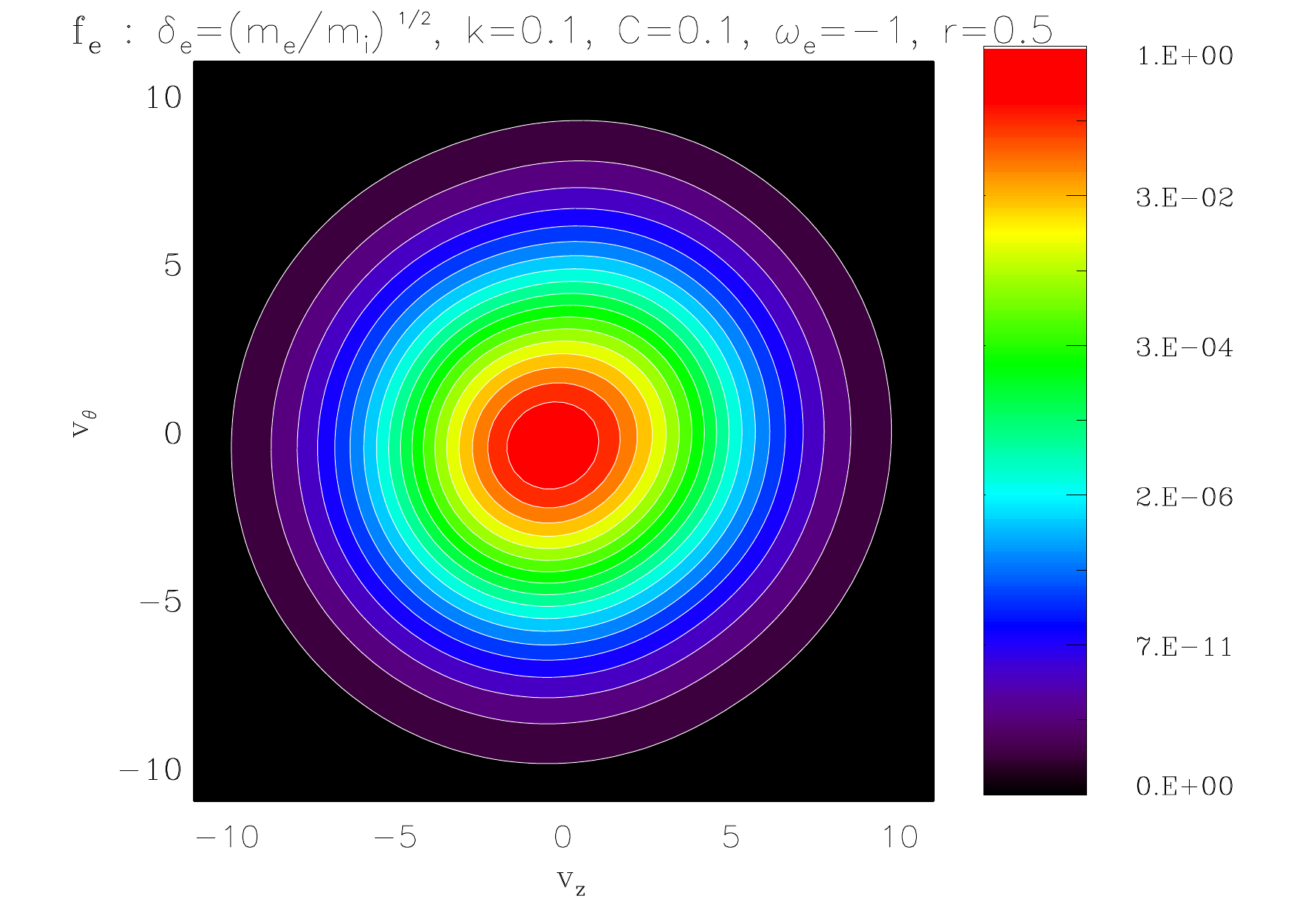}
        \caption{$(\tilde{\omega}_e,\tilde{r},C_e)=(-1,0.5,0.1)$}
        \label{fig:}
    \end{subfigure}
       \begin{subfigure}[b]{0.245\textwidth}
        \includegraphics[width=\textwidth]{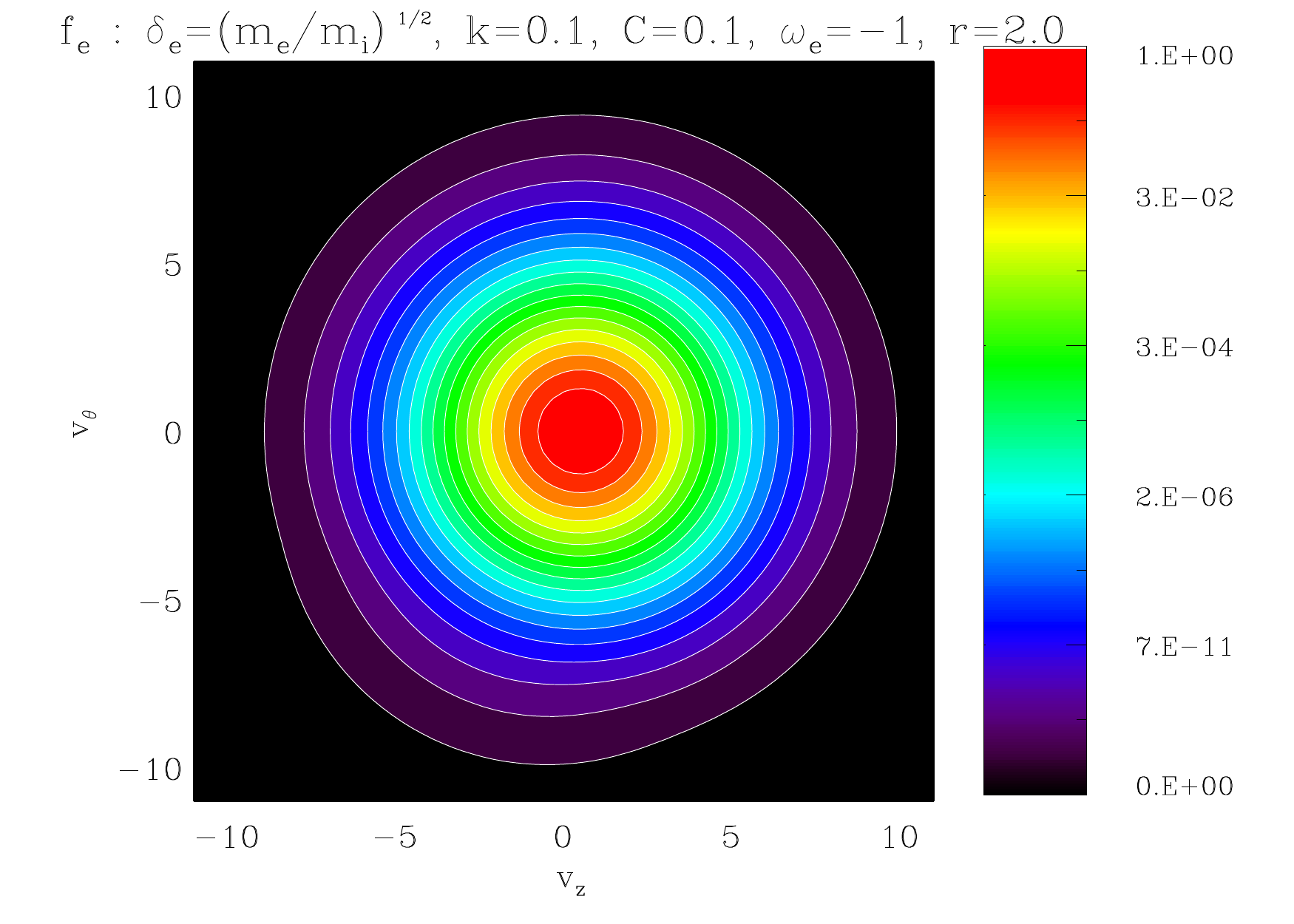}
        \caption{$(\tilde{\omega}_e,\tilde{r},C_e)=(-1,2,0.1)$}
        \label{fig:}
    \end{subfigure}
        \begin{subfigure}[b]{0.245\textwidth}
        \includegraphics[width=\textwidth]{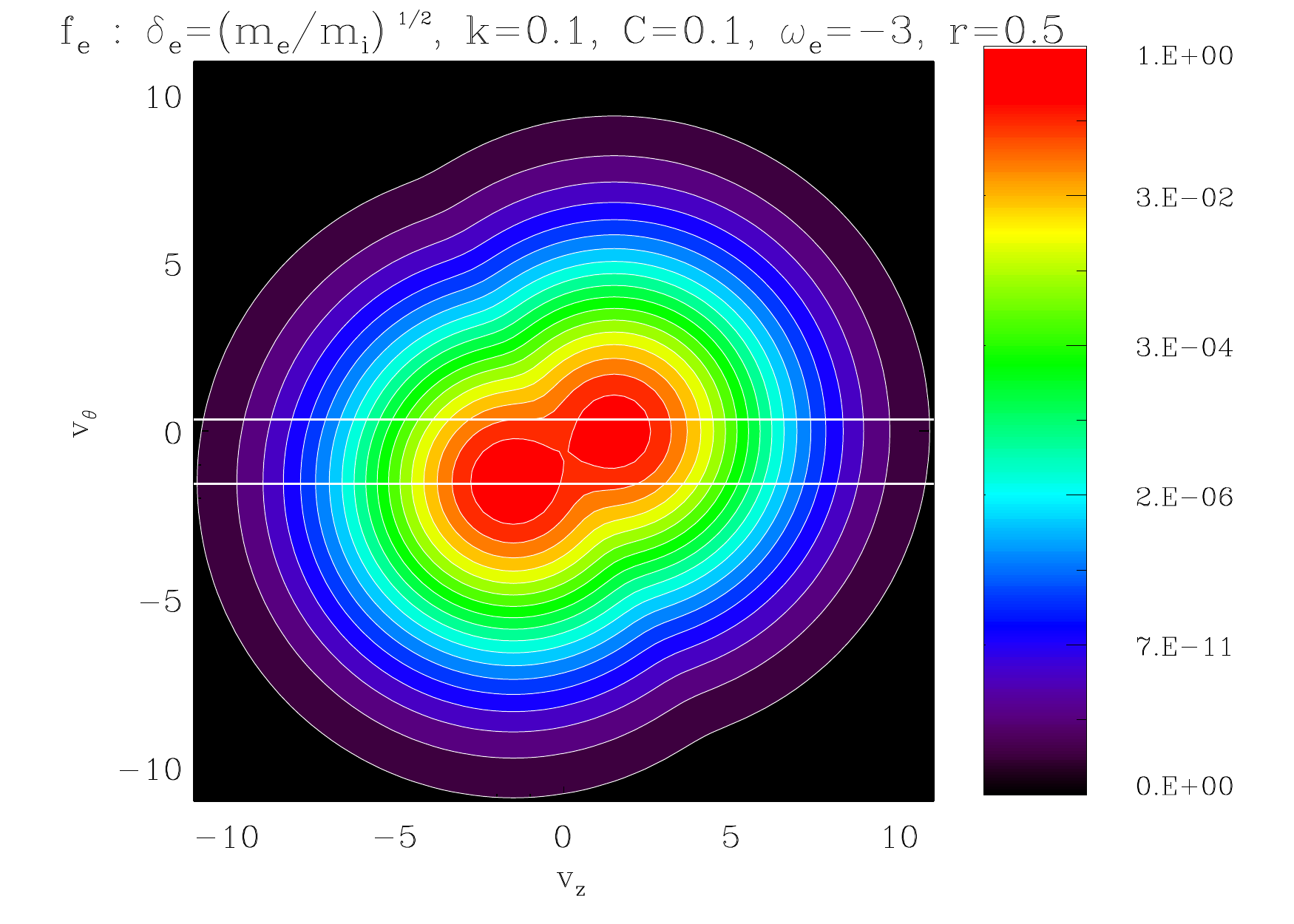}
        \caption{$(\tilde{\omega}_e,\tilde{r},C_e)=(-3,0.5,0.1)$}
        \label{fig:}
    \end{subfigure}
    \begin{subfigure}[b]{0.245\textwidth}
        \includegraphics[width=\textwidth]{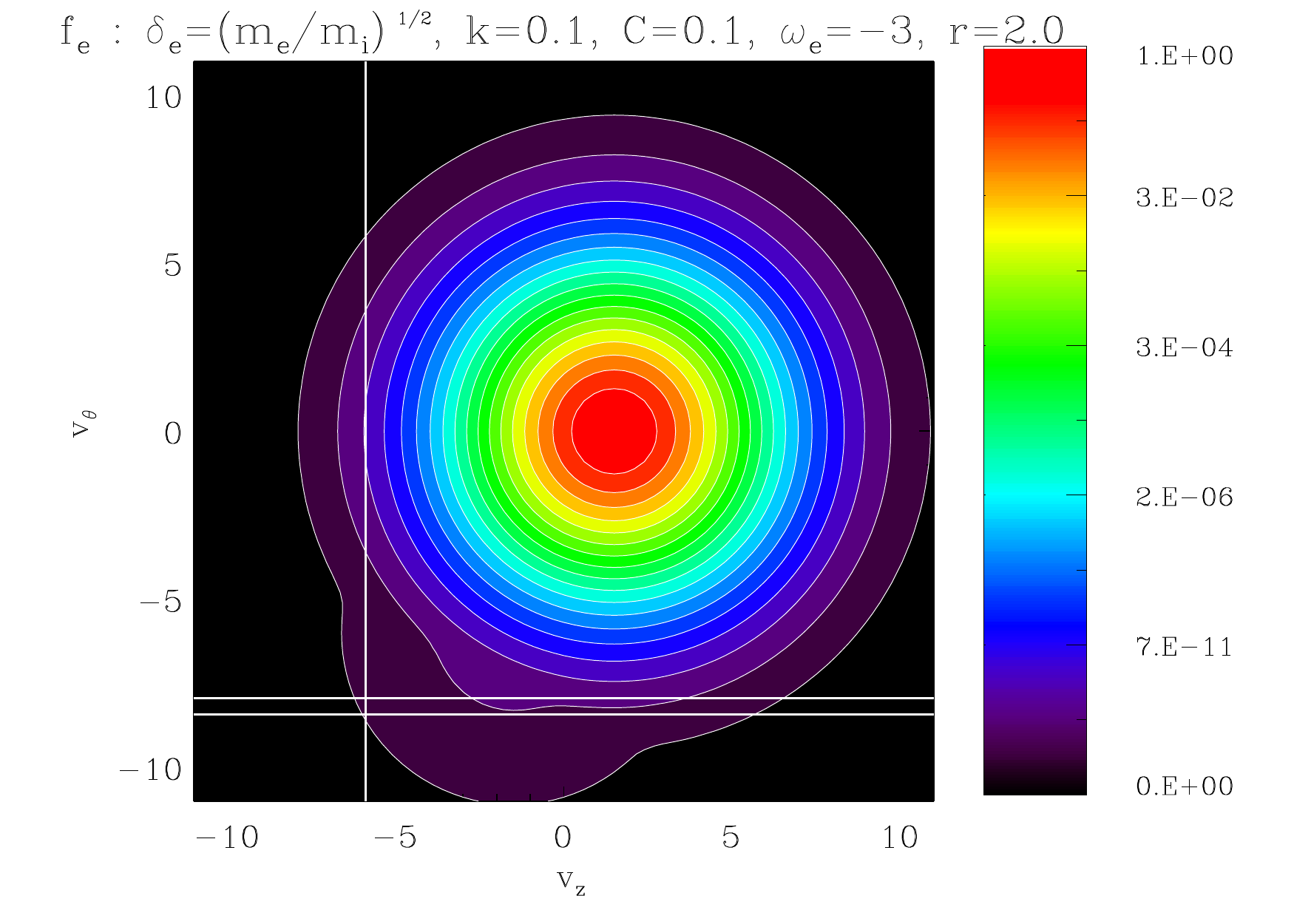}
        \caption{$(\tilde{\omega}_e,\tilde{r},C_e)=(-3,2,0.1)$}
        \label{fig:}
    \end{subfigure}
        \begin{subfigure}[b]{0.245\textwidth}
        \includegraphics[width=\textwidth]{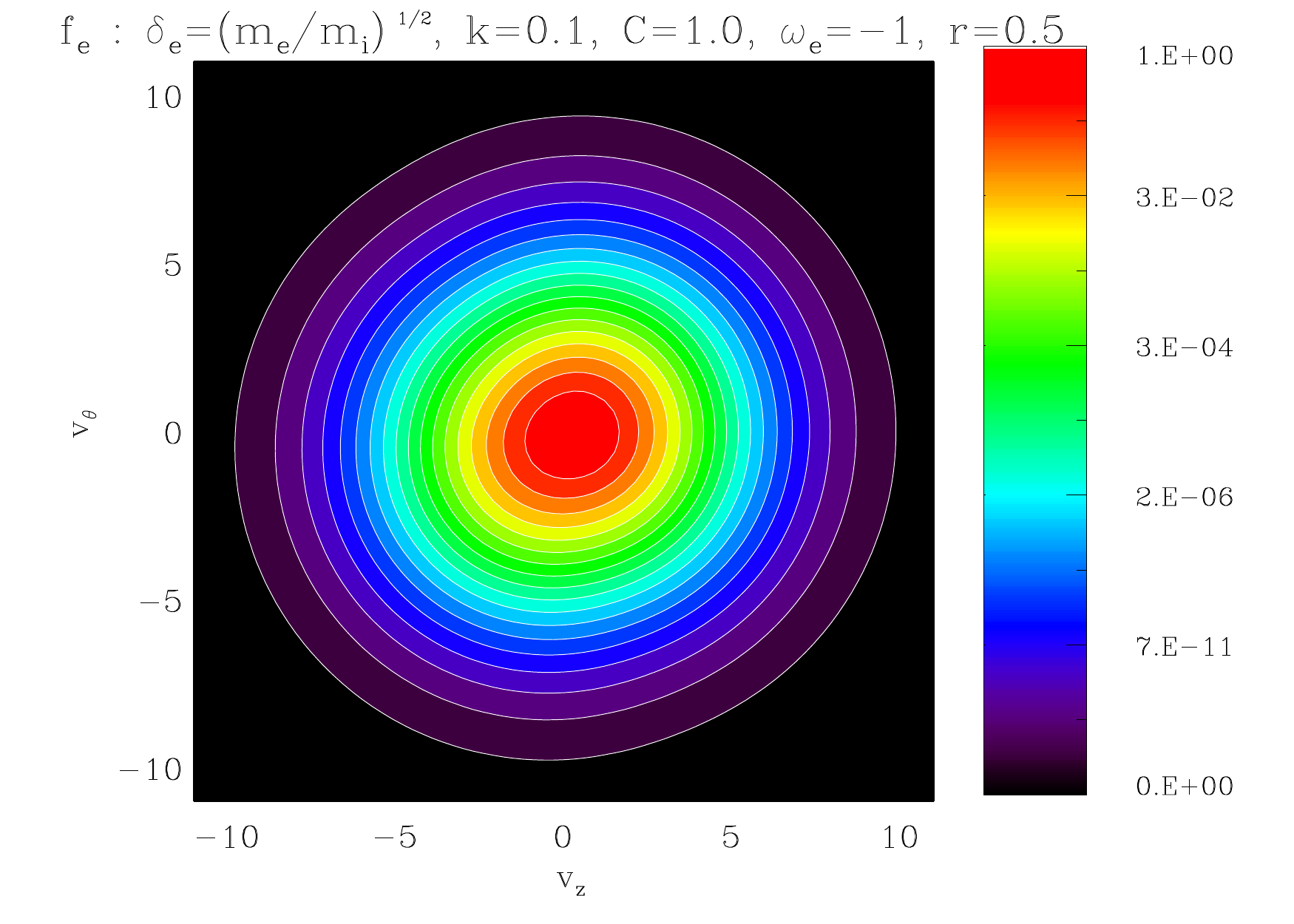}
        \caption{$(\tilde{\omega}_e,\tilde{r},C_e)=(-1,0.5,1)$}
        \label{fig:}
    \end{subfigure}
    \begin{subfigure}[b]{0.245\textwidth}
        \includegraphics[width=\textwidth]{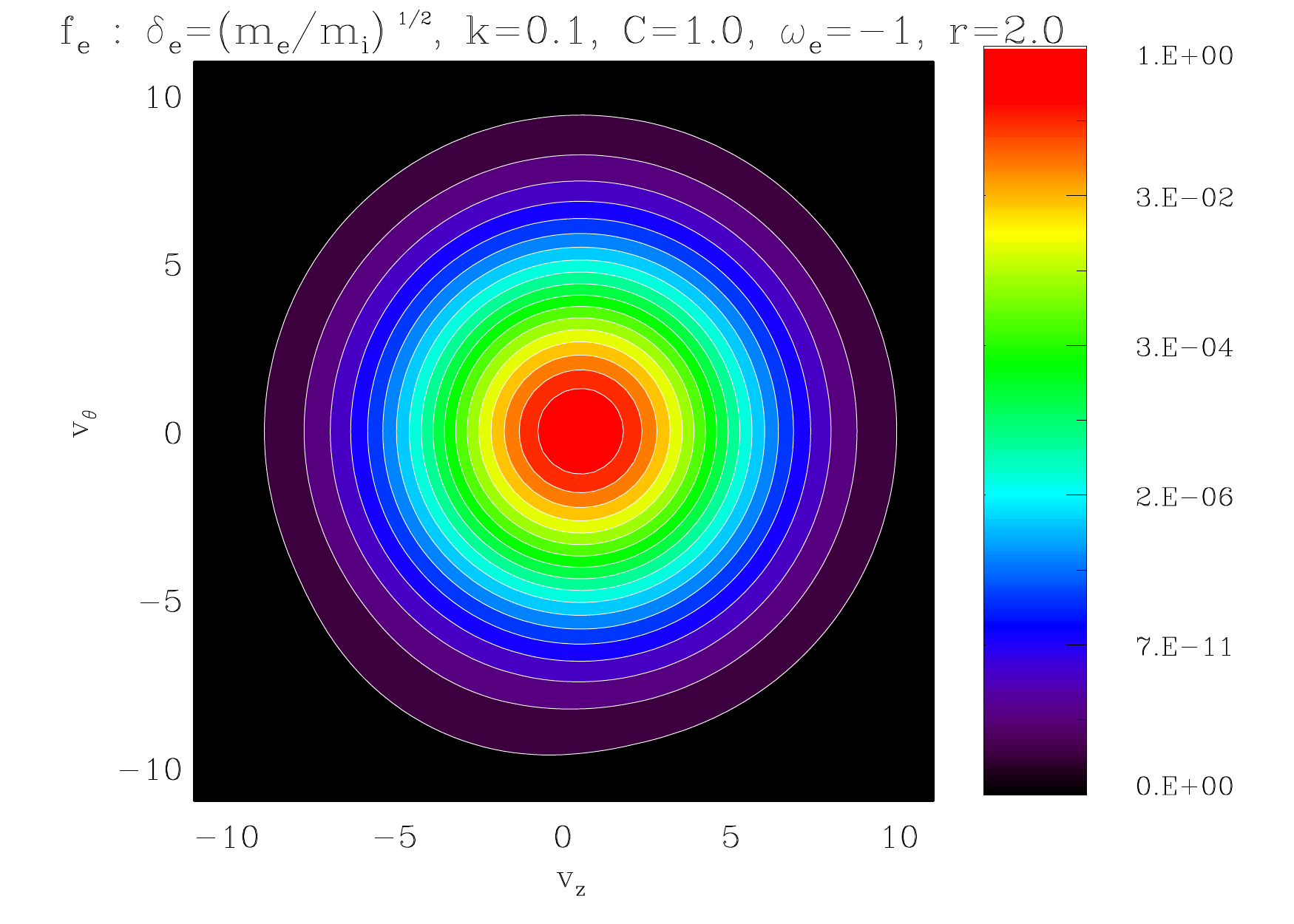}
        \caption{$(\tilde{\omega}_e,\tilde{r},C_e)=(-1,2,1)$}
        \label{fig:}
    \end{subfigure}
        \begin{subfigure}[b]{0.245\textwidth}
        \includegraphics[width=\textwidth]{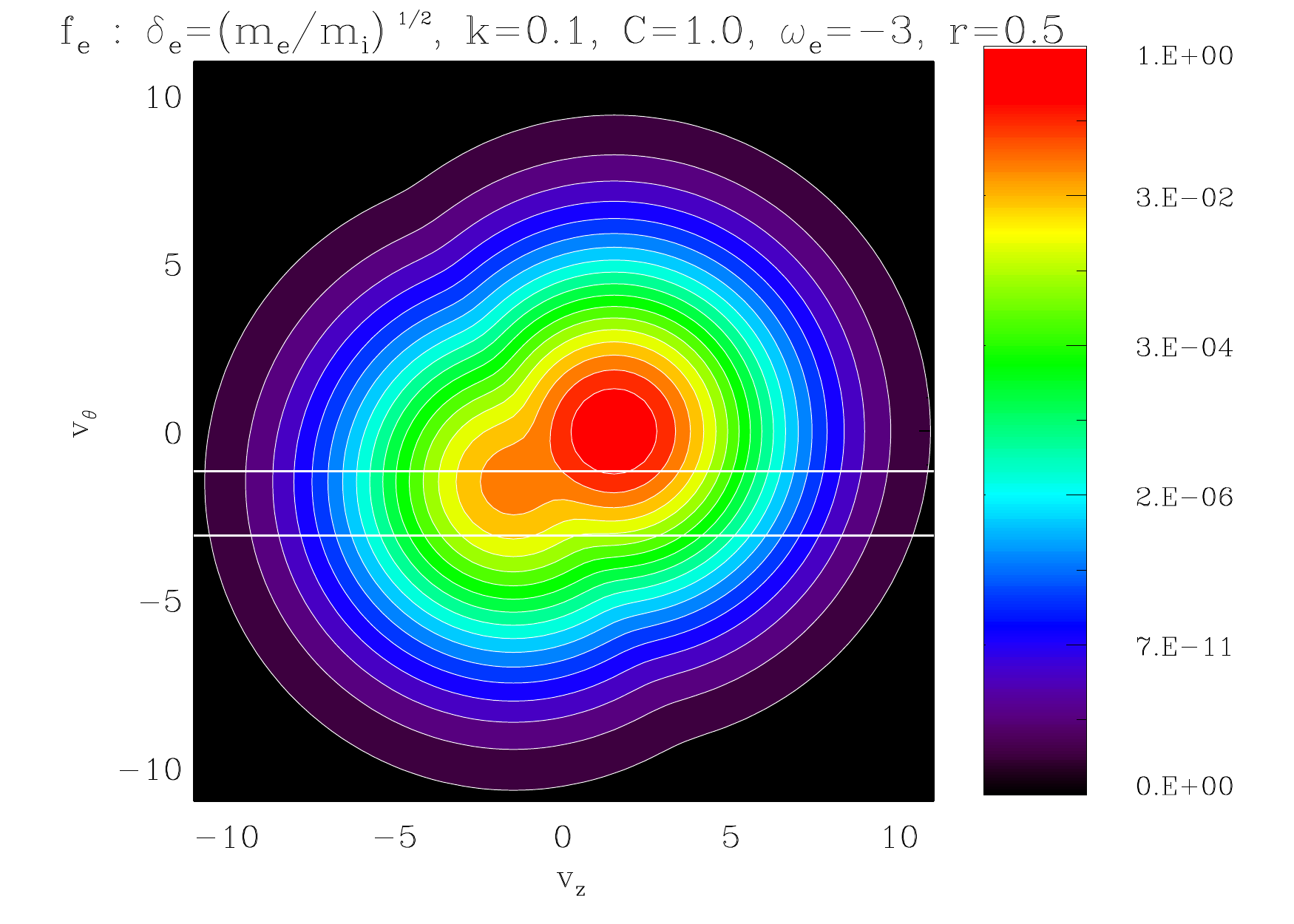}
        \caption{$(\tilde{\omega}_e,\tilde{r},C_e)=(-3,0.5,1)$}
        \label{fig:}
    \end{subfigure}
     \begin{subfigure}[b]{0.245\textwidth}
        \includegraphics[width=\textwidth]{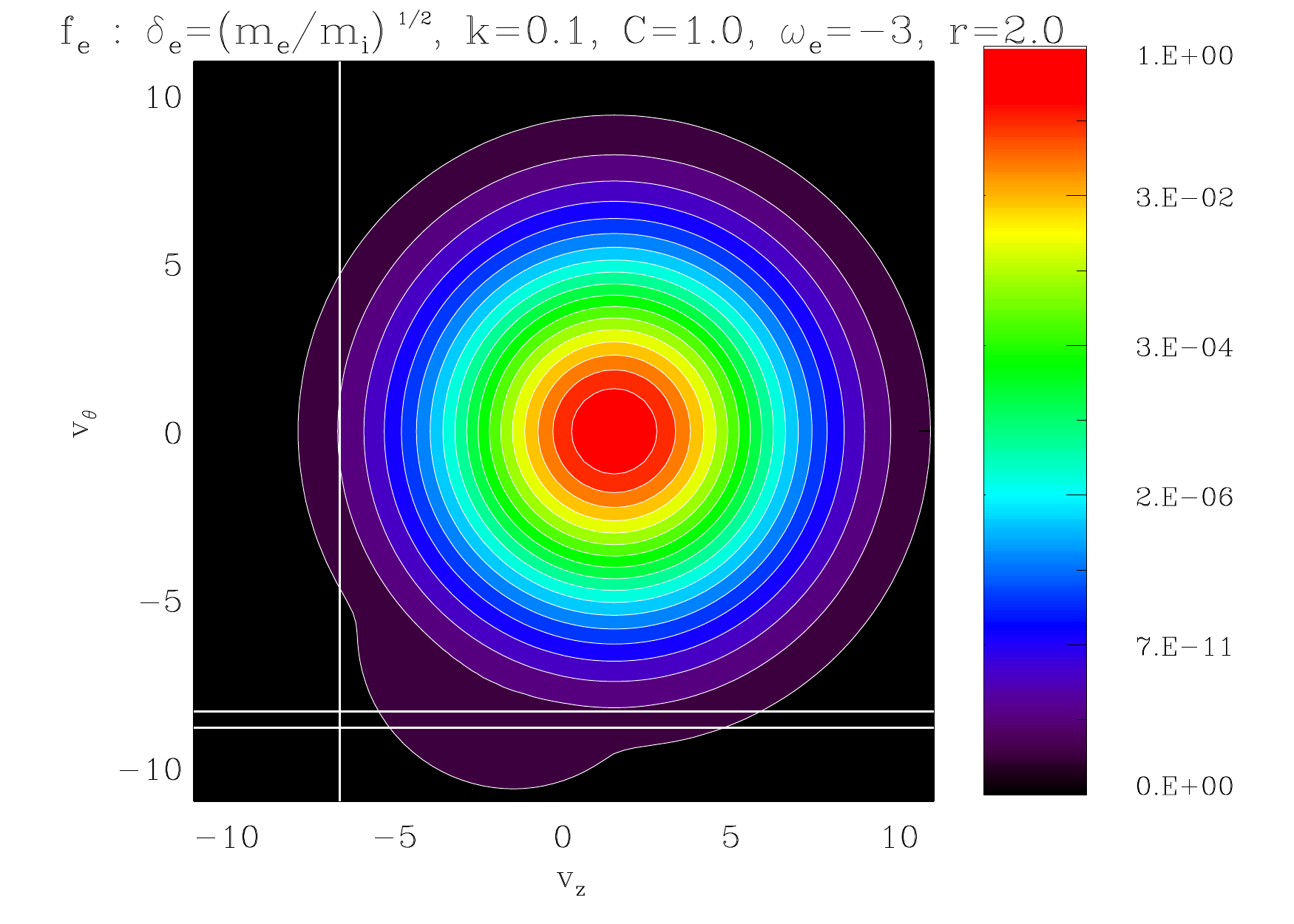}
        \caption{$(\tilde{\omega}_e,\tilde{r},C_e)=(-3,2,1)$}
        \label{fig:}
    \end{subfigure}
    \caption{    Contour plots of $f_e$ in $(\tilde{v}_z,\tilde{v}_{\theta})$ space for an equilibrium with field reversal ($k=0.1<0.5$), for a variety of parameters ($\tilde{\omega}_e,\tilde{r},C_e$) and $\delta_e\approx1/\sqrt{1836}$. The white horizontal/vertical lines indicate the regions in which multiple maxima in either the $\tilde{v}_z$ or $\tilde{v}_{z}$ directions can occur, if at all. A single line indicates that the `region' is a line.   }\label{fig:6}
\end{figure}

\begin{figure}
    \centering
    \begin{subfigure}[b]{0.245\textwidth}
        \includegraphics[width=\textwidth]{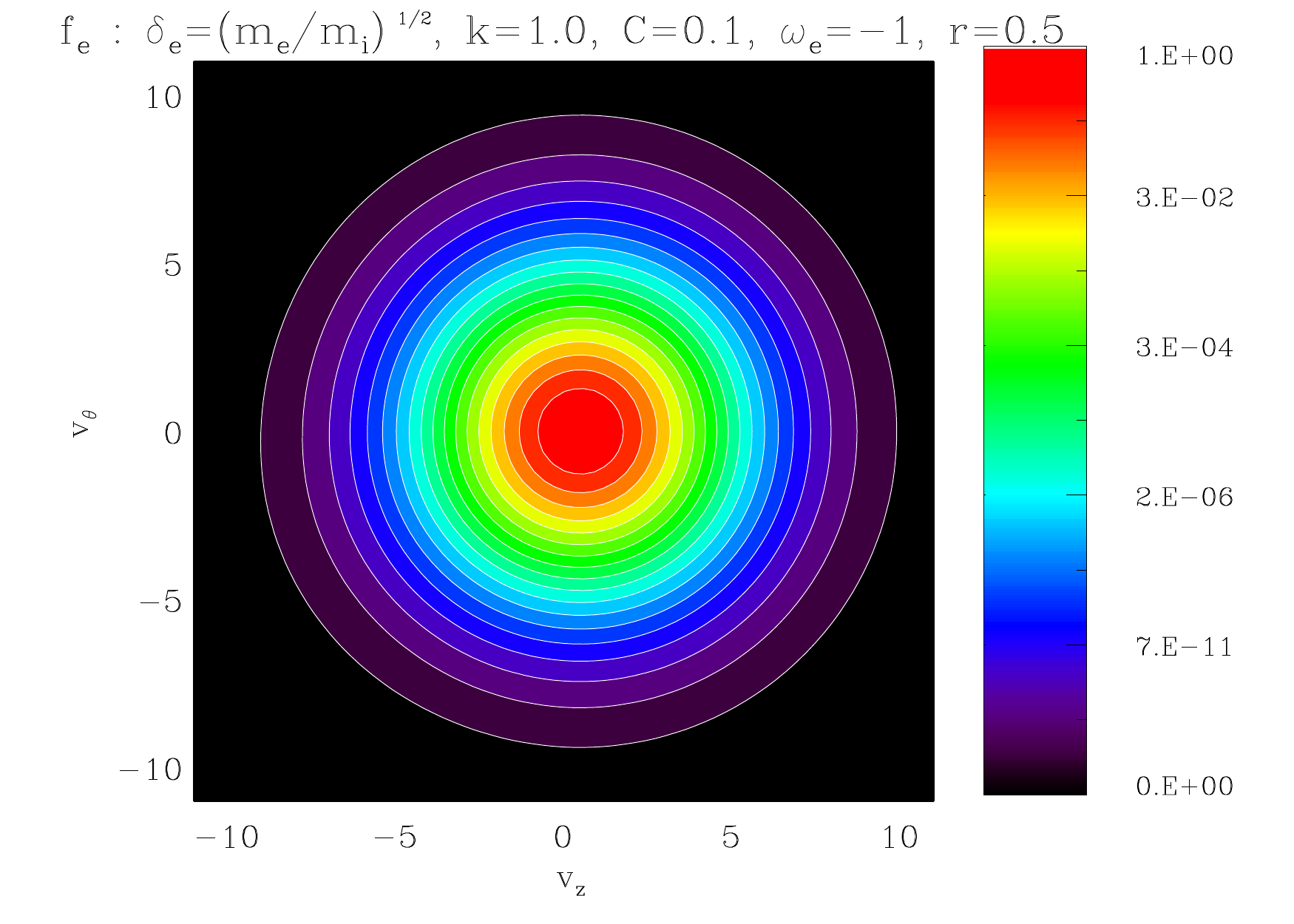}
        \caption{$(\tilde{\omega}_e,\tilde{r},C_e)=(-1,0.5,0.1)$}
        \label{fig:}
    \end{subfigure}
       \begin{subfigure}[b]{0.245\textwidth}
        \includegraphics[width=\textwidth]{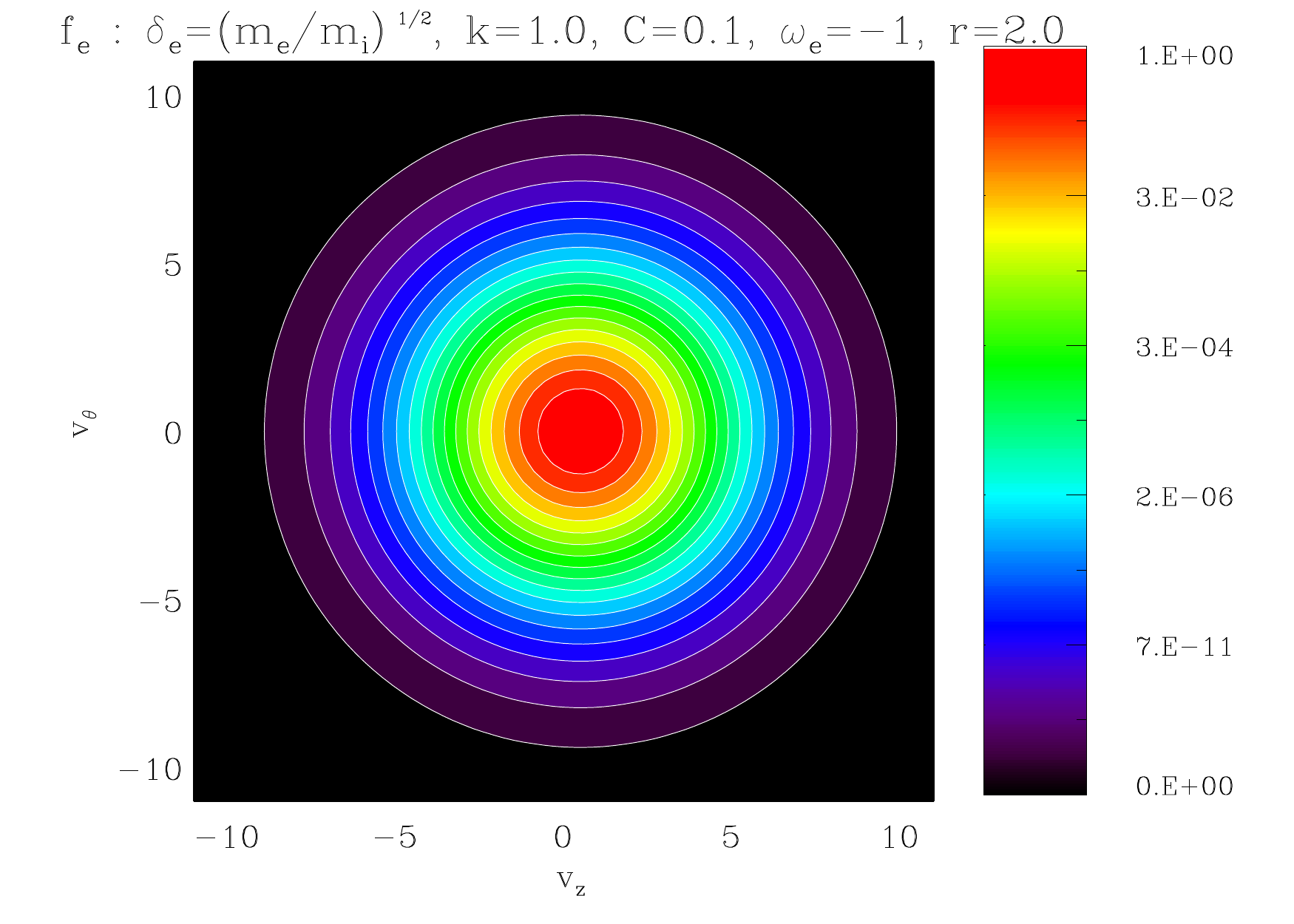}
        \caption{$(\tilde{\omega}_e,\tilde{r},C_e)=(-1,2,0.1)$}
        \label{fig:}
    \end{subfigure}
        \begin{subfigure}[b]{0.245\textwidth}
        \includegraphics[width=\textwidth]{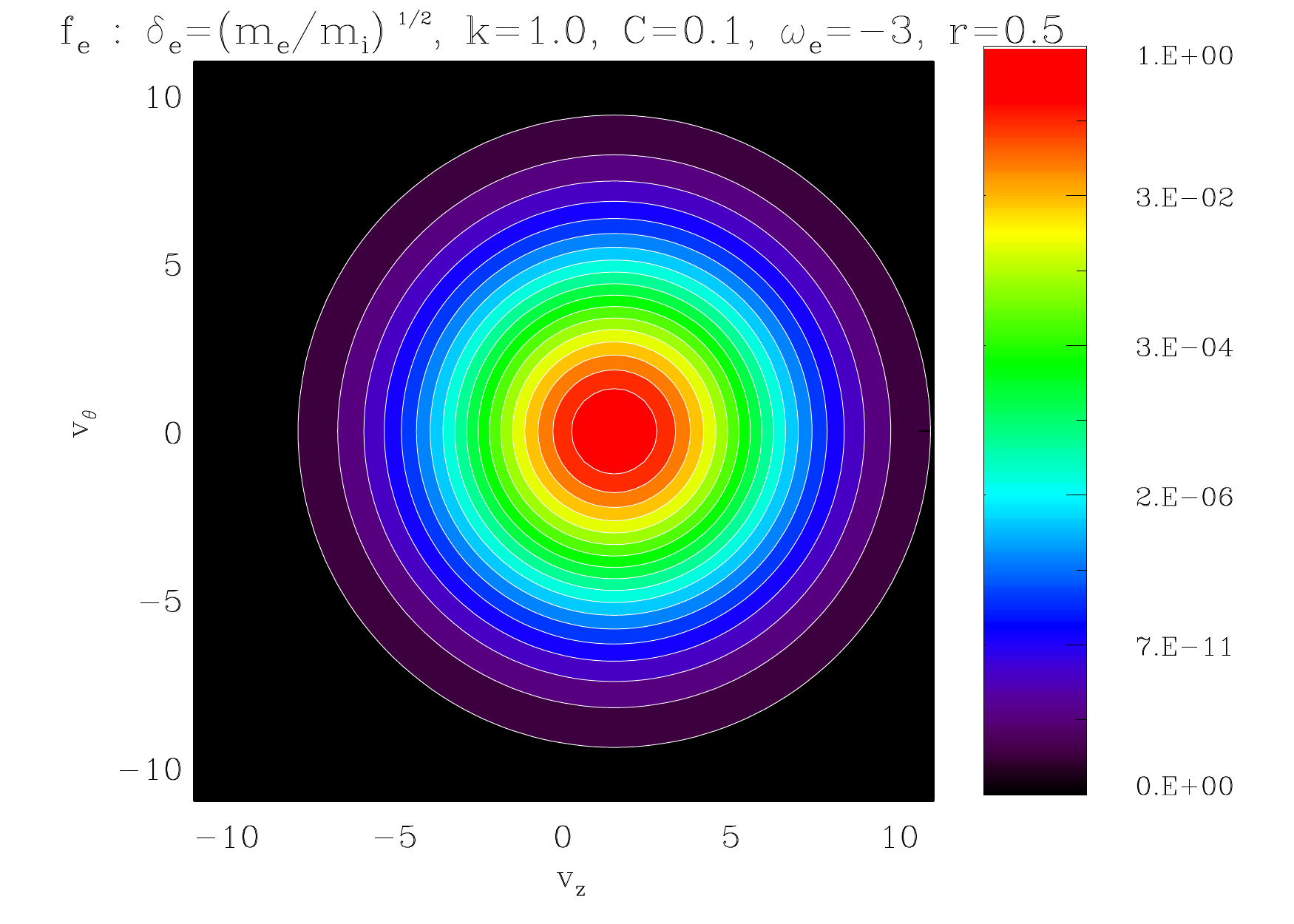}
        \caption{$(\tilde{\omega}_e,\tilde{r},C_e)=(-3,0.5,0.1)$}
        \label{fig:}
    \end{subfigure}
    \begin{subfigure}[b]{0.245\textwidth}
        \includegraphics[width=\textwidth]{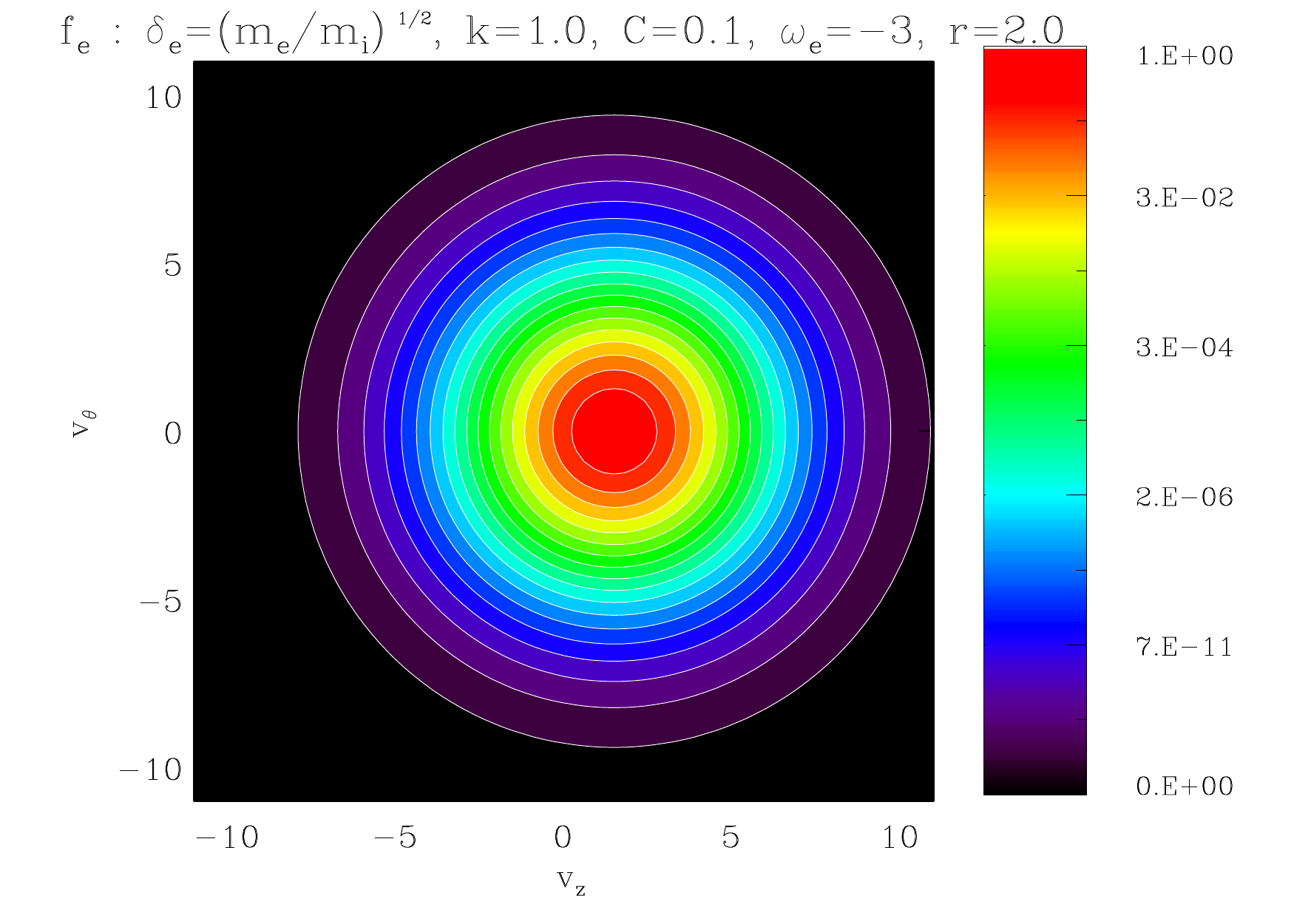}
        \caption{$(\tilde{\omega}_e,\tilde{r},C_e)=(-3,2,0.1)$}
        \label{fig:}
    \end{subfigure}
        \begin{subfigure}[b]{0.245\textwidth}
        \includegraphics[width=\textwidth]{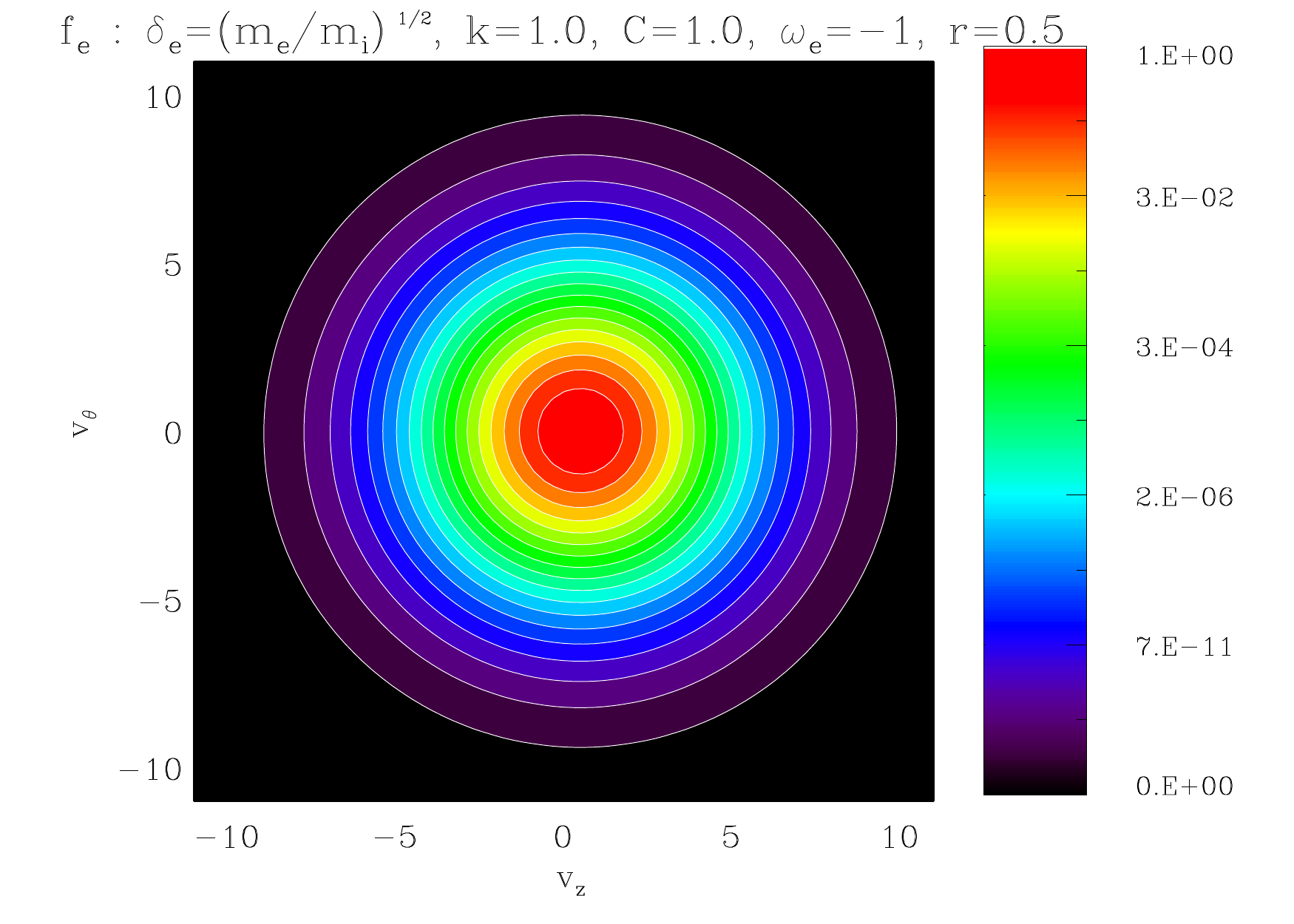}
        \caption{$(\tilde{\omega}_e,\tilde{r},C_e)=(-1,0.5,1)$}
        \label{fig:}
    \end{subfigure}
    \begin{subfigure}[b]{0.245\textwidth}
        \includegraphics[width=\textwidth]{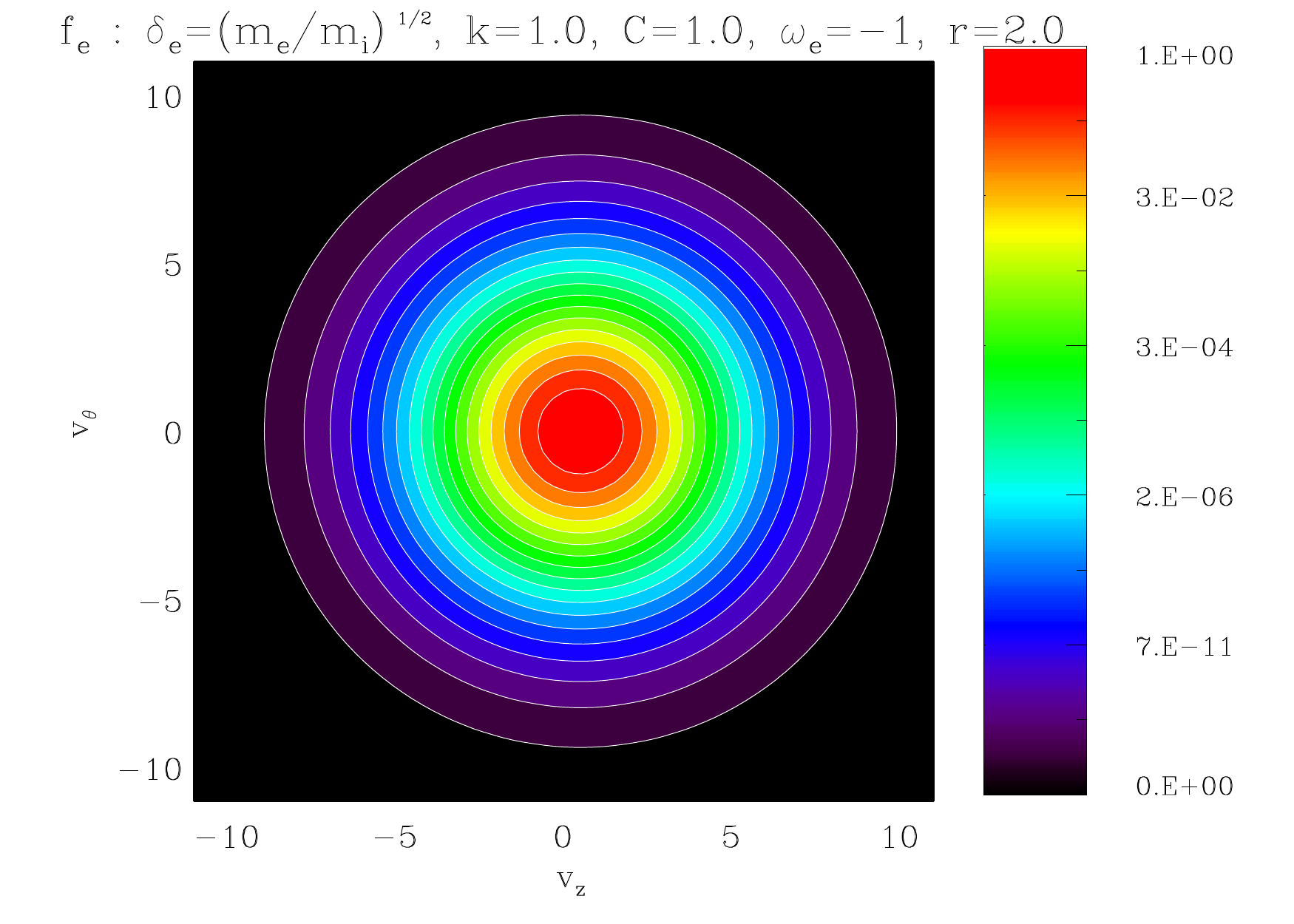}
        \caption{$(\tilde{\omega}_e,\tilde{r},C_e)=(-1,2,1)$}
        \label{fig:}
    \end{subfigure}
        \begin{subfigure}[b]{0.245\textwidth}
        \includegraphics[width=\textwidth]{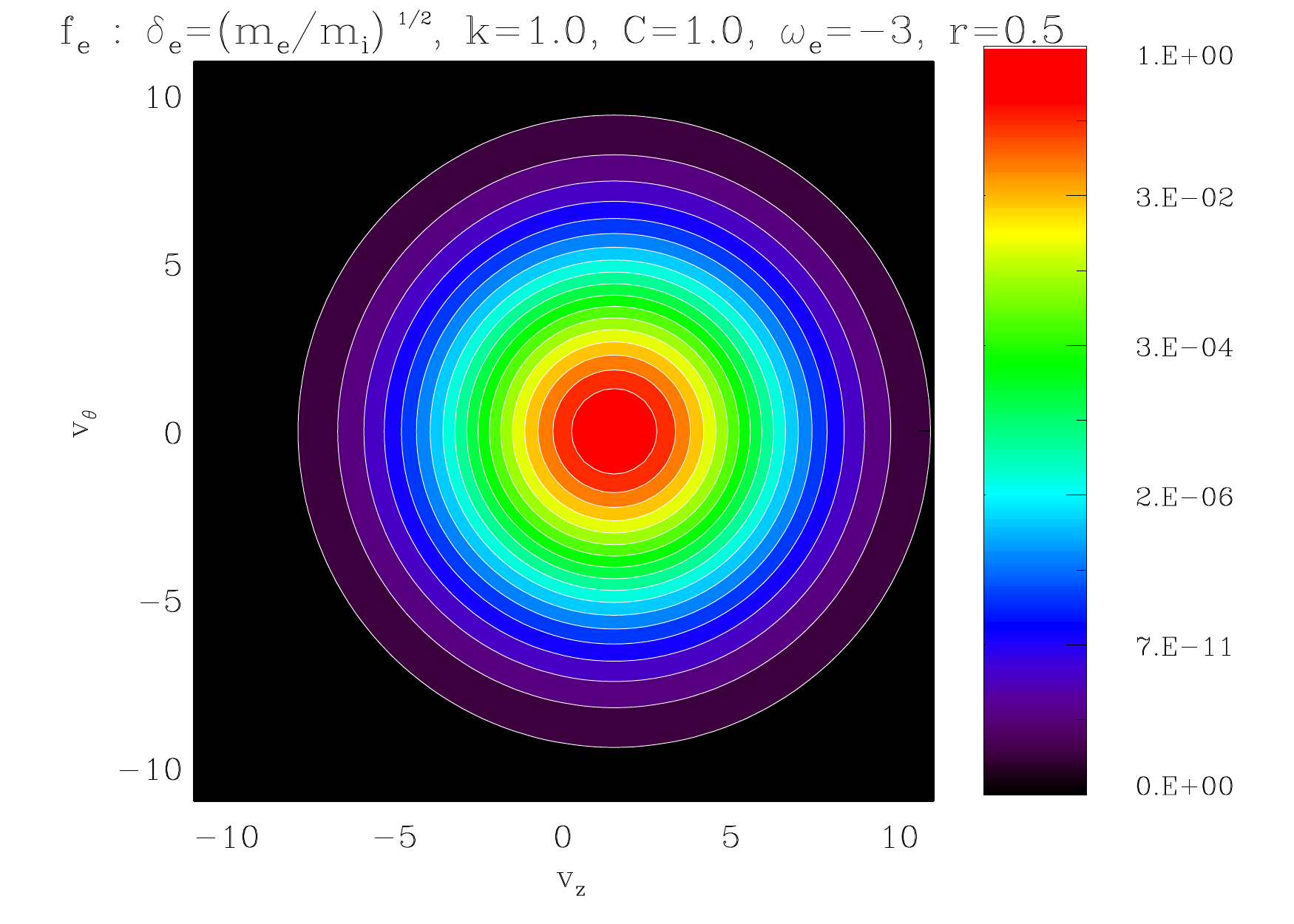}
        \caption{$(\tilde{\omega}_e,\tilde{r},C_e)=(-3,0.5,1)$}
        \label{fig:}
    \end{subfigure}
     \begin{subfigure}[b]{0.245\textwidth}
        \includegraphics[width=\textwidth]{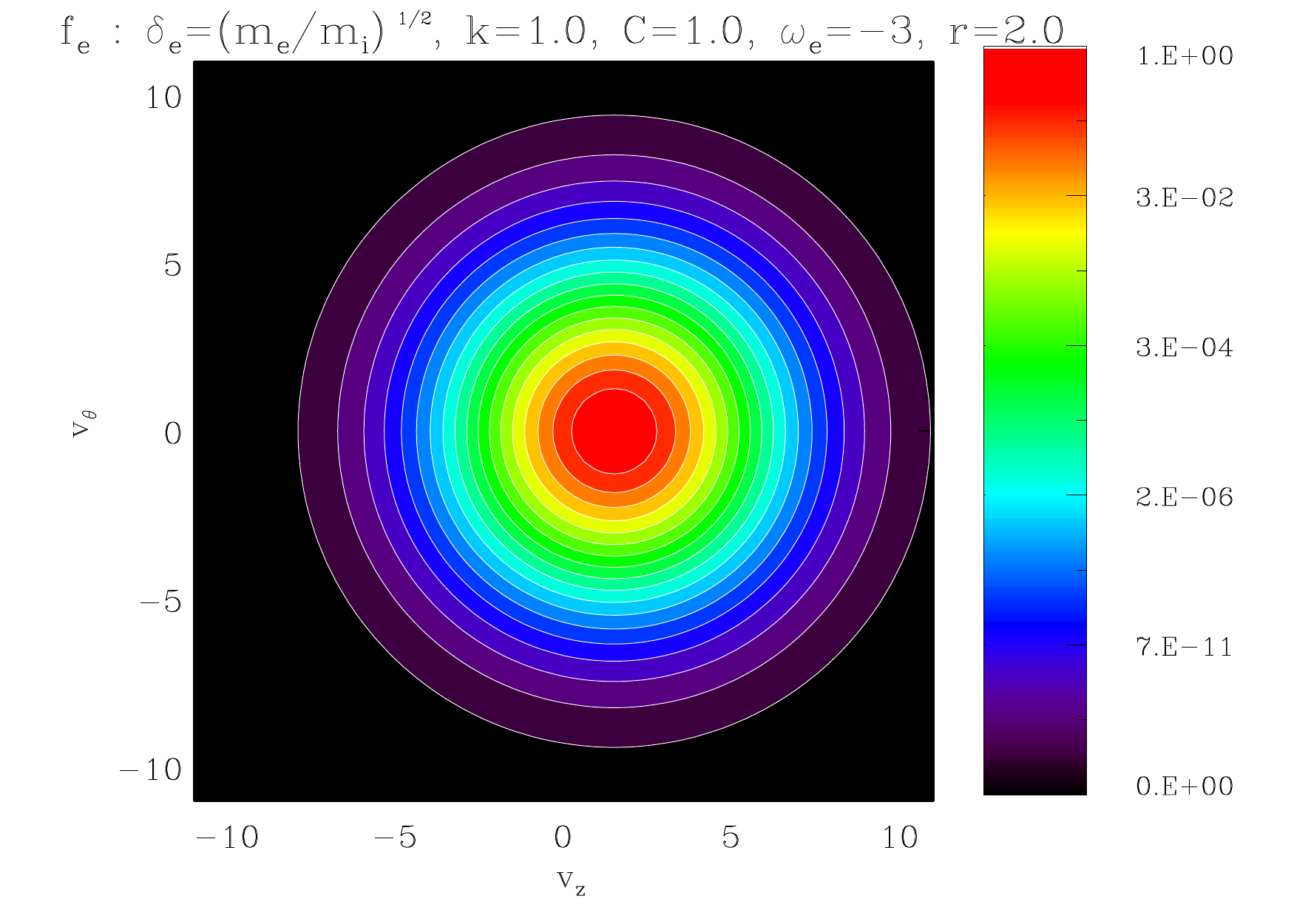}
        \caption{$(\tilde{\omega}_e,\tilde{r},C_e)=(-3,2,1)$}
        \label{fig:}
    \end{subfigure}
    \caption{      Contour plots of $f_e$ in $(\tilde{v}_z,\tilde{v}_{\theta})$ space for an equilibrium without field reversal ($k=1>0.5$), for a variety of parameters ($\tilde{\omega}_e,\tilde{r},C_e$) and $\delta_e\approx1/\sqrt{1836}$. Note that there are not any multiple maxima in this case.   }\label{fig:7}
\end{figure}

\end{widetext}

\clearpage
% Create the reference section using BibTeX:
%\bibliography{/user/oliver/Dropbox/biblio} %Maths
%\bibliography{/Users/Oliver/Dropbox/biblio} %Home windows
%\bibliography{/Users/OAllanson/Dropbox/biblio} %home mac

%merlin.mbs aipnum4-1.bst 2010-07-25 4.21a (PWD, AO, DPC) hacked
%Control: key (0)
%Control: author (8) initials jnrlst
%Control: editor formatted (1) identically to author
%Control: production of article title (-1) disabled
%Control: page (0) single
%Control: year (1) truncated
%Control: production of eprint (0) enabled
%

\end{document}